\documentclass{article}

\usepackage{graphicx} 
\usepackage{xcolor}
\usepackage[square]{natbib}
\usepackage{authblk}
\usepackage{lineno}
\usepackage[normalem]{ulem} 
\usepackage{siunitx} 
\usepackage[T1]{fontenc}
\usepackage[utf8]{inputenc} 
\usepackage{lmodern}

\newif\ifshowchanges
\showchangesfalse  

\ifshowchanges
  \newcommand{\add}[1]{\textbf{#1}}      
  \newcommand{\delete}[1]{\sout{#1}}     
\else
  \newcommand{\add}[1]{#1}               
  \newcommand{\delete}[1]{}              
\fi

\begin{document}



\title{Physicochemical controls on the compositions of the Earth and planets}

\author[1]{Paolo A. Sossi*$^,$}
\author[2]{Remco C. Hin}
\author[3]{Thorsten Kleine}
\author[4,5]{Alessandro Morbidelli}
\author[6]{Francis Nimmo}

\affil[1]{\small Institute of Geochemistry and Petrology, ETH Zürich, Clausiusstrasse 25, CH-8092, Zürich, Switzerland}
\affil[2]{\small Institute of Environmental Geology and Geoengineering, Consiglio Nazionale delle Ricerche (CNR), Via Sandro Botticelli 23, 20133, Milan, Italy}
\affil[3]{\small Max Planck Institute for Solar System Research, Justus-von-Liebig-Weg 3, 37077, Göttingen, Germany}
\affil[4]{\small Laboratoire Lagrange, Université Côte d'Azur, Observatoire de la Côte d'Azur, CNRS, CS 34229, F-06304, Nice, France}
\affil[5]{\small Collège de France, CNRS, PSL Université, Sorbonne Université, F-75014, Paris, France}
\affil[6]{\small Department of Earth \& Planetary Sciences, University of California Santa Cruz, CA 95064, Santa Cruz, USA}
\affil[*]{corresponding author. email: paolo.sossi@eaps.ethz.ch}

\maketitle


\Large \textbf{Abstract} \normalsize

Despite the fact that the terrestrial planets all formed from the protoplanetary disk, their bulk compositions show marked departures from that of material condensing from a canonical H$_2$-rich solar nebula. Metallic cores fix the oxygen fugacities ($f$O$_2$s) of the planets to between $\sim$5 (Mercury) and $\sim$1 log units below the iron-wüstite buffer, orders of magnitude higher than that of the nebular gas. Their oxidised character is coupled with a lack of volatile elements with respect to the solar nebula. Here we show that condensates from a canonical solar gas at different temperatures ($T_0$) produce bulk compositions with Fe/O (by mass) ranging from $\sim$0.93 ($T_0$ = 1250~K) to $\sim$0.81 ($T_0$ = 400~K), far lower than that of Earth at 1.06. Because the reaction Fe(s) + H$_2$O(g) = FeO(s) + H$_2$(g) proceeds only below $\sim$600 K, temperatures at which most moderately volatile elements (MVEs) have already condensed, oxidised planets are expected to be rich in volatiles, and vice-versa. That this is not observed suggests that planets $i)$ did not accrete from equilibrium nebular condensates and/or $ii)$ underwent additional volatile depletion/$f$O$_2$ changes at conditions distinct from those of the solar nebula. Indeed, MVEs in small telluric bodies (Moon, Vesta) are consistent with evaporation/condensation at $\Delta$IW-1 and $\sim$1400-1800 K, while the extent of mass-dependent isotopic fractionation observed implies this occurred near- or at equilibrium. On the other hand, the volatile-depleted elemental- yet near-chondritic isotopic compositions of larger telluric bodies (Earth, Mars) reflect mixing of bodies that had themselves experienced different extents of volatile depletion, overprinted by accretion of volatile-undepleted material. On the basis of isotopic anomalies in Cr- and Ti in the BSE, such undepleted matter has been proposed to be CI chondrites, which would comprise 40~\% by mass if the proto-Earth were ureilite-like. However, this would result in an overabundance of volatile elements in the BSE, requiring significant loss thereafter, which has yet to be demonstrated. On the other hand, 6~\% CI material added late to an enstatite chondrite-like proto-Earth would broadly match the BSE composition. \add{However,} because the Earth is an end-member in isotopic anomalies of heavier elements, no combination of existing meteorites alone can account for its chemical- and isotopic composition. 
Instead, the Earth is most likely made partially or \delete{wholly}\add{essentially entirely} from an NC-like missing component. If so, the oxidised-, yet volatile-poor nature of differentiated bodies in the inner solar system, including Earth and Mars, is a property intrinsic to the NC reservoir.


\section{Introduction}

The present-day locations and properties of the terrestrial planets, Mercury, Venus, Earth and Mars (section \ref{sec:observations}), are the culmination of chemical- and physical processes occurring through space and time that followed the collapse of the molecular cloud. 
Observations of young stars show that these systems are governed by the interaction of an H$_2$-rich gas with the central star (section \ref{sec:disks}).  Temporal evolution from a `protostellar' or Class 0 stage in which molecular cloud material infalls to the central star occurs within $\sim$10$^4$ to 10$^5$ yr (Class I) and forms envelopes extending from $\sim$10$^3$ to 10$^4$ au. Over timescales of 10$^6$ to 10$^7$ yr, these envelopes develop into the more common T-Tauri stars, which are invariably surrounded by a circumstellar disk of 1 -- 10 Jupiter masses \citep{montmerle2006}. The lifetimes of such disks range from $\sim$1 -- 10 Myr \citep{hartmann1998,hillenbrand2008}, with that of our Sun having been inferred to have dissipated by 3.8 Myr after the condensation of the first solids by measurements of magnetic fields preserved in angrite meteorites \citep{wang2017lifetime}. \\

The theory for the evolution of disks was initially developed assuming that turbulence is the main source of viscosity in disks, so that viscosity is proportional to the product of the square of scale height (the size of the largest turbulent eddies) and of the orbital frequency \citep{lynden1974evolution}. However, it now appears that turbulent viscosity can be high only in the initial phase of the disk, due to ionising temperatures \citep{BalbusHawley1998}, Reynolds stresses due to gas infalling onto the disk \citep{Kuznetsova_2022} and/or gravitational instabilities \citep{rafikov2016}. Once this initial phase is over, turbulence becomes too weak in most parts of the disk to account for the observed gas accretion rates \citep[e.g.,][see section \ref{sec:disks}]{Cassen2006}. This mismatch implies an alternative driver of accretion in protoplanetary disks. The strong X-ray emission of T-Tauri stars is testament to the magnetisation of both the star and disk \citep{feigelsonmontmerle1999}. It is now understood that the magnetic effects not only govern the collapse of gas from the initial cloud onto the protostar and its disk \citep{matsumototomisaka2004,joos2012protostellar,krumholzfederrath2019}, but also the transport of gas through the disk thanks to the removal of angular momentum in disk winds \citep{suzuki2009disk,bai_stone2013,Gressel2015}. Disk evolution under disk winds \citep{tabone2022secular} leads, on a typical timescale of $\sim$5~Myr, to accretion rates that are sufficiently low so as to result in the removal of the remaining gas by photoevaporation \citep{alexander2006photoevaporation}. \\


Chemically, these protoplanetary disks are composed predominantly of H$_2$ ($\sim$91 mol.~\%) and He ($\sim$8 mol~\%), with diminishing quantities of other major components, namely, in order of decreasing abundance O, C, Ne, N, Mg, Si and Fe \citep{lodders2010}. Relative to the protoplanetary disk (and hence the Sun), the terrestrial planets are impoverished in H, C, N, He and other rare gases \citep{aston1924} that attests to their accretion from predominantly condensed matter rather than from capture of large quantities of nebular gas. The composition of this material has typically been reconciled in the framework of equilibrium condensation \citep[][see section \ref{sec:nebular_condensation}]{larimer1967,grossmanlarimer1974,lodders2003}. Chemical equilibrium, though not mandated, is invoked for these calculations throughout the entire temperature range over which condensation occurs at an assumed, and constant, pressure of 10$^{-4}$ bar, taken to be representative of the midplane\add{ for a Sun-like star}. \\ 

The first solids predicted to have condensed in these calculations are analogous to calcium-aluminium-rich inclusions (CAIs) preserved in some meteorites \citep{macpherson2014}. They are also the oldest materials to have originated within the Solar System \citep[establishing the $t_0$ age in relative chronology,][]{amelin2002lead,connelly2012pb}, lending credence to the concept of equilibrium condensation. The `matrix' component of chondrites may also result from near-complete condensation of the nebular gas (save for the ice-forming elements; H, C, N, O and the noble gases) \citep{grossmanlarimer1974}. The third component, chondrules, may have originally condensed from the solar nebula, but subsequently experienced partial melting \citep[e.g.,][]{hewins2005experimental,ebel2018,hellmann2020}. Mixtures between these components likely gave rise to the chemical variability observed in chondrites \citep{humayuncassen2000, braukmuller2018, alexander2019}, yet the nature of the processes producing such mixtures, and whether they reflect nebular-wide gradients, remains unclear \citep{li2023tempdisc,boyce2024large}. \\

The importance of the chondrites lies in their adoption, to first order, as the building blocks of the planets \citep{ringwood1966chemical,safronov1969}. While the constancy of the ratios of refractory lithophile elements (RLEs) to within $\sim$10~\% among chondrites substantiates this approximation, the abundances of the major planet-forming elements (Fe, Mg, Si, O) vary markedly \citep[up to a factor $\sim$2,][]{yoshizaki_mcdonough2020}. In detail, 
the core/mantle ratios of rocky bodies, which range from $\sim$0.65 for Mercury to 0.01 for the Moon by mass \citep[section \ref{sec:observations},][]{ringwood1966chemical,righter2006}, contrasts with the range expected from chondrites, at roughly 0.2 -- 0.3 \citep[e.g.,][]{bercovici2022effects}. Indeed, no set of mixtures of chondrites is readily able to reproduce the chemical compositions of the planets \citep{drakerighter2002,campbelloneill2012}, which may instead reflect chemical transformations unique to growing planets (section \ref{sec:planet_acc}).  \\


One such macroscopic difference is the differentiation of planetary bodies into a core and mantle, implying they experienced melting. 
Moreover, the terrestrial planets, as well as small telluric bodies (such as Vesta or the Moon), are depleted in volatile elements in a manner unlike that observed in chondritic meteorites \citep{witt-eickschen2009,mezger2021accretion}. \cite{sossi2022stochastic} noted that the depletion of elements as a function of their volatilities approximates a cumulative normal distribution, suggesting that rocky bodies accreted by the stochastic combination of many smaller bodies that each experienced volatile depletion at different temperatures. Whether a heliocentric gradient in the temperatures, and hence compositions of bodies is required is unclear \citep[see also][]{palme2000heliocentric}. 
This observation only supports the notion that mixing of material with disparate thermal histories led to the formation of the planets, but does not speak to the locus or conditions at which volatile depletion occurred. \\

An enduring but as yet inconclusive thesis states that the volatile depletion that affected small telluric bodies occurred under different physicochemical conditions to those prevailing during the formation of chondrites \citep[][section \ref{sec:chemistry_accretion}]{kreutzberger1986,wulf1995,oneillpalme2008, visscherfegley2013,norris_wood2017,sossi2019evaporation}. The observation that at least the Earth, with a single-stage Hf-W age of 34$\pm$3 Myr \citep{kleine_walker2017tungsten} experienced protracted growth permits the possibility that melting and evaporation took place \textit{after} dispersal of the nebular gas and hence under more oxidised (i.e., higher $f$O$_2$) conditions. Evidence for such `post-nebular' evaporation may be found in 
the Mn/Na ratio, which is elevated in all small telluric bodies with respect to chondrites, while the bulk silicate Earth falls within the chondritic range \citep{oneillpalme2008,siebert2018}. \\

The mass-dependent variations in the stable isotopic compositions of these elements provide another means with which to interrogate the origin of volatile depletion among planetary materials (section \ref{sec:chemistry_accretion}). 
The coupled isotopic variations in Fe, Si and potentially Mg to superchondritic (i.e., heavy) values suggest that volatile depletion processes, rather than core formation, were most likely responsible for the range of isotopic compositions of planetary materials \citep{poitrasson2004iron,pringle2014silicon,dauphas2015, sossi2016iron,hin2017magnesium, bourdon2018isotope,sossi_shahar2021}. 
On the other hand, the mass-dependent isotopic ratios of moderately volatile elements, such as K and Zn, are near-chondritic in the Earth and Mars \citep{wang2016, sossi2018zinc,paquet2023origin}, whereas small telluric bodies, such as Vesta, are isotopically heavy \citep[or extremely light in the case of the angrite parent body,][]{hu2022potassium}. These observations support the inference of \cite{sossi2022stochastic}, in which volatile element depletion affecting small telluric bodies generates isotopic fractionation, which is then progressively overprinted by mixing as the body accretes isotopically unfractionated, volatile-undepleted material.   \\

Such undepleted material is often equated with CI chondrites, an hypothesis reinforced by the discovery, in 2011 \citep{warren2011stable}, of two discrete clusters of meteorites in their isotopic compositions of Cr and Ti \citep{trinquier2007cr,trinquier2009origin}, when normalised to mass-dependent variations. These observations prompted the proposal that two \delete{physically isolated }reservoirs \delete{evolved contemporaneously}\add{existed} in the early Solar System; the NC and CC groups (section \ref{sec:provenance}). \add{That this dichotomy is likely spatial rather than temporal in nature comes from the overlap in accretion ages of both NC- and CC bodies \citep{kruijer2017age,bollard2017}, with the former presumed to represent the inner- and the latter the outer solar system. Furthermore, within these groups, isotopically distinct reservoirs exist, implying that mixing, should it have taken place during the accretion of planetary bodies, was likely local in nature. Whether this  decreases the likelihood of dynamical scenarios in which the bodies of the early solar system were agitated and resulted in eccentric, crossing orbits, such as in the Grand 
Tack \citep{mahbrasser2021,mah2022evidence,woo2022terrestrial}, remains to be seen.} \\

Earth and Mars fall close to- or within the NC group for these two elements \citep[e.g.,][]{warren2011stable, dauphas2017isotopic}. 
Some small, telluric bodies also occupy the NC group space, and, for Cr, Ti and other iron-peak elements, together define an `NC trend' that passes linearly through\add{ or near to} the composition of CI chondrites \citep{trinquier2009origin,williams2020chondrules,palmemezger2024}. 
On this basis, some models argue that the Earth is made predominantly ($\sim$95~\%) of EH-like material \citep{kleine2023inner,dauphas2024}, whereas others suggest a ureilite-like NC end-member and require up to $\sim$40~\% CC material \citep{schiller2018calcium}. However,\add{ the perceived necessity for the addition of carbonaceous chondrites may be circumstantial, since CCs lie on fractionation lines distinct from terrestrial in triple O isotope space \citep{claytonmayeda1999}, implying that they, in combination with ureilites or EH-like material, are unlikely to represent a significant portion of the Earth on a mass basis. Moreover,} for the isotopes of the heavy elements, notably Mo and Zr, there is no suitable member of the extant CC group that would produce the composition of the Earth from an existing NC member. Possible solutions state that the Earth accreted some `missing' material not represented in the meteorite collection for an EH-like proto-Earth \citep{burkhardt2021terrestrial} or preferentially lost $s$-process-depleted material via envelope processing for a ureilite-like end-member \citep{onyett2023}. \\


Here, we synthesise the major physical properties of the terrestrial planets and other differentiated telluric bodies and compare them with chondrites on an empirical basis (section \ref{sec:observations}). Section \ref{sec:disks} describes the thermophysical states of protoplanetary disks and their evolution in time so as to identify possible loci and times of planet formation. In section \ref{sec:nebular_condensation}, we characterise the chemistry of solids that condense from the solar nebula under the assumption of equilibrium, before exploring the effects of non-equilibrium condensation on the stability and compositions of condensates. Whether mixtures of such condensates are consistent with the bulk compositions of telluric bodies are evaluated in section \ref{sec:planet_acc}, and the influence of chemical equilibrium between core- and mantle at a range of pressure-temperature conditions on the gross structure of differentiated bodies is examined. The effect of planetary mass on the chemical and isotopic effects of volatile depletion processes are characterised in section \ref{sec:chemistry_accretion}. Finally, a critical assessment of the distribution of nucleosynthetic anomalies, both across planetary materials and among different elements in the Earth and Mars is undertaken in section \ref{sec:provenance}, with a view to developing a coherent astrophysical model for planetary formation.



\section{Physical- and chemical nature of terrestrial planets}
\label{sec:observations}

\subsection{Physical structure and distribution}

The distribution of the mass of rocky material in the inner Solar System is concentrated in the terrestrial planets, with two, Earth and Venus, together comprising 91.8~\% (see Table \ref{tab:chemphys_prop}). The integrated mean of the mass therefore is nearly symmetrically centred about 0.897 AU, an observation that is used to inform models of the initial mass distribution of the protoplanetary disk \citep{weidenschilling1977}. Despite the symmetry in mass distribution, the constitution of the planets is not uniform. The internal structures of the terrestrial planets are \textit{differentiated} and can be approximated by silicate mantles and iron-rich cores, but the core/mantle ratio of the terrestrial planets broadly decreases with increasing semi-major axis \citep[e.g.,][]{ringwood1966chemical}, ranging from 0.67$\pm$0.06 in Mercury \citep{hauck2013} to 0.204$\pm$0.015 in Mars \citep{khan2023} and 0.008--0.015 in the Moon \citep{sossi2024moon} In detail, however, Venus has a slightly lower uncompressed density (by 2 \%) than does the Earth, but its moment of inertia \citep[0.337$\pm$0.024,][]{margot2021venus} overlaps with that of the Earth \citep[0.3307,][]{williams1994moi}, meaning its core/mantle ratio is indistinguishable, within error, from that of the Earth (0.325). 

\begin{table}[!ht]
    \centering
    \scriptsize
    \caption{Selected physical and chemical properties of planetary bodies.}
    \begin{tabular}{lcccccccccccccc}
    \hline
        ~ & Semi-major axis (AU) & ± & log(mass, kg) & ± & CMF & ± & FeO (wt.~\%) & ± & Mg\# & ± & $\Sigma$Fe & ± & $\Delta$IW & ±  \\ \hline
        Mercury  & 0.387 & - & 23.519 & - & 0.67 & 0.06 & 0.21 & 0.10 & 0.98 & 0.02 & 0.604 & 0.054 & -5.40 & 0.40  \\
        Venus & 0.723 & - & 24.687 & - & 0.27 & 0.05 & 9.0 & 3.0 &  0.88 & 0.04 & 0.294 & 0.056 & -2.11 & 0.35  \\ 
        Earth & 1.000 & - & 24.776 & - & 0.325 & - & 8.1 & 0.2 & 0.890 & 0.005 & 0.335 & - & -2.20 & 0.02  \\ 
        Moon & 1.000 & - & 22.866 & - & 0.012 & 0.003 & 11 & 3 & 0.84 & 0.04 & 0.095 & 0.025 & -1.58 & 0.26  \\ 
        Mars & 1.524 & - & 23.807 & - & 0.204 & 0.015 & 13.8 & 1.0 & 0.81 & 0.02 & 0.298 & 0.029 & -1.74 & 0.07  \\ 
        Vesta & 2.36 & - & 20.413 & - & 0.15 & 0.07 & 24.4 & 3.0 & 0.68 & 0.03 & 0.296 & 0.068 & -1.24 & 0.12  \\ 
        UPB & 2.7* & 0.5 & 21* & 1 & 0.20 & 0.09 & 22.5 & 7.0 & 0.70 & 0.07 & 0.31 & 0.10 & -1.36 & 0.25  \\ 
        APB & 2.7* & 0.5 & 21* & 1 & 0.17 & 0.04 & 22 & 4 & 0.71 & 0.04 & 0.295 & 0.048 & -1.33 & 0.11  \\ \hline
    \end{tabular}
    \label{tab:chemphys_prop}

CMF = core mass fraction. $\Sigma$Fe = fraction of planetary mass comprised by total Fe (FeO+Fe). $\Delta$IW = Oxygen fugacity with respect to the iron-wüstite buffer, $\Delta$IW = log$f$O$_2$ - log$f$O$_2$(IW).
References: Mercury - \citep{hauck2013,namur2016}, Venus \citep{BVSP1981,helbert2021venusfe,shah2022, rodriguez2022}, Earth \citep{frost_mccammon2008redox,palme2014treatise}, Moon \citep{sossi2024moon}, Mars \citep{khan2022geophysical,khan2023}, Vesta - \citep{zuber2011,toplis2013}, Ureilite Parent Body (UPB) - \citep{goodrich2007}, Angrite Parent Body (APB) - \citep{longhi1999,mckibbinoneill2018, tissot2022}. The semi-major axes and masses of the UPB and APB are unknown. They are thought to reside (or have resided) in the asteroid belt, and to have (or have had) masses similar to that of Vesta. Nominal ranges of 2.7$\pm$0.5 AU and 10$^{21\pm1}$ kg for semi-major axis and mass, respectively, are given here.   
\end{table}

\subsection{Chemical composition}
\subsubsection{Major rock-forming elements}

The core/mantle ratio, is, to first order, an indicator of the relative proportions of Fe/O accreted by the planetary body \citep[as seen for chondritic meteorites;][]{ureycraig1953}. \add{There is no requirement for planets to adhere to the chondritic range, not only due to potential chemical gradients in the solar nebula \citep[e.g.,][]{larimerbartholomay1979,palme2000heliocentric} but also owing to giant impacts \citep[e.g.,][]{fegleycameron1987vaporization,oneillpalme2008}.} 
\delete{Consequently}\add{Irrespective of the pathway via which differentiated bodies accreted, at equilibrium}, the amount of oxygen that is available to oxidise metallic iron \delete{influences the equilibrium}\add{is described by};
    
\begin{equation}
    Fe (core) + \frac{1}{2}O_2 = FeO (mantle),
    \label{eq:IW}
\end{equation}

in which FeO is dissolved in a multicomponent silicate (e.g., liquid phase), and metallic iron forms a discrete phase (including minor Ni, Co and light elements), that then constitutes the core of the body. Equation \ref{eq:IW} represents the iron-wüstite (IW) buffer, and 
can be used to constrain the oxygen fugacity, $f$O$_2$, relative to IW, which is given by;

\begin{equation}
    \Delta \mathrm{IW} = \mathrm{log_{10}}fO_2(\mathrm{body}) - \mathrm{log_{10}}fO_2(\mathrm{IW}) = 2\mathrm{log_{10}}\left(\frac{aFeO}{aFe.K_{(\ref{eq:IW})}}\right) - 2\mathrm{log_{10}}\left(\frac{1}{K_{(\ref{eq:IW})}}\right) = 2\mathrm{log_{10}}\left(\frac{aFeO}{aFe}\right),
    \label{eq:def_dIW}
\end{equation}

\begin{figure}[!ht]
    \centering
    \includegraphics[width=1\linewidth]{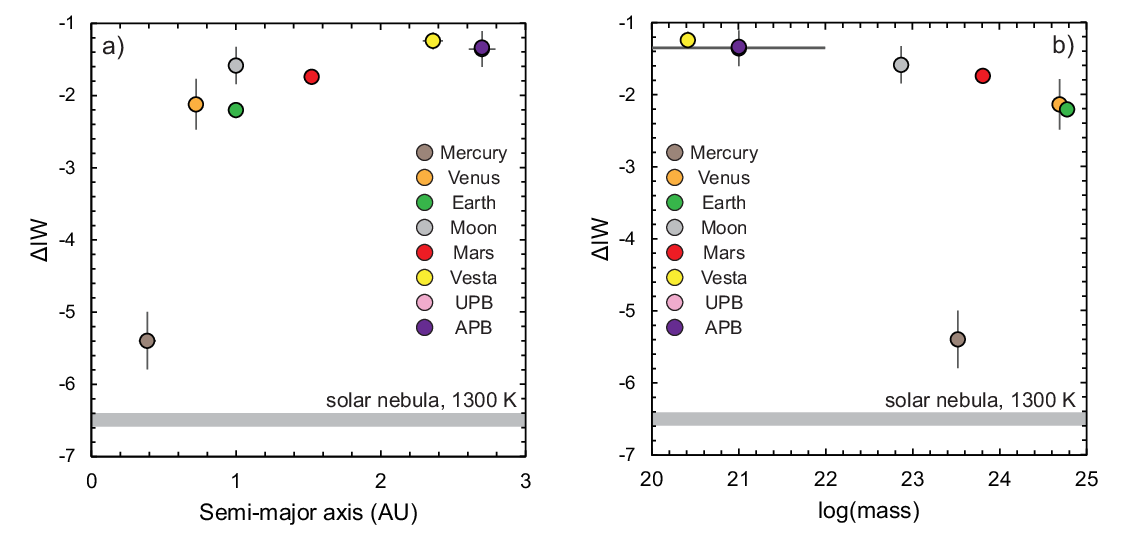}
    \caption{Oxygen fugacities, with respect to the iron-wüstite (IW) buffer \citep{oneill_pownceby1993} of the planets as a function of a) heliocentric distance (in AU) and b) the log$_{10}$ of their masses. Also shown is the $f$O$_2$ defined by a solar gas at 1300 K and 10$^{-4}$ bar \citep{grossman2008redox}. Source data is shown in Table  \ref{tab:chemphys_prop}.}
    \label{fig:fO2_planets}
\end{figure}

where $a$ refers to the activity and $K_{(\ref{eq:IW})}$ the equilibrium constant of eq. \ref{eq:IW} \citep[see, for example,][]{oneill_pownceby1993}. If the mantle FeO content of a given planetary body (see Table \ref{tab:chemphys_prop}) is assumed to be well-mixed and to represent a quantity determined by equilibrium with the core, there is a general tendency for the $f$O$_2$ to increase as a function of heliocentric distance (Fig. \ref{fig:fO2_planets}a), but also with decreasing mass (Fig. \ref{fig:fO2_planets}b). The small telluric bodies in the asteroid belt define values near $\Delta$IW-1.3, Mars has an intermediate value of $\Delta$IW-1.7, and the Earth the most reduced, $\Delta$IW-2.2 \citep[see also][]{wadhwa2008redox,frost_mccammon2008redox}. The Moon is displaced to higher values with respect to the Earth \citep[see][]{oneill1991moon,sossi2024moon}, whereas Mercury has a conspicuously lower $f$O$_2$ \citep{namur2016}. Even so, the $f$O$_2$ inferred for core formation on Mercury is similar to that for a gas of solar composition at 1300 K, which has an H$_2$/H$_2$O ratio of 5 $\times$ 10$^{-4}$, corresponding to $\Delta$IW = -6.5 \citep[e.g.,][]{grossman2008redox}. \add{Thus, the $f$O$_2$s recorded by all differentiated planetary bodies, with the possible exception of Mercury, are more oxidised than that expected for equilibrium with the solar nebula.} \\

\begin{figure}[!ht]
    \centering
    \includegraphics[width=0.5\linewidth]{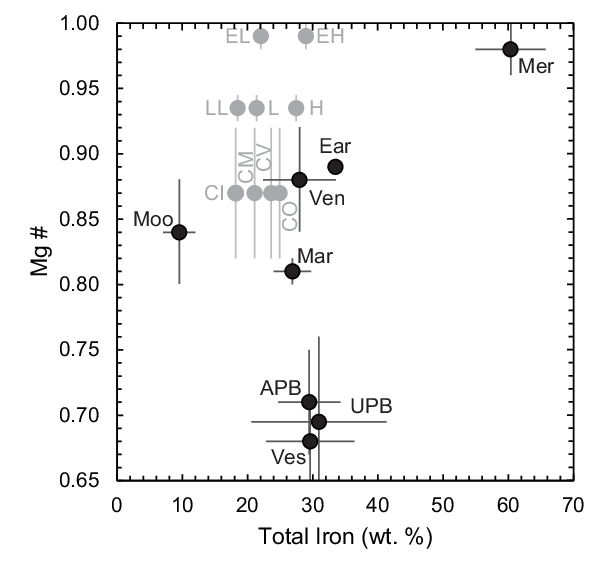}
    \caption{ The Mg \# of differentiated (black) and undifferentiated (grey) bodies as a function of their total iron contents. Source data is from \cite{wassonkallemeyn1988} for chondritic meteorites.}
    \label{fig:mg_no}
\end{figure}

Most bodies are consistent with an amount of total iron ($\sum$Fe $\sim$ 28 - 33 wt. \%, Table \ref{tab:chemphys_prop}) that overlaps with those found in chondrites \citep[18 - 30 wt. \%,][]{wassonkallemeyn1988}, though it should be noted that estimates for core mass fractions for the small telluric bodies are made on the assumption of a chondritic bulk composition \citep[e.g.,][]{toplis2013}. This notwithstanding, Mercury and the Moon are again exceptions and require roughly 60 wt. \% and 10 wt. \% total Fe, respectively (Table \ref{tab:chemphys_prop}; Fig. \ref{fig:mg_no}). Comparisons of the oxygen fugacities of differentiated bodies with those of chondritic meteorites are hampered because the most primitive examples (i.e., type 3 chondrites) are a collection of unequilibrated grains and therefore the $f$O$_2$ is undefined. A relative scale of oxidation may be instead provided by the Mg\# of silicates contained therein (calculated on a molar basis);

\begin{equation}
    Mg \# = \frac{Mg}{[Mg+Fe^{2+}]}.
    \label{eq:mg_no}
\end{equation}

Because Mg is lithophile and Ni is siderophile, plots of the Fe/Mg and Ni/Mg ratios of groups of bulk chondrites can be used in conjunction with eq. {\ref{eq:IW}} to infer the oxygen fugacity at which their silicates formed \citep{larimer_anders1970}. 
As Ni is almost entirely hosted in metal grains, the Fe/Mg ratio of the silicate fraction of a given chondrite group is recovered at Fe/Ni = 0. Carbonaceous chondrites yield higher intercepts (Fe/Mg)$_0$ = 0.35$\pm$0.16 than do the ordinary chondrites (Fe/Mg)$_0$ = 0.16$\pm$0.03, while that for enstatite chondrites cannot be reliably assessed from only two groups, but their constituent silicates have Fe/Mg very near 0 \citep[see also][]{oneill_palme1998} and here we adopt 0.01$\pm$0.01. These are compared to the range of Mg\#s observed among planetary mantles\add{, as a proxy for $f$O$_2$,} in Fig. \ref{fig:mg_no}. \\

Chondritic meteorites occupy but a limited parameter space, down to Mg\#s $\sim$0.82, within uncertainty. Strikingly, all small telluric bodies (STBs) and Mars have mantle Mg\#s far below even silicate phases in carbonaceous chondrites (CI, CM, CV, CO in Fig. \ref{fig:mg_no}). It should be noted that chemical equilibration in the presence of S could result in $f$O$_2$s higher than the IW buffer \citep{bercovici2022effects}. Hence, the STBs must have experienced additional oxidation relative to that observed in at least the ordinary and enstatite meteorites, and potentially also the carbonaceous chondrites, considering some cores of STBs are thought to be S-poor \citep{steenstra2020}\add{, though others are inferred to contain up to 19~wt.\% S \citep{hirschmann2021}}. To first order, the vertical trend in Fig. \ref{fig:mg_no} implies differences in Fe/O ratios at constant total iron contents. This could not have been produced through physical, binary mixtures of metal and silicate, which would instead give rise to coupled changes in Fe/O and $\Sigma$Fe, as vaguely defined by the trend passing through the Moon, Venus, Earth and Mercury. Therefore, and unlike refractory elements, there is no \textit{a priori} reason to expect chondritic ratios (which differ even between themselves) among the major rock-forming elements (Fe, Mg, Si, O) in the terrestrial planets \citep[see also][]{yoshizaki_mcdonough2020}.

\subsubsection{Volatile elements}

\begin{figure}[!ht]
    \centering
    \includegraphics[width=0.5\linewidth]{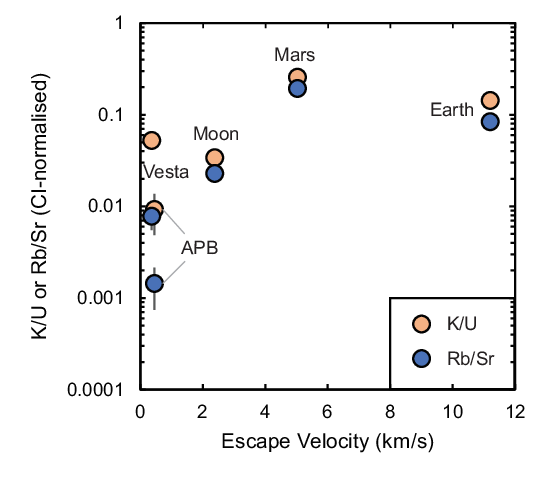}
    \caption{The K/U and Rb/Sr ratios, normalised to those in CI chondrites, of differentiated planetary bodies. References: Angrite parent body - \citep{dauphas2022alkali}, Vesta - \citep{sossi2022stochastic}, Moon - \citep{sossi2024moon}, Mars and Earth - \citep{dauphas2022alkali}.}
    \label{fig:KU_RbSr}
\end{figure}

Unlike the major, rock-forming elements, abundances of the volatile elements vary by orders of magnitude among planetary materials. In order to derive their abundances in planetary mantles, they are typically referenced to a refractory element of similar geochemical character during partial melting of the mantle, such as La or U for the moderately volatile K, or Sr for Rb \citep[e.g.,][]{halliday_porcelli2001}. Furthermore, both K and U are natural gamma-ray emitters and their abundances can thus be inferred via remote sensing \citep[e.g.,][]{prettyman2006elemental,peplowski2012}. Ratios of these elements for the mantles of bodies in which they are well constrained, are plotted as a function of the escape velocity of the body, $v_e = \sqrt{\frac{2GM}{r}}$, in Fig. \ref{fig:KU_RbSr}.

In broad terms, the planets, Mars and Earth, are more volatile-rich, as attested to by their markedly higher K/U and Rb/Sr ratios, than are the small telluric bodies, the Moon, Vesta and the APB \citep[see also][]{mccubbin_barnes2019}. 
In detail, however, Mars appears to have higher quantities of volatile elements than does the Earth \citep{yoshizaki_mcdonough2020,khan2022geophysical,dauphas2022alkali}, a conclusion reinforced from higher S content of the Martian core \citep[$\sim$10 wt.~\%][]{khan2023} relative to that of Earth \citep[$\sim$2 wt. \% S,][]{dreibus_palme1996,suer2017sulfur}. Moreover, despite the greater mass of the Moon relative to Vesta, both exhibit similar degrees of volatile depletion (Fig. \ref{fig:KU_RbSr}). \add{The exception is Mn, which is essentially undepleted with respect to Mg and CI chondrites in Vesta, but is lower, by nearly a factor of 4, in the Moon \citep{sossi2024moon}. This observation was used as evidence to suggest that the Moon was derived from the Earth's mantle after core formation \citep{ringwoodkesson1977,sossi2024moon}. The APB remains the most highly volatile-depleted body known in the solar system, and thus defines a hierarchy from (most volatile-rich) Mars $>$ Earth $>$ Moon $\sim$ Vesta $>$ APB (least volatile-rich).} Therefore, differences in the present-day escape velocities of planetary bodies (proportional to mass for constant density) may only partially account for the observed range in their volatile element contents. Available observations do not support any trend in K/U with heliocentric distance \citep{mccubbin_barnes2019}, either, though volatile abundances in Mercury and Venus remain essentially unconstrained at the resolution required to draw such conclusions. \add{The surface of Mercury, as revealed by MESSENGER, compared to the Earth and even Mars, is rich in C \citep[1.4$\pm$0.9 wt.~\%,][]{peplowski2016remote,nittler2018} and S \citep[$\sim$2.5 wt.~\%,][]{nittler2011major,weider2016evidence}, which, in the canonical nebular condensation sequence, are regarded as ice-forming and moderately volatile elements, respectively. This leaves at least three possibilities; i) the Hermean surface is richer in these elements than is its interior, ii) the region of the inner disk from which Mercury formed did not experience extensive volatile depletion and/or iii) C and S did not behave in a volatile manner as expected for a typical solar nebular composition. } 

\section{Thermal states of accreting disks}
\label{sec:disks}

The first steps of planet formation -- condensation of the first solids, aggregation of dust into planetesimals, formation of protoplanets -- occur in a protoplanetary disk. The temperature of the protoplanetary disk dictates which chemical species are in the solid or gaseous forms, and therefore influences the volatile element content (both moderately and highly) of the bodies depending on where and when they form. At a given time and location, gas in the disk can be moving either outwards or inwards. \add{The following discussion details physics applicable to typical solar-mass (G-type) stars.} \\


\subsection{Disks dominated by internal viscosity}
\label{sec:disk_viscous}

When modelling viscous disks, following \cite{shakura_sunyaev1973,shakura_sunyaev1976}, it is customary to assume that the viscosity is

\begin{equation}
    \nu=\alpha H^2 \Omega = \alpha c_s H,
    \label{eq:disk_visc}
\end{equation}

where $\Omega$ is the orbital frequency at distance $r$ from the central star, $c_s$ is the sound speed and

\begin{equation}
H = \frac{c_s}{\Omega} = \sqrt { \frac{\gamma_{ad}RTr^3}{\mu GM_*} }
\label{eq:disk_scaleheight}
\end{equation}

is the pressure scale height of the disk assumed to be vertically isothermal and in hydrostatic equilibrium, $T$ is the temperature and $\gamma_{ad}$, $R$, $\mu$ are the adiabatic index-, constant- and mean molecular weight of an ideal gas, with  $G$ gravitational constant, and $M_*$ the mass of the star. The coefficient $\alpha$ is a proportionality parameter with values between 0 and 1. Its precise value is not well known as it depends on the physical conditions inside the disk (degree of ionisation, coupling with the magnetic field, as discussed below in section \ref{sec:disk_winds}). 
A viscous disk spreads radially. Over most of the radial extension of the disk, the gas is transported towards the central star at a speed $v_r=-{{3}\over{2}} {{\nu}\over{r}}$, but near the disk's outer edge, the gas radial velocity becomes positive. This allows the disk to grow in radial extent over time, while transporting gas inwards (hence the name ``accretion disk") and angular momentum outwards. The radius where the radial velocity of the gas changes sign increases over time \citep{lynden1974evolution} (see Fig.~\ref{fig-temp-infall}); as a consequence, a fixed point in the disk (relative to the star) will see outwards-then-inwards gas motion.

The accretion rate on the star is given by:

\begin{equation}
    \dot{M}=2 \pi r v_r \Sigma
    \label{eq:disk_accretionrate}
\end{equation}

where $\Sigma$ is the surface density of the disk at distance $r$. The friction among different ``rings" of the disk in differential rotation generates heat on the disk's midplane at a rate 

\begin{equation}
    Q_+^{visc}= {{9}\over{4}} \nu \Sigma \Omega^2 = {{3}\over{4\pi}} \dot{M} \Omega^2
    \label{eq:disk_viscQ}
\end{equation}

The irradiation from the central star deposits heat at the disk's effective surface $H_s$, where it is absorbed at a rate
\begin{equation}
Q_+^{irr}= \frac{\phi L_*}{2 \pi r^2},
\label{eq:disk_irr}
\end{equation}

where $L_*$ is the stellar luminosity and $\phi$ is the incidence angle between the $H_s$ surface and the stellar rays. Notice that viscous heating is proportional to the accretion rate $\dot{M}$, hence to the product $\nu \Sigma$ while irradiation heating is independent of both. The disk cools at the $H_s$ surface by radiating in space as a black body, at the rate 

\begin{equation}
    Q_-=2\sigma_{SB} T_{eff}^4,
    \label{eq:disk_cool}
\end{equation}

where $\sigma_{SB}$ is the Stefan-Boltzmann constant  and $T_{eff}$ is the temperature of the disk at the surface $H_s$.  The energy balance equation

\begin{equation}
    Q_-=Q_+^{visc}+Q_+^{irr}
    \label{eq:disk_energybal}
\end{equation}

sets $T_{eff}$ assuming steady state has been reached. The temperature on the midplane $T$ is then related to $T_{eff}$ via the equation

\begin{equation}
    T=\left[\frac{(Q_+^{irr} +  {{3 \tau}\over{4}} Q_+^{visc})}{Q_-}\right]^{1/4} T_{eff}
\end{equation}

where $\tau=\kappa\Sigma$ is the optical depth of the disk and $\kappa$ its opacity  (see \cite{DullemondHouches2013} for more details). These simplified analytical calculations are useful to get a general view of the structure of the disk and its dependence on physical parameters. A more precise calculation can be obtained with hydrodynamical simulations, where $Q_+^{visc}$ is imposed on each cell of the disk (substituting $\Sigma$ with $\rho dz$, where $\rho$ is the local density and $dz$ the vertical extent of the cell) and $Q_+^{irr}$ and $Q_-$ are computed for each cell accounting for the local gas optical depth (see for instance \cite{bitsch2014stellar,bitsch2015structure}). 

\begin{figure}
    \centering
    \includegraphics[width=1\linewidth]{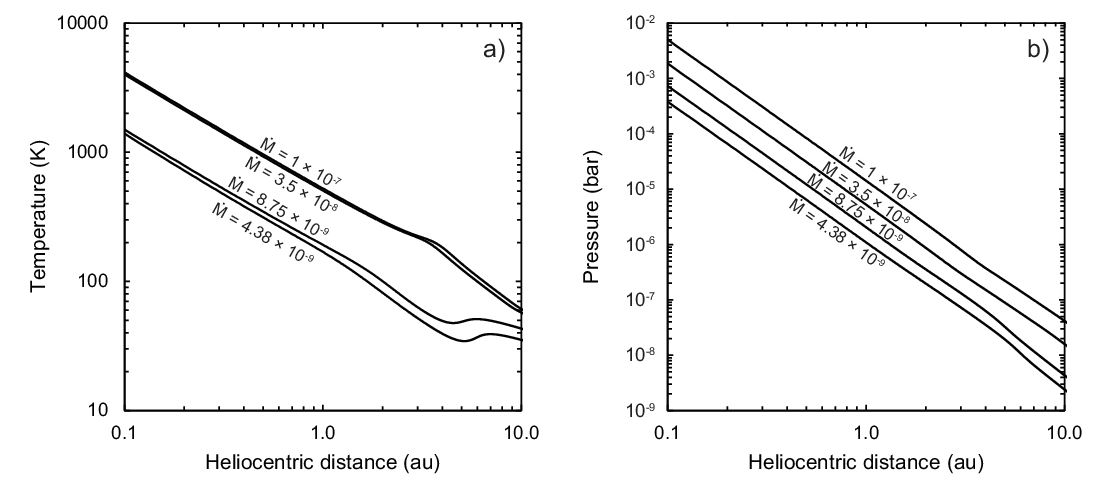}
    \caption{Radial distribution of \textbf{a)} temperature in Kelvin and \textbf{b)} pressure in bar in the midplane of the disk for different stellar mass accretion rates ($\dot{M}$) in solar masses/yr. Here $\alpha= 0.0054$. The temperature of the snowline (i.e., the temperature at which the reaction H$_2$O(g) = H$_2$O(s) proceeds to the right) is roughly between 220 -- 150 K, depending on pressure. NB: Temperatures and pressures inward of 1 au are extrapolated. Constructed with data from \cite{bitsch2015structure}.}
    \label{fig:Bitsch15}
\end{figure}

Fig.~\ref{fig:Bitsch15} shows the radial distribution of the midplane disk temperature as a function of the stellar accretion rate. The latter is constrained by observations \cite{hartmann1998} to be dependent on the age of the star-disk system, as

\begin{equation}
    \mathrm{log} \dot{M}= -8 \pm 0.1 -(1.4\pm 0.3) \mathrm{log} t,
    \label{eq:hartmann_time_acc}
\end{equation}

where $\dot{M}$ is measured in $M_\odot/y$ and $t$ is the time in Myr. The transition from a viscous-dominated to an irradiation-dominated disk is in the 5--10 au region. The calculation presented in the figure is made for a specific value of $\alpha=5.4\times 10^{-3}$. In the region dominated by the viscous heating the midplane temperature is proportional to $(\dot{M}^2/\alpha)^{1/5}$ \citep{BatyginMorby2022self}. Also, notice that the temperature at 1 au barely exceeds 500~K, even in the highest accretion case of $\dot{M}=10^{-7} M_\odot$/y. We return to this in section~\ref{disk-evol}.

\subsection{Dust dynamics and disk chemistry}
\label{sec:disk_dynamics}

By looking at Fig.~\ref{fig:Bitsch15} one could naively think that, as the disk cools over time (i.e. with decreasing $\dot{M}$) the gas condenses increasingly volatile species and that the chemical composition of the disk in its solid and gaseous parts simply depends on the local temperature. This is not true in general because (i) the radial motion of gas towards the central star is typically faster than the rate of radial displacement of a condensation line (i.e. the disk cooling rate) and (ii) the solid material is carried by dust particles that rapidly grow to sizes at which grain dynamics partially decouples from gas dynamics \citep{Cassen2001,Morby15}. \\

\begin{figure}
    \centering
    \includegraphics[width=.8\linewidth]{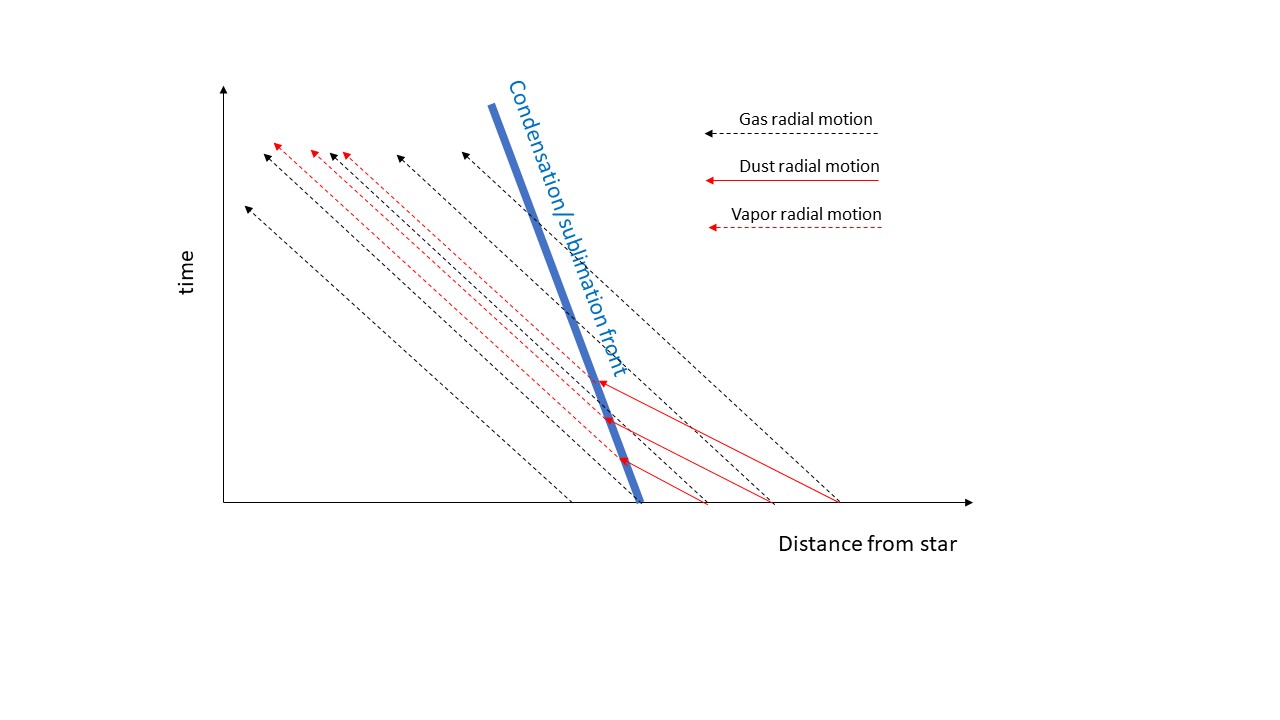}
    \includegraphics[width=.8\linewidth]{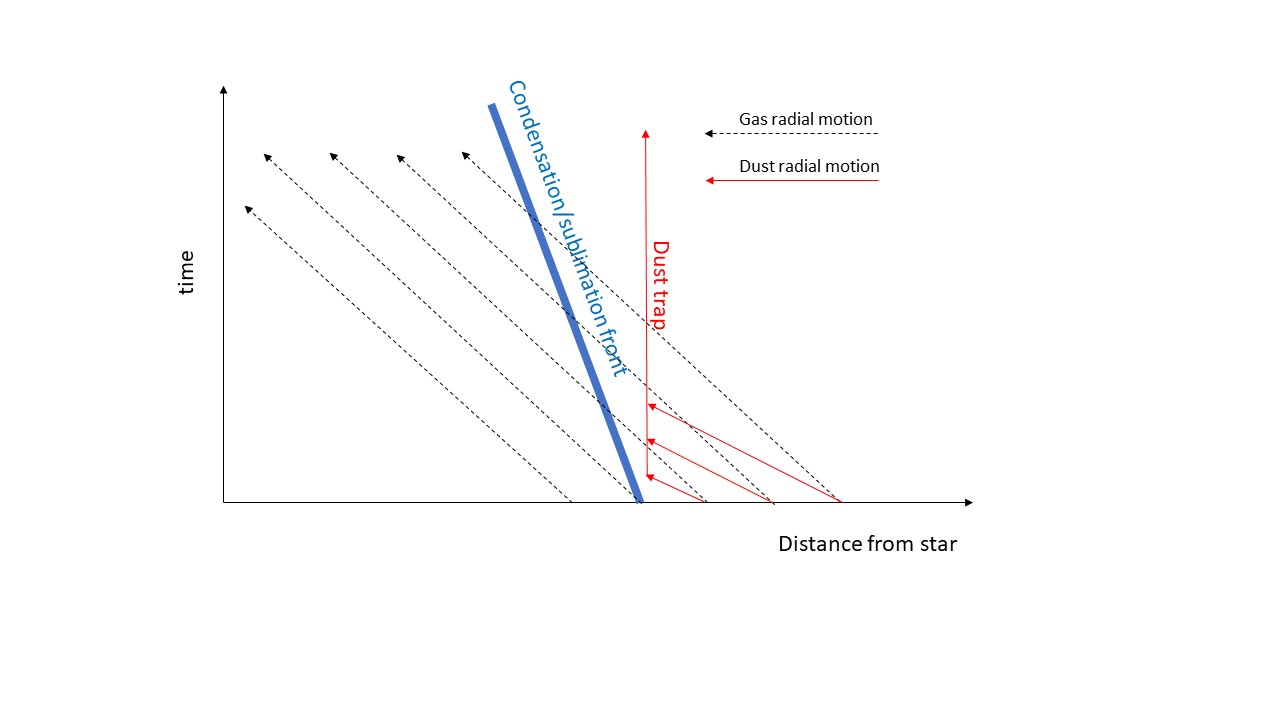}
    \caption{A schematic illustration of the radial motion of the evaporation/condensation front of gas and dust. The top panel shows the situation prevailing in a continuous disk. The radial motion of the gas (mostly hydrogen) towards the central star (dashed black arrows) is faster than the radial displacement of the evaporation/condensation front (thick blue line) that is due to disk cooling over time (cf. eq. \ref{eq:hartmann_time_acc}). The dust moves towards the star (solid red arrows) even faster than does the gas. When crossing the evaporation front, the released vapour then moves inward (red dashed arrows) at the same speed as the gas. As a consequence, the vapour/hydrogen ratio is increased with respect to the its initial ratio (i.e. the stellar abundance). Indeed, the red dashed arrows are denser in space than the black dashed lines, which indicates an enhanced density of the evaporated species. The bottom panel shows the case where the dust is trapped at some location beyond the evaporation front, due to the formation of a pressure maximum. In this case, the dust does not reach the evaporation front. 
    Therefore, even if the disk cools, there is no further condensation possible. }
    \label{scheme}
\end{figure}

The consequence of (i) is that ``all [gas] trajectories pass from the cool side of the [condensation] front to the hot side; no condensation occurs" \citep[][their Fig. 13]{Cassen2001}; the only exception being in the region where the gas radial motion is outward or near the radius where the radial velocity of the gas changes sign and so is small, as illustrated schematically in Fig.~\ref{scheme}.  Condensation can occur only if there is a continuous presence of the concerned species in gaseous form just inward of the condensation front, sustained by inward radial drift of dust and sublimation.  


The consequence of (ii) is that, because, typically the inward radial flow of dust is faster than the flow of gas, the evaporated species has, in general, a higher concentration, relative to hydrogen, at the locus of evaporation than in the initial stellar composition (i.e., the mean solar nebula). This is illustrated in the top panel of Fig.~\ref{scheme}. 

On the other hand, if the radial drift of the dust is blocked by a pressure maximum appearing in the disk (for instance due to the formation of a gap-opening planet, but also to the formation of rings due to magnetic effects \cite{Bethune2017}), there is no evaporation at the condensation front and therefore no possible recondensation during disk cooling 
(bottom panel, Fig.~\ref{scheme}). \cite{Morby15} proposed that the blockage of the radial drift of icy grains, possibly due to the formation of Jupiter, prevented the inner disk from becoming enriched in ice relative to the solar value, even when the snowline was eventually able to pass inward of 1 au, as expected from the decay of $\dot{M}$ over time and the related effect on the disk's temperature (Fig.~\ref{fig:Bitsch15}, eq. \ref{eq:hartmann_time_acc}). This scenario can explain the deficit in water (and other volatile elements) in bodies formed in the inner solar system \citep[][see also section \ref{sec:provenance}]{mccubbin_barnes2019}. \\

One can imagine also more complicated scenarios. For instance, if the dust trap is located between the H$_2$O and CO condensation lines, CO(g) would be released but water vapour would not. As a result, the inner disk would have an enhanced C/O ratio relative to the original ratio in the disk and the star. Thus, depending on the location, number and efficacy of dust traps, various scenarios can arise that influence the chemistry of the disk inward of the trap(s). Predicting the actual evolution of the dust, which, in turn, depends on dust growth, radial extension of the disk, formation of pressure maxima, and so on, is currently an active field of research.

\subsection{Disks dominated by magnetised winds}
\label{sec:disk_winds}

Up to this point, we have considered that the disk evolves and is heated under the effect of its own viscosity (cf. eq.~\ref{eq:disk_visc}). However, the origin of the disk's viscosity is elusive. In the $\alpha$-disk model expounded in section \ref{sec:disk_viscous}, the disk evolution timescale is given \citep{hartmann1998,Cassen2006}:

\begin{equation}
    t = \frac{{r_d}^2}{\nu}
    \label{eq:disk_timescale}
\end{equation}

where $r_d$ is the characteristic length scale of the disk. For a typical disk extent of 100 AU at 1 Myr, as constrained by observations, $\nu$ is 10$^{15}$ cm$^{2}$/s. Molecular viscosity would be several orders of magnitude smaller, $\sim$ 10 $^{6}$ cm$^2$/s  \citep{Cassen2006}, than the viscosity considered in the simulations of \cite{bitsch2015structure}, as shown in Fig. \ref{fig:Bitsch15}. Thus, an additional source of viscosity is needed to reconcile the viscous disk model with observed disk properties. For a decade or so, it was generally believed that this additional viscosity contribution arises from turbulence within the disk, itself generated by the so-called magneto-rotational instability \citep{BalbusHawley1998}. However, this instability occurs only if there is a strong coupling between the magnetic field and the gas, which requires a large amount of gas ionisation. Ionisation is expected to occur only at locations in the disk where it is very hot and the alkali elements have sublimed, i.e. above $\sim 1,000$~K or so \citep{Umebayashi1983,Turner_2007,Flock2017}. \\

A strong effective viscosity can also be generated if there are hydro-dynamical stresses in the disk. These may be the consequence of the formation of spiral waves in the disk, such as those generated when the disk is so dense so as to become nearly gravitationally unstable \citep{rafikov2016}. The accretion of gas from the molecular cloud onto the disk can also generate large effective viscosities \citep{Kuznetsova_2022}. In the absence of these non-generic mechanisms, the viscosity in the disk is now expected to be small, with a value $\alpha$ of the order of $10^{-4}$ or smaller, set by weak disk instabilities such as the vertical shear instability \citep{Stoll_Kley2014} or the convective overstability \citep{Klahr_2014}. \\

If the transport of gas towards the central star were solely due to disk's (i.e., molecular) viscosity, such low values of $\alpha$ would imply very massive disks (i.e. large $\Sigma$, large $r_d$, eq. \ref{eq:disk_timescale}) in order to account for the values of $\dot{M}$ that are observed. But the observed disk masses do not seem as high as required \citep{manara2016,manara2018}. This suggests that another mechanism than internal viscosity is responsible for the mass transport towards the central star. \\

The emerging view is that, in weakly ionised disks \add{(i.e., beyond $\sim$1 au)}, the interaction with magnetic fields generates a magnetised wind, which ejects ions and extracts angular momentum from the disk, promoting the inward radial motion of the remaining gas \add{\citep{Bethune2017,Lesur2023}}. \add{These magnetic fields are thought to have been inherited from the molecular cloud collapse stage, and subsequently lace the protoplanetary disk \citep{stephens2017alma,fu2021}. Though there is no consensus as to how such fields evolve \citep{Lesur2021},} stellar accretion due to magnetised winds breaks the relationship between $\dot{M}, \alpha$ and $\Sigma$ that is typical of viscous disks (section \ref{sec:disk_viscous}). The main consequence, for what concerns this chapter, is that $\dot{M}$ is no longer related to the disk's temperature, unlike in the curves shown in Fig. \ref{fig:Bitsch15}. In absence of viscous heating, disks are solely heated by stellar irradiation and therefore are expected to be cold \citep{ChiangYoudin2010}, with a snowline ($\sim$170 K) inwards of 1 au. Because such temperatures would have resulted in water-rich bodies in the inner solar system, which are not observed, disks clearly must have been much warmer than this, at least initially, owing to the high viscosity induced by one of the sources of stress in the disk (see above). 
An additional source of heating could come from the existence of a current sheet at about 3$H$ \citep{Latter2010,Gressel2015}, but the effect is expected to be small \citep{Mori_2019}.

\subsection{A view of disk evolution}
\label{disk-evol}

\begin{figure}
    \centering
    \includegraphics[width=0.5\linewidth]{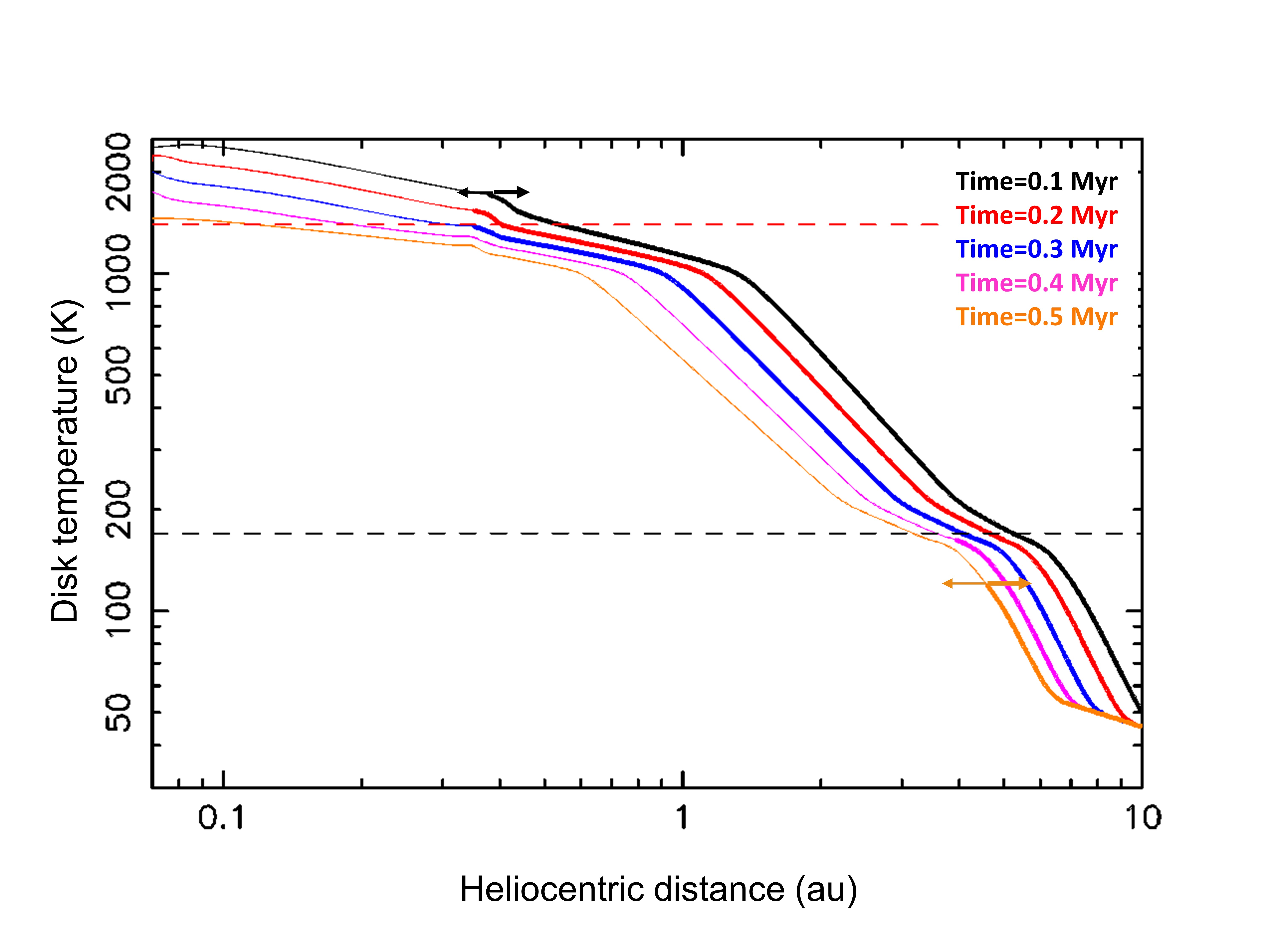}
    \caption{Temperature in the disk as a function of time during the epoch when the disk is accreting material from the envelope surrounding the forming star. The thick part of each curve shows the region where the radial velocity of the gas is positive (outward), whereas the thin part depicts the accretion part of the disk (negative radial velocity), as also indicated by the black and orange arrows. The horizontal dashed lines mark the condensation temperature of water ($T$ = 170~K, black), and rocks ($T$ = 1400~K, red). The intersection of these lines with the various coloured curves identify the location of the condensation/sublimation fronts of these elements as a function of time. From \cite{Morby22}.}
    \label{fig-temp-infall}
\end{figure}

In light of the foregoing discussion, we can envision a general view of disk evolution. Initially, when stellar collapse starts and the disk forms, the disk is small and dense. Its effective viscosity can be very high (even $\alpha=0.01-0.1$) due to stresses exerted on it by the infalling gas and by the anisotropy in the gas distribution generated by conditions close to gravitational instability. The disk is therefore very hot (it is also strongly irradiated by the accretion shock at the stellar surface) and spreads very rapidly outwards \citep{HuesoGuillot2005, Pignatale2017, Morby22, MarschallMorbidelli2023}; see Fig.~\ref{fig-temp-infall}. At this time, the disk is at its hottest; significantly higher than that of a standard accretion disk (compare Fig.~\ref{fig-temp-infall} with Fig.~\ref{fig:Bitsch15}), reaching $\sim 1,000$~K at 1 au. During the radial expansion stage, the gas flows from the hot side of the condensation fronts towards the cold side (inward to outward), whereas this is reversed in the later evolutionary stage. Thus, the outward expansion stage is likely to result in direct condensation of gas, such as that leading to the formation of \add{primitive grains, namely calcium-aluminium-rich inclusions,} CAIs and \add{amoeboid-olivine-aggregates}, AOAs \citep{Pignatale2017, MarschallMorbidelli2023}. Notice that, from Fig.~\ref{fig-temp-infall}, the CAI condensation temperature is met at 0.3--0.6 au during the early stages of the disk\add{, where these grains are  thought to have formed \citep[e.g.,][]{gounelle2001}}. It should be noted that such refractory grains show evidence for having formed by very rapid cooling \citep[within $\sim$10$^{-1}$ -- 10$^{-2}$ yr,][]{marrocchi2019rapid}, and it is therefore unlikely that the thermal regime under which they condensed is representative of gross disk structure.

When the rate of infall of new gas onto the protoplanetary disk wanes, typically after a few $10^5$~yr according to observations, and once the disk has spread sufficiently in the radial direction to be far from gravitationally unstable conditions, the viscosity rapidly drops to low values, corresponding to $\alpha < 10^{-4}$. Consequently, the temperature also drops very rapidly to approach that of a passively irradiated disk \citep{ChiangYoudin2010}. We remind readers that such a disk has a temperature of 120~K at 1 au and decaying as $1/r^{3/7}$. The transport of gas towards the central star is now dominated by the process of angular momentum removal in magnetised disk's winds.\\ 

The relative timing of the radial spreading of the disk and subsequent accretion (inward transport) stages is unclear and the formation ages of planetary bodies are also used to inform models. For instance, \cite{lichtenberg2021bifurcation} assume an outflow phase that results in condensation from $\sim$0.2 Myr and persists until 0.7 Myr over 1.5 -- 7 au, in order to explain the inferred iron meteorite (and other achondrite parent body) formation times. The disk cools and contracts thereafter, resulting in the formation of a second generation of planetesimals from 3 -- 15 au at 0.7 -- 5 Myr after $t_0$. By contrast, models which approximate alpha disks with a decreasing value of $\alpha$ 
lead to the contemporaneous generation of two \citep{Morby22} or three \citep{izidoro2022planetesimal} planetesimal populations from $\sim$0.1 -- 0.5 Myr at 1, 5 and, in \cite{izidoro2022planetesimal}, 30 au for nominally `rock', `ice' and `soot', respectively. \\

While the formation times and locations detailed above are based on the temperature in the disk (which evolves with time), they do not describe how these dust grains migrate in the disk thereafter. One should expect a generalised condensation of volatiles during the rapid cooling phase of the disk. However, if an obstacle to the radial drift of dust appears in the disk before the rapid cooling phase starts, as suggested by the preservation of the so-called isotopic dichotomy (see section \ref{sec:provenance}) established during the disk's radial expansion phase \citep{kruijer2017age, Nanne2019, BrasserMojzsis2020, Morby22}, the inner disk may remain volatile depleted because 
volatile-rich dust cannot penetrate the inner disk to be accreted by the local planetesimals \citep{Morby15}. \\

\add{The efficiency with which particles are able to traverse any such barrier between the outer- and inner disks is inversely proportional to their size, owing to the better coupling of smaller particles to the surrounding gas \citep{kalyaan2023effect,stammler2023leaky}. Consequently, such barriers are expected to have been permeable to small grains. Empirical constraints on plausible threshold sizes were first provided by \cite{haugbolle2019probing} on the basis of the size distribution of CAI grains in an ordinary chondrite, concluding that only grains smaller than 100 -- 300 $\mu$m would be able to penetrate the barrier. This estimate was subsequently revised downward to 46$\pm$48 $\mu$m by \cite{dunham2023calcium} after having examined some 76 sections across a range of chondrites, both NC and CC. This figure overlaps with the 50$\mu$m size distribution peak observed by \cite{hezel2008} in CCs. Furthermore, the region occupied by CAIs, on average, in NC chondrites is 0.0097$\pm$0.0002 \%, compared to an average of 3-5 \% for the most CAI-rich CCs \citep[CK and CV,][]{hezel2008,ebel2024}. Therefore, CAIs are up to 500$\times$ less abundant in the NC-forming region than they were in the CC-forming region. Although these CAIs likely formed in close proximity to the Sun, their abundance in CCs implies they were rapidly transported to regions of the disk at which they formed, plausibly the outer solar system \citep{shu1997,desch2010,jacquet2024}. If true and such CAIs resided in the outer solar system, then, at most, 1~\% of these particles were able to (re-)enter the inner solar system.} Interestingly, the NC chondritic planetesimals, which are thought to have formed in the inner solar system (parent bodies of ordinary and enstatite chondrites), are highly depleted in water \citep{mccubbin_barnes2019}, but only mildly depleted in more moderately volatile elements such as Na, Zn, K \citep{wassonkallemeyn1988}. This may suggest that the drift of dust was blocked after that the temperature at 1-2 au had decreased below $\sim 800$~K, but before it decreased to $170$~K.

\section{Condensation of the solar nebula}
\label{sec:nebular_condensation}

\subsection{Equilibrium condensation}
\label{sec:neb_eq_cond}

The discussion in section \ref{sec:disks} has highlighted the dynamic nature of the protoplanetary disk, both in time and in space. This contrasts with the `static' picture afforded by the canonical cosmochemical understanding of disk evolution, which is frequently tied to nebular condensation temperatures of the elements, $T_c$, from a gas of fixed composition (the Sun) and pressure \citep[10$^{-4}$ bar; ][]{lodders2003,wood2019}. The utility of this volatility scale lies in its simplicity; it quantifies the temperature at which half of the mass of a given element exists in its condensed (liquid or solid) state \citep{larimer1967}. However, this simplicity comes at the expense of information as to the condensing mineral assemblage, and thus the temperature range over which condensation occurs. For oxides and metals, condensation stoichiometries take the general form:

\begin{equation}
   M^{x+n}O_{\frac{x+n}{2}} (l,s) = M^xO_{\frac{x}{2}} (g) + \frac{n}{4}O_2 (g)
    \label{eq:reaction_stoichiometry}
\end{equation}

where $x$ is the oxidation state of the metal, $M$, in the gas phase and $n$ the number of electrons exchanged. The condensation temperature of $M$ in the above reaction can be described by \citep[e.g.,][]{sossi2019evaporation,ebel2023}:

\begin{equation}
    T = \frac{- \Delta H}{R \left( \frac{n}{4}\mathrm{ln}fO_2 + \mathrm{ln} P + \mathrm{ln}f^{i,T}_{vap} \right) - \Delta S }
    \label{eq:T_cond}
\end{equation}

where $P$ is the total pressure, $\Delta H$ and $\Delta S$ are the enthalpy and entropy change of eq. \ref{eq:reaction_stoichiometry}, respectively, and $f^{i,T}_{vap}$ is the fraction of element $i$ remaining in the vapour;

\begin{equation}
   f^{i,T}_{vap} = N^{i,T}_{vap}/N^{i,0}_{vap}
   \label{eq:def_fvap}
\end{equation}

where $N$ is the number of moles of $i$ in the gas phase at any given temperature, $T$, relative to the total abundance, $0$. Values therefore range between $0 < f^{i,T}_{vap} < 1$. Condensation curves of the elements for any given reaction occur over a similar temperature interval, relating to the fact that gas-solid reactions have similar $\Delta G/RT$ among one another (cf. eq. \ref{eq:T_cond}), where $\Delta G$ is the change in Gibbs energy. Because $\Delta G$ is large relative to $RT$, the temperature intervals over which condensation occurs are small ($\sim$50 -- 100 K) \citep{larimer1967}. This is typically the case for the major elements (Mg, Fe, and Si), whose condensation reactions are \textit{not} limited by the availability of other elements in the gas phase. However, the 50 \% condensation temperature is not an intrinsic thermodynamic property of the element, but is related to the nature of the particular condensation reaction(s) for a given composition, pressure and temperature. When a single equilibrium predominates, a constant $\Delta H$ with respect to $T$ is a good approximation, and the first derivative of eq. \ref{eq:T_cond} with respect to $f^{i,T}_{vap}$ yields the condensation rate;

\begin{equation}
    dT = \frac{\Delta H R}{f^{i,T}_{vap} \left(R \left( \frac{n}{4}\mathrm{ln}fO_2 + \mathrm{ln} P + \mathrm{ln}f^{i,T}_{vap} \right) - \Delta S\right)^2 } df^{i}_{vap}
   \label{eq:T_cond_f'x}
\end{equation}

\begin{figure}[!ht]
    \centering
    \includegraphics[width=1\linewidth]{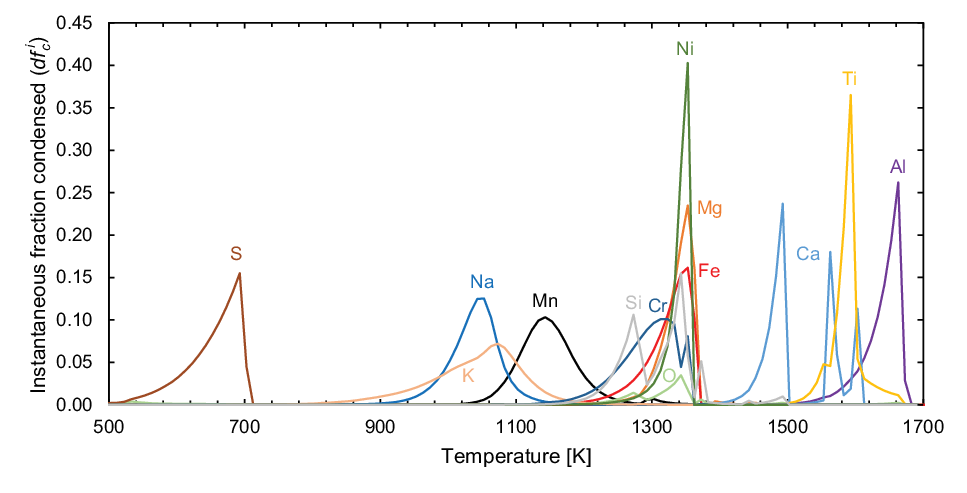}
    \caption{Condensation curves of the elements from a gas of solar composition \citep{lodders2010} at 10$^{-4}$ bar total pressure as calculated with FactSage 8.2. It shows curves of the instantaneous fraction condensed ($df^i_{c}$) of a given element, $i$, as a function of temperature computed at 10~K intervals ($dT$). The quantity $df^i_{c}$ = $f^{i, T_2}_{c} - f^{i, T_1}_{c}$, where $f^{i,T}_{c} = 1 - f^{i,T}_{vap}$. Elements with multiple peaks (notably Ca and Si) condense via a series of reactions. See text for details.}
    \label{fig:cond_curves}
\end{figure}

The functional form of eq. \ref{eq:T_cond_f'x} indicates that the temperature interval is proportional to $1/[f_{vap}\mathrm{ln}f_{vap}^2$], leading to decreasing condensation rates as temperature declines.  These phenomena are best understood by examining the equilibrium condensation curves of the elements from a solar gas at 10$^{-4}$ bar in Fig. \ref{fig:cond_curves}.

It is evident that some elements condense largely into a single phase, such as Al into Al$_2$O$_3$ (corundum) or S into FeS (troilite). The behaviour of others, however, is dictated by several condensation reactions occurring at different temperatures, the most prominent example being Si, which condenses into olivine at $\sim$1345 K via the reaction;

\begin{equation}
\mathrm{
2Mg(g) + SiO(g) + 3/2O_2(g) = Mg_2SiO_4(s)},
\label{eq:cond_olivine}
\end{equation}

which then reacts with remaining SiO(g) to form orthopyroxene at $\sim$1275 K, according to:

\begin{equation}
\mathrm{
Mg_2SiO_4(s) + SiO(g) = 2MgSiO_3(s)}.
\label{eq:cond_opx}
\end{equation}

This gives rise to a $T_c^{50}$ of roughly 1320 K, which, thermodynamically, does not correspond to any particular reaction, but rather the arithmetic average of the two major reactions. The root of the second derivative of the instantaneous fraction condensed with respect to temperature, $\mathrm{d}^2{f}/\mathrm{d}T^2$ gives the peak of the condensation curve, and each of the major reactions (contributing $>$5 \% to the total condensed fraction of an element, $f_c > 0.05$) are shown, together with $f_c$ at that temperature, in Table \ref{tab:cond_phases}.

\begin{table}[!ht]
    \centering
    \caption{Major condensation reactions in the solar nebula}
    \begin{tabular}{lccccc}
    \hline
        \textbf{Element} & \textbf{Major Gas(es)} & \textbf{Mineral} & \textbf{Mineral Name} & \textbf{$T$ [K]} & \textbf{$f^{i,T}_c$} \\ \hline
        \textbf{Al} & Al $>$ AlOH $>$ Al$_2$O & Al$_2$O$_3$ & Corundum & 1663 & 0.29  \\ 
        \textbf{Ti} & TiO $>>$ Ti & CaTiO$_3$ & Perovskite & 1593 & 0.54  \\ 
        \textbf{Ca} & Ca $>>$ CaOH & CaAl$_{12}$O$_{19}$ & Hibonite & 1603 & 0.11  \\ 
        \textbf{Ca} & Ca $>>$ CaOH & CaAl$_4$O$_7$ & Grossite & 1563 & 0.38  \\ 
        \textbf{Ca} & Ca $>>$ CaOH & Ca$_2$Al$_2$SiO$_7$ & Melilite & 1493 & 0.63  \\ 
        \textbf{Ni} & Ni $>>$ NiH, NiS & Fe alloy & Fe alloy & 1353 & 0.40  \\ 
        \textbf{Mg} & Mg & Mg$_2$SiO$_4$ & Olivine & 1343 & 0.24  \\      
        \textbf{Fe} & Fe & Fe alloy & Fe alloy & 1340 & 0.35  \\ 
        \textbf{Cr} & Cr $>>$ Cr(OH)$_2$ & Fe alloy & Fe alloy & 1353 & 0.08  \\ 
        \textbf{Cr} & Cr $>>$ Cr(OH)$_2$ & MgCr$_2$O$_4$ & Spinel & 1318 & 0.37  \\ 
        \textbf{Si} & SiO $>$ SiS & CaAl$_2$Si$_2$O$_8$ & Feldspar & 1373 & 0.09  \\ 
        \textbf{Si} & SiO $>$ SiS & Mg$_2$SiO$_4$ & Olivine & 1343 & 0.31  \\ 
        \textbf{Si} & SiO $>$ SiS & MgSiO$_3$ & Orthopyroxene & 1273 & 0.76  \\ 
        \textbf{O} & CO $\sim$ H$_2$O $>>$ SiO & Mg$_2$SiO$_4$ & Olivine & 1343 & 0.08  \\ 
        \textbf{O} & CO $\sim$ H$_2$O $>>$ SiO & MgSiO$_3$ & Orthopyroxene & 1273 & 0.17  \\ 
        \textbf{Mn} & Mn & Mg$_2$SiO$_4$ & Olivine & 1143 & 0.48  \\ 
        \textbf{Na} & Na $>$ NaCl & CaAl$_2$Si$_2$O$_8$ & Feldspar & 1048 & 0.49  \\ 
        \textbf{K} & K $>$ KCl & CaAl$_2$Si$_2$O$_8$ & Feldspar & 1083 & 0.40  \\ 
        \textbf{S} & H$_2$S & FeS & Troilite & 693 & 0.20  \\ \hline
    \end{tabular}
    \label{tab:cond_phases}
    
    $T$ refers to the root of the second derivative of ${f^{i,T}_{c}}$ (the fraction condensed) with respect to temperature (i.e., the peak temperature in Fig. \ref{fig:cond_curves}). 
\end{table}

\begin{figure}[!ht]
    \centering
    \includegraphics[width=0.5\linewidth]{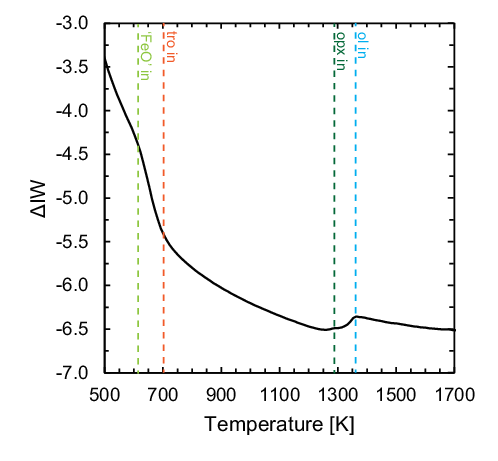}
    \caption{Evolution of oxygen fugacity ($f$O$_2$) relative to the iron-wüstite buffer \citep{oneill_pownceby1993} during cooling and equilibrium condensation of the solar nebula gas at 10$^{-4}$ bar. Dashed vertical lines show the onset of condensation of major phases; ol = olivine (blue), opx = orthopyroxene (dark green), tro = troilite (orange) and FeO as a component in olivine, `FeO' (light green). }
    \label{fig:cond_fo2}
\end{figure}

Equation \ref{eq:cond_olivine} proceeds through the consumption of O$_2$ gas. As a result, the $f$O$_2$ of the solar nebula decreases relative to those defined by solid-solid buffers as condensation of the major elements (notably Mg and Si) occur (Fig. \ref{fig:cond_fo2}). That of Fe (and Ni) is oxygen-neutral, as its reaction involves no electron exchange, yet would result in net increase in $f$O$_2$ of the gas phase by increasing its O/Fe ratio, given that $f$O$_2$ is proportional to $x\mathrm{O}_2P$, and the total pressure is held constant. This does not arise, however, because this effect is not sufficient to offset the decrease in $f$O$_2$ caused by reaction \ref{eq:cond_olivine} occurring at the same temperature (Fig. \ref{fig:cond_fo2}). Little condensation occurs thereafter, and oxygen fugacity then increases monotonically below 1250 K owing to internal gas phase reactions (chiefly H$_2$O = 0.5O$_2$ + H$_2$ and CO$_2$ = 0.5O$_2$ + CO) that favour the left-hand-side down-temperature. This increase in $f$O$_2$ relative to the iron-wüstite buffer is accelerated upon troilite precipitation (Fig. \ref{fig:cond_fo2}). 

Mineralogically, oxidation is manifest in the redistribution of Fe from its host predominantly in Fe metal above $\sim$700 K into increasingly fayalitic olivine and troilite (FeS) below this temperature threshold. The latter transformation is described by the well-known equilibrium \citep{lewis1972}:

\begin{equation}
Fe(s) + H_2S(g) = FeS(s) + H_2(g)   
    \label{eq:Fe-FeS}
\end{equation}

which, owing to the equal number of molecules of gas in the reactants and products, is independent of total pressure. Although O is not shown in eq. \ref{eq:Fe-FeS}, because the gas is in chemical equilibrium, it can equally be written;

\begin{equation}
Fe(s) + H_2S(g) + 1/2O_2(g) = FeS(s) + H_2O(g)   
    \label{eq:Fe-FeS-O2}
\end{equation}

explicitly highlighting the increasing $f$O$_2$ that attends FeS condensation. This reaction results in a reduction of the modal abundance of Fe metal by a factor 2, from $\sim$27 \% (by mass) to 13 \% by 600 K, causing a concomitant increase in FeS to 15 \%. The consumption of the remaining condensed Fe$^0$ proceeds due to increasing $f$O$_2$ ($f$H$_2$O/$f$H$_2$) below $\sim$600 K \citep{lewis1972};

\begin{equation}
2Fe(s) + H_2O(g) = FeO(s) + H_2(g)
    \label{eq:Fe-oxidation}
\end{equation}

here, FeO is treated as a component in olivine, and reaction \ref{eq:Fe-oxidation} tempers the $f$O$_2$ increase shown in Fig. \ref{fig:cond_fo2}. The Mg\# (molar Mg/[Mg + Fe$^{2+}$]) in olivine decreases from $\sim$0.98 at 700 K to 0.71 by 520 K in our computation \citep[see also][]{grossman1972}, consuming almost the entirety of the remainder of the Fe$^0$ by this temperature. Like eq. \ref{eq:Fe-FeS}, eq. \ref{eq:Fe-oxidation} is pressure-independent. Consequently, and because reactions \ref{eq:Fe-FeS} and \ref{eq:Fe-oxidation} occur at similar temperatures, the abundance of S and the FeO/Fe ratio of condensed material are correlated. The small fractions of remaining metal are oxidised below $\sim$380 K by:

\begin{equation}
3Fe(s) + 4H_2O(g) = Fe_3O_4(s) + 4H_2(g)
    \label{eq:magnetite-oxidation}
\end{equation}

though diffusion-controlled growth rates mean that reaction \ref{eq:magnetite-oxidation} is unlikely to proceed far \citep{hongfegley1998}. At lower temperatures still (below $\sim$350 K), olivine and orthopyroxene may be consumed by the condensation of H$_2$O to form serpentine, talc and brucite, however, the low temperatures mean that reaction kinetics are sluggish, and such minerals are unlikely to form by direct condensation \citep{fegley2000}.


\subsection{Non-equilibrium and fractional condensation}
\label{sec:neb_noneq_cond}

The idealised condensation sequence presented above is subject to the implicit assumption that all condensed material remains in contact with the same parcel of gas over its entire cooling history. This assumption characterises all such equilibrium models \citep[e.g.,][]{grossman1972,woodhashimoto1993,lodders2003,wood2019}. For this condition to hold, it implies a static, slow-cooling solar nebula, in which the gas-solid equilibration timescale ($t_{eq}$) is at least as rapid as the cooling timescale ($t_{cool}$). Observations of T-Tauri stars indicate mean nebular lifetimes of $\sim$1 -- 10 Myr \citep[e.g.,][]{armitage2003}. Kinetic considerations show that $t_{eq}~<~t_{cool}$ is not met, even in a static disk, below $\sim$ 400 K owing to the exponential dependence of diffusion coefficients with reciprocal temperature, putting a lower limit on the equilibrium assumption \citep{fegley2000}.  \\

Temperatures of 400 K represent a lower limit because the analysis in section \ref{sec:disks} emphasises the dynamic nature of disks. Gas and dust not only move radially inward or outward depending on their location in the disk, but also vertically as condensed grains coagulate and settle to the midplane \citep{Cassen2001}. The degree to which settling occurs depends on the relative timescales of coagulation ($t_{coag}$) to the dynamical evolution of the disk ($t_{dyn}$), with the former of the order of 10$^{4}$ yr \citep{weidenschillingcuzzi1993} and the latter $\sim$10$^5$ yr (and increasing with time). Perfect equilibrium is therefore unlikely on dynamical grounds. Settling of coagulated dust to the midplane has several consequences; \textit{i)} it reduces disk opacities, leading to enhanced cooling rates (see section \ref{sec:disks}), \textit{ii)} it fractionates volatile- from refractory elements and \textit{iii)} it enhances the dust/gas ratio of the midplane. \\

The presence of CAIs in contact with chondrules and matrix is evidence of disequilibrium among chondritic meteorites; the condensation sequence illustrated in Fig. \ref{fig:cond_curves} predicts that refractory phases (corundum, perovskite, hibonite, melilite) are no longer present below $\sim$1400 K, having been replaced by feldspar, pyroxenes and olivine \citep[e.g.,][]{grossmanlarimer1974}. Settling to the midplane and subsequent isolation from the evolving gas is one plausible mechanism for preserving CAIs. Alternatively, the cooling rate of the disk could have been so rapid (see section \ref{sec:disk_dynamics}) so as to prevent equilibration of newly condensed phases down-temperature. This occurs when the reciprocal cooling rate $dt^{-1}$ is of the order of the condensation rate, $dN_i/dt$, which is in turn approximated by the Hertz-Knudsen equation:

\begin{equation}
   \frac{dN_i}{dt} = A\frac{\gamma_{cond}(p_{i,eq} - p_i)}{\sqrt{2 \pi m_i k_B T}},
   \label{eq:HK}
\end{equation}

where $N$ is the number of gaseous particles of $i$, $t$ is the time, $A$ is the surface area, $\gamma_{cond}$ the dimensionless condensation coefficient with value 0 $<~\gamma_{cond}~<$ 1, $m$ is the mass of $i$ and $k_B$ the Boltzmann constant. This process would engender supersaturation (i.e., $p_{i} > p_{i,eq}$) of the remaining gas phase components that are unable to condense at their equilibrium $P-T$ conditions, thereby resulting in condensation of new solids at higher-than-equilibrium temperatures \citep[see][]{grossman2012formation}. More sluggish are reactions that involve the conversion of an already condensed phase plus a gas to a new condensed phase;

\begin{equation}
    solid(1) + gas = solid(2).
    \label{eq:solid-gas-general}
\end{equation}

Equation \ref{eq:cond_opx} is an example of one such reaction. Simple collision theory (SCT) has been developed to predict the rates of these reactions \citep{fegley1988cosmochemical,fegley2000}, which scale the Hertz-Knudsen equation (eq. \ref{eq:HK}) by an additional activation energy, $E_a$,

\begin{equation}
    \frac{dN_i}{dt}_{rxn} = \frac{dN_i}{dt} e^{-\frac{-E_a}{RT}},
    \label{eq:arrhenian}
\end{equation}

resulting in a linear scaling for evaporation/condensation kinetics with time. After some time, however, the growth rim of the new solid(2) becomes diffusion-limited, producing a parabolic growth curve where $dN_i$ is proportional to $\sqrt{dt}$ \citep{fegley2000}. A detailed investigation of how cooling rates affect the mineralogy of the resulting phases is underway \citep{charnoz2025}. \\

Chemical evidence for deviations from canonical solar nebula conditions and/or equilibrium comes from the low Mg\# (0.86$\pm$0.003) of olivine and pyroxene in chondrules \citep[][Fig. \ref{fig:mg_no}]{larimer_anders1970,huang1996,oneill_palme1998}. For the canonical case of equilibrium condensation of a solar gas, these Mg\#s would be achieved by $\sim$ 575 K. At these temperatures, chondrules should have their full complement of S, Zn and other moderately volatile elements, which is not observed \citep{wassonkallemeyn1988}. \cite{ebelgrossman2000} suggest high FeO contents could arise from CI-like dust enrichment in the disk (either at the midplane or globally), resulting in $f$O$_2$ of $\Delta$IW-3.1, $\Delta$IW-1.7, and $\Delta$IW-1.2 for dust enrichment factors of 100×, 500× and 1000×, respectively. Resulting olivine Mg\#s at  1200 K are 0.96, 0.8 and 0.65-0.60, 
respectively \citep{ebelgrossman2000}, suggesting $\sim$ 300 -- 400× dust enrichment is required. \add{Using the FactSage model, \cite{mokhtaribourdon2025} find slightly lower Mg\#, 0.90, at 1200~K and 100$\times$ CI dust enrichment, implying roughly 150$\times$ dust enrichment is required to achieve Mg~\# = 0.86.} 
Enrichment factors above $\sim$100 at the midplane are difficult to obtain, however, in dynamical models of the solar nebula \citep{Cassen2001}. Appealing to the accretion of ices to oxidise iron requires comparable degrees of enrichment ($\sim$250$\times$) to achieve a similar H$_2$O/H$_2$ ratio \citep[7.6 $\times$ 10$^{-2}$,][]{fegleypalme1985} which are, in turn, higher than thought plausible by transport of ice-rich planetesimals into the inner disk (5--100 $\times$) in the $\alpha$-disk-based model of \cite{cieslacuzzi2006}.

\section{Mixing and differentiation during planet formation}
\label{sec:planet_acc}

\subsection{Estimating bulk compositions}
\label{sec:planet_comp}

In predicting the bulk compositions of the planets, $T_c$ is of limited use, as it is undefined for elements that do not condense to more than 50 \% relative to their initial mass. This is particularly relevant in the case of O, whose abundance in the planets is much higher than its $T_c$ (180 K) would suggest. Roughly 20 \% of the total nebular O budget is condensed by 1200 K, owing to the formation of solids in which it is bound with metals, yet only a further $\sim$ 3 \% condenses by 500 K, associated with the oxidation of Fe to FeO (eq. \ref{eq:Fe-oxidation}). Hence, the remaining $\sim$ 77 \% condenses as H$_2$O($s$) at much lower temperatures \citep[$\sim$ 180 K;][]{lodders2003}. \\

\add{Therefore, the use of nebular condensation temperatures alone is insufficient to define the bulk compositions of planetary bodies.} To test the extent to which the composition of the Earth and planets can be reproduced through mixing of components formed by nebular condensation, a Gaussian distribution (eq. \ref{eq:gauss}) is employed to sample the compositions of the equilibrium condensates over a range of temperatures. This model, shown to provide a sound match to the volatile element composition of the Earth \citep{sossi2022stochastic}, takes the form here:

\begin{equation}
    N_{i,body} = \sum_T \left[ \frac{N^T_{i,body}}{\sigma \sqrt{2\pi}} \exp \left( -\frac{1}{2}\left[\frac{T-T_0}{\sigma}\right]^2 \right)\right]
\label{eq:gauss}
\end{equation}

where $N_{i,body}$ is the number of moles of element $i$ in the mixture (planetary body), $T_0$ is the mean temperature of the Gaussian distribution and $\sigma$ its standard deviation, and is therefore the sum across all temperatures ($T$) of the product of the number of moles in the body at each temperature step ($N^T_{i,body}$) and the Gaussian probability distribution calculated at $T$. The mole fraction is then given:

\begin{equation}
    x_{i,body} = \frac{N_{i,body}}{\sum_i N_{i,body}}
\label{eq:gauss_molfraction}
\end{equation}

\begin{figure}[!ht]
    \centering
    \includegraphics[width=0.5\linewidth]{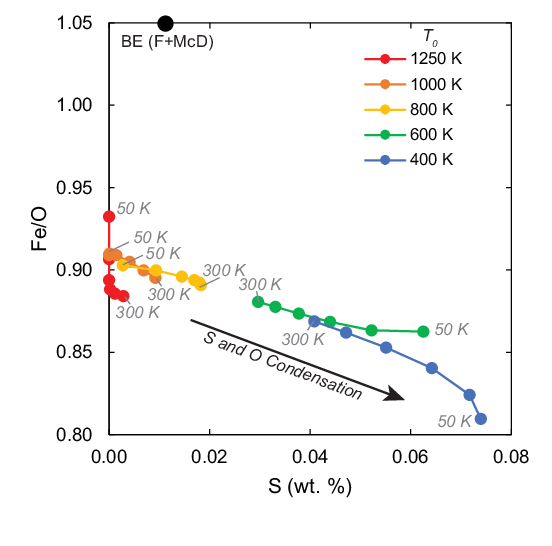}
    \caption{Modelled bulk planetary compositions emerging from mixing of equilibrium nebular condensates from a Gaussian distribution as a function of temperature at 10$^{-4}$ bar. The coloured lines correspond to different values of $T_0$ (red = 1250 K, orange = 1000 K, yellow = 800 K, green = 600 K, blue = 400 K) and grey numbers denote the standard deviation of the Gaussian distribution, $\sigma$. The black point illustrates the composition of the Bulk Earth \citep[BE,][]{fischermcdonough2025}.}
    \label{fig:nebular_mixing}
\end{figure}

The values of $T_0$ and $\sigma$ are varied between ranges of 400 -- 1250 K and 50 -- 300 K, respectively and results shown in Fig. \ref{fig:nebular_mixing}. 







The Earth's composition \citep[BE,][]{fischermcdonough2025} is not captured by such models; falling to higher Fe/O than any mixture of equilibrium nebular condensates. This results from both a high Fe abundance (nearly 32~\% by mass) \textit{and} a low O abundance ($\sim$30~\% by mass) in the BE compared to condensate mixtures; 27--29.5~\% for Fe and 32--33~\% for O. Such iron enrichment in the Earth has been cited as evidence of its `non-chondritic' bulk composition \citep[cf.][]{oneillpalme2008}. 
Mechanisms causing departures from chondritic compositions include, but are not limited to the physical `sorting' of nebular condensates, such as the preferential incorporation of early-condensed forsterite into feedstocks for the Earth relative to later-condensing enstatite, as inferred on the basis of Si isotopes \citep{dauphas2015}. 
Erosion of mantle- relative to core material is another plausible process, mooted to have given rise to the Earth's superchondritic Fe/Mg ratio \citep[2.1 by mass,][]{oneillpalme2008}. Numerical simulations support the dynamical plausibility of collisional erosion in producing superchondritic Fe/Mg ratios \citep{Carte-etal:2015,carter2018collisional}. The potential for evaporative loss at a post-nebular stage to fractionate Fe/Mg is explored in section \ref{sec:chemistry_accretion}. \\


Importantly, the Fe/O ratios and S contents of mixtures of nebular condensates are anticorrelated. This reflects the co-condensation of S, starting near 700~K at 10$^{-4}$ bar, and additional O into the fayalite component of olivine near 550~K (eq. \ref{eq:Fe-oxidation}, Fig. 
\ref{fig:cond_curves}). At $T_0 > 700$~K, increasing $\sigma$ at constant $T_0$ leads to increasing S contents, whereas the opposite is true for $T_0 < 700$ K.  Thus, the more volatile-rich the body, the more oxidised it is expected to be. Concretely, values of $T_0$ around 900 K would be required \citep[at constant $\sigma$ = 225~K, see][]{sossi2022stochastic} in order to condense quantities of S sufficient to match its estimated BE abundance \citep[$\sim$1 wt.~\%][]{fischermcdonough2025}. 
However, at this $T_0$ value, the Earth would be too enriched in moderately volatile elements (Na, K, Zn, among others) relative to observations \citep[which indicate $T_0$ for the Earth of $\sim$1150 K,][]{sossi2022stochastic}. This problem is exacerbated in the small telluric bodies; Vesta, APB and the Moon are more oxidised than the Earth (Fig. \ref{fig:fO2_planets}) yet more depleted in volatile elements (Fig. \ref{fig:KU_RbSr}). Together, these considerations are grounds to reject the hypothesis that the composition of the Earth (and, by extension, those of the other terrestrial planets) reflect mixing of components produced by equilibrium condensation of a canonical solar nebula alone. 


\subsection{Internal differentiation}
\label{sec:internal_diff}

In order to translate planetary bulk compositions into those of their observable mantle and core, internal differentiation must be considered \citep[e.g.,][]{ringwood1966chemical}. Core-mantle differentiation in growing planets is thought to have occurred during transient periods of (partial) melting at high pressures (several -- to tens of GPa) and temperatures ($>$2000 K) informed by experiments and \textit{ab-initio} simulations \citep[e.g.,][]{ringwood1991,thibaultwalter1995,oneill1998,gessmann2001,rubie2004,wadewood2005,ricolleau2011,fischer2015,fischer2020,suer2017sulfur,huangbadro2018,li2020,huang2021,blanchard2022}. 

To leverage these constraints, here we develop a model that takes as input parameters the pressure and temperature of core-mantle equilibrium, and the bulk composition of the material (section \ref{sec:planet_comp}) in the system Si-Al-Ca-Mg-Fe-Ni-O, which, together, comprise $\sim$98.5~wt.~\% of the Earth \citep{palme2014treatise}. This approach was first formalised by \cite{rubie2004} and expanded by \cite{frost2008redox} and \cite{rubie2011}. Our treatment is formally incomplete as it neglects other, minor core-forming elements, notably Co, S, C and H, which, together, likely comprise $\sim$2.5~wt.\% of the core \citep{hirose2021}. The partitioning of the alloying elements - Ni, O and Si - are described according to eq. (\ref{eq:Fe_exchange}) and in Table \ref{tab:KDs} over a range from 1 bar to 100 GPa at a temperature corresponding to the peridotite liquidus of \cite{andrault2011}. Three compositions are taken from mixtures of nebular condensates at different $T_0$ and $\sigma$ values from section \ref{sec:planet_comp}, and the fourth is the Bulk Earth composition (Table \ref{tab:abundances}).

\begin{table}[!ht]
    \centering
    \caption{Parameters for metal-silicate partitioning exchange reactions.}
    \begin{tabular}{lcccc}
    \hline
        \textbf{} & \textbf{log$K$} & \textbf{ln$\gamma$ } & \textbf{$n$} & \textbf{Reference} \\ \hline
        \textbf{Si} & 0.52-13000/$T$ & 0.5-5500/$T$ & 4 & S13, F15 \\ 
        \textbf{Fe} & - & 0  & 2 & Ri11 \\ 
        \textbf{O} & 0.7-5000/$T$+(22$P$)/$T$ & 0  & - & Ri11, S13, F15 \\ 
        \textbf{Ni} & 1.06+1553/$T$-(98$P$)/$T$ & 0  & 2 & Ru11 \\ \hline
        \multicolumn{5}{l}{Ri11 = \cite{ricolleau2011}, Ru11 = \cite{rubie2011}, S13 = \cite{siebert2013}, F15 = \cite{fischer2015}}
    \end{tabular}
    \label{tab:KDs}
\end{table}

As iron is the dominant element in planetary cores, and is typically present at several weight percent in the coexisting mantles (cf. Table \ref{tab:chemphys_prop}), element partitioning is described by an exchange coefficient with Fe \citep[e.g.,][]{wadewood2005},

\begin{equation}
    \frac{n}{2}Fe + MO_{\frac{n}{2}} = \frac{n}{2}FeO + M
    \label{eq:Fe_exchange}
\end{equation}

which circumvents the definition of $f$O$_{2}$ but requires that $n$, the number of electrons transferred in the reaction, be known. The abundances of Ni, O and Si in the mantle are allowed to vary in order to satisfy the equation;

\begin{equation}
    x(M)_{core} = \frac { K_{(\ref{eq:Fe_exchange})} (aFe)^{\frac{n}{2}} aMO_{\frac {n}{2} } } { \gamma M (aFeO)^{\frac{n}{2}} } = \frac{(xM)_T - (xM)_{mantle}(1-CMF)  } {CMF}
    \label{eq:KD_mass_balance}
\end{equation}

where the subscript $T$ is the total mass and $CMF$ the core mass fraction. The values of $(xM)_{mantle}$ and $CMF$ are iterated until the left-hand side is equivalent to the right-hand side. The iteration is performed with the additional constraints that;

\begin{equation}
    x(O)_{mantle} = 1.5[x(Al)_{mantle}] + x(Ca)_{mantle} + x(Mg)_{mantle} + 2[x(Si)_{mantle}] + x(Fe)_{mantle} + x(Ni)_{mantle}
    \label{eq:O_mass_balance}
\end{equation}

and 

\begin{equation}
\sum_M x(M)_{mantle} = 1
\label{eq:mantle_mass_balance}
\end{equation}

The remaining elements, Al, Ca and Mg, are assumed to be perfectly lithophile, hence their mantle abundances are given analytically by $C_{mantle} = C_T/(1-CMF)$. This leaves \textit{four unknowns} - $x(Fe)_{mantle}$, $x(Ni)_{mantle}$, $x(Si)_{mantle}$ and $CMF$, which are soluble with four constraints (eq. \ref{eq:KD_mass_balance} for M = Ni, Si and O and eq. \ref{eq:mantle_mass_balance}). 

\begin{table}[!ht]
    \centering
    \caption{Modelled bulk compositions using the parameters for eq. \ref{eq:gauss} as described in section \ref{sec:planet_comp}. All elements in wt. fraction. Bulk Earth from \cite{fischermcdonough2025}.}
    \begin{tabular}{lcccc}
    \hline
        \textbf{} & \textbf{Fully reduced} & \textbf{Bulk Earth} & \textbf{Intermediate} & \textbf{Fully oxidised} \\ \hline
        \textbf{Parameters} &  &  &  &   \\ \hline                
        \textbf{$T_0$} & 1250 & - & 600 & 400  \\ 
        \textbf{$\sigma$} & 50 & - & 50 & 50  \\ \hline
        \textbf{Elements} &  &  &  &   \\ \hline        
        \textbf{Al} & 0.0166 & 0.0159 & 0.0141 & 0.0138  \\ 
        \textbf{Ca} & 0.0174 & 0.0167 & 0.0148 & 0.0145  \\ 
        \textbf{Mg} & 0.1668 & 0.1578 & 0.1524 & 0.1491  \\ 
        \textbf{Si} & 0.1636 & 0.1567 & 0.1721 & 0.1684  \\ 
        \textbf{Fe} & 0.2971 & 0.3268 & 0.2913 & 0.2850  \\ 
        \textbf{O} & 0.3187 & 0.3073 & 0.3377 & 0.3520  \\ 
        \textbf{Ni} & 0.0198 & 0.0187 & 0.0176 & 0.0172  \\ \hline
        \textbf{Fe/O} & 0.932 & 1.063 & 0.863 &  0.809  \\ \hline        
    \end{tabular}
    \label{tab:abundances}
\end{table}

\begin{figure}[!ht]
    \centering
    \includegraphics[width=1\linewidth]{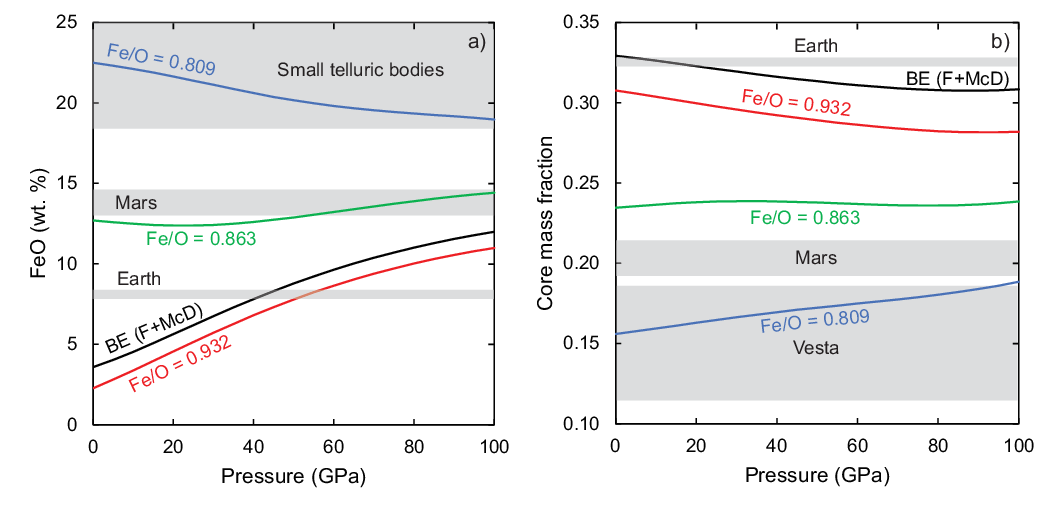}
    \caption{The equilibrium a) FeO content in the mantle and b) core mass fraction of synthetic rocky planets calculated along the peridotite liquidus of \cite{andrault2011} plotted as a function of pressure for different bulk compositions with Fe/O (by weight) ratios ranging from fully reduced (all Fe as Fe$^0$), 0.932 (red), 0.863 (green) and 0.809 (blue, fully oxidised, all Fe as FeO). These bulk compositions are derived from the model presented in section \ref{sec:planet_comp} at different $T_0$ and $\sigma$ values. The black curve has Fe/O = 1.063 and is the Bulk Earth composition of \cite{fischermcdonough2025}. All compositions are shown in Table \ref{tab:abundances}. Observed FeO contents and core mass fractions of selected telluric bodies are shown as grey horizontal bars (see Table \ref{tab:chemphys_prop} for data sources).}
    \label{fig:core_fm_P}
\end{figure}

As first proposed by \cite{javoy1995}, even for precursor materials that contain all Fe as Fe$^0$ (i.e., in reduced form), the process of core-mantle equilibration at increasing $P-T$ results in progressively higher FeO contents in the mantle (Fig. \ref{fig:core_fm_P}). 
This behaviour arises due to the competition between the incorporation of Si in the core via;

\begin{equation}
    SiO_2(mantle) + 2Fe(core) = Si(core) + 2FeO(mantle)
    \label{eq:si_fe}
\end{equation}

and that of O by;

\begin{equation}
    FeO(mantle) = Fe(core) + O(core)
    \label{eq:o_fe}
\end{equation}

Both reactions are self-limiting \citep[cf.][]{oneill1998}: FeO contents tend to converge as $P-T$ increase along the peridotite liquidus\add{, which reflects the tendency for both Si and O to partition into the core as $P-T$ increases} (Fig. \ref{fig:core_fm_P}a).\add{ At a given $P-T$, whether Si or O predominantly enters the core depends on the Fe/O ratio.} For initially high Fe/O, the low $a$FeO promotes Si dissolution in the core, which, in turn, engenders an increase in $a$FeO via eq. \ref{eq:si_fe}; a negative feedback that results in a keeling over of mantle FeO content (red and black curves, Fig. \ref{fig:core_fm_P}). 
Because mass is conserved, any change in mantle FeO content must be compensated for by a concomitant adjustment in the core mass fraction (Fig. \ref{fig:core_fm_P}b). At low Fe/O, the $CMF$ increases because both Fe and O enter the core (eq. \ref{eq:o_fe}). Conversely, the $CMF$ decreases when eq. \ref{eq:si_fe} prevails over eq. \ref{eq:o_fe}, as the reduction of 1 mole of SiO$_2$ to Si in turn oxidises 2 moles of Fe to FeO. A corollary of the model is that the $f$O$_2$ of core formation changes as a function of $P$ and $T$ at \textit{constant} bulk composition \citep[see also][their Fig. 2]{rubie2011}.\\


For single-stage core formation with the BE composition, we find FeO = 8.33 wt.~\% and Ni = 1881 ppm in the mantle with a CMF of 0.315 at $P$ = 40 GPa and $T$ = 3150 K. These values compare well with the 8.1$\pm$0.2 wt.~\% FeO and 1860$\pm$93 ppm Ni in the BSE \citep{palme2014treatise} and a CMF of $\sim$0.325 (Table \ref{tab:chemphys_prop}). At these conditions, our model predicts Si = 3.6 wt.~\% and O = 1.2 wt.~\% in the core, consistent with other estimates \citep{oneill_palme1998,badro2014,dauphas2015,hirose2021}. That there is no single composition that fits the observed $CMF$ and FeO at a given pressure should not be taken as evidence against single-stage core formation. At 40 GPa, this is likely remedied by the additional S ($\sim$2~wt.~\%), Co + P + Cr ($\sim$1.5~wt.~\%) in the core, which are not included in our model. Addition of these quantities to the $CMF$ value of 0.315 at 40 GPa would result in the observed $CMF$ of 0.325. \\

Such estimates also match the 30--50 GPa and 3000 K inferred from $D_{Ni}$ and $D_{Co}$ \citep{thibaultwalter1995,liagee1996,bouhifdjephcoat2003,wadewood2005,kegler2008}. \cite{wadewood2005} and \cite{rubie2011,rubie2015accretion} emphasised that, although homogeneous accretion models are readily able to satisfy the observed FeO, Ni and Co abundances in Earth's mantle (as also shown here), they lead to elevated mantle Cr and V contents, which would instead require temperatures $\sim$650~K higher. A notable caveat to their models lies in the implicit assumption that the extant families of chondrites \textit{de facto} formed Earth. Should this assumption be relaxed, it remains to be seen whether heterogeneous accretion is required from a chemical standpoint. Moreover, \cite{siebert2013} demonstrated V and Cr abundances could be fit, independent of whether accretion were homogeneous or heterogeneous with respect to composition, at (mean) pressures of equilibrium between 40 and 60 GPa, by virtue of the negative interaction parameters between Cr-O and V-O in iron-rich alloys. In either case, as is evident from Fig. \ref{fig:core_fm_P} and highlighted by \cite{rubie2015accretion}, the composition of Earth-forming material would need to have been initially somewhat reducing (Fe/O $>$ 0.9) and high in total Fe ($\sim$32 wt.~\%; Table \ref{tab:abundances}), as lower Fe/O ratios lead to mantle FeO contents that are too high relative to that observed (8.1$\pm$0.02 wt.~\%; Table \ref{tab:chemphys_prop}). \\

On the other hand, small telluric bodies have mantle FeO contents and core mass fractions (Table \ref{tab:chemphys_prop}) that appear to require oxidised starting materials (Fe/O $<$ 0.86; Fig. \ref{fig:core_fm_P}) because the low-pressure ($<$ 5 GPa) core formation that must have occurred on these bodies would have led to temperatures insufficient \citep[e.g.,][]{cartier2024}, for self-oxidation to operate to the extent to reach the $\sim$20 wt.~\% FeO observed \textit{via} eq. \ref{eq:si_fe} in their mantles. 
Mantle FeO contents inferred for the silicate mantles of iron meteorite parent bodies are almost invariably high, with all but the IIABs lying between 10--25 wt.~\% \citep{grewal2024}. Rather, the volatile-poor nature of STBs, such as Vesta and the angrite parent body \citep[e.g.,][]{oneillpalme2008} together with the wide range of volatile contents in iron meteorites \citep{scottwasson1975,hirschmann2021} at near constant mantle FeO \citep{grewal2024} indicates that the relationship between $f$O$_2$ and volatile element content expected for nebular condensation (Figs. \ref{fig:cond_fo2}, \ref{fig:nebular_mixing}) decoupled during the formation of STBs.
Finally, even at the lowest Fe/O ratios considered here, FeO contents relevant to Mercury's mantle \citep[0.2$\pm$0.1 wt.~\%][Table \ref{tab:chemphys_prop}]{namur2016,nittler2018} are not obtained, suggesting it accreted material more reduced and with higher metal/silicate ratios than that produced through equilibrium condensation and/or contains significant S or C that are not considered in this model.

\section{Physicochemical conditions of volatile depletion}
\label{sec:chemistry_accretion}

One possible explanation for the divergence in the compositions of the terrestrial planets and STBs relative to chondrites is that they (or their components) underwent chemical fractionation under conditions that differed from those set by the nebular gas \citep[e.g.,][]{oneillpalme2008}. 


\subsection{Chemical fractionation}
\label{sec:chem_consequences}

Here we examine the \textit{relative} fractionation between two volatile elements, $M1$ and $M2$, such that the physical mechanism by which the two elements were lost cancels from the operation 
and their relative fractionation is dependent on the thermodynamics of the presumed vaporisation reaction alone.\\

For most metals, vaporisation reactions are simple because one condensed component (e.g., MnO) and one gas species (e.g., Mn) predominate over a range of planetary- and nebular conditions \citep[Table \ref{tab:cond_phases},][]{oneill1991moon,sossi2019evaporation}. When comparing the fraction of an element vaporised ($fM1$) relative to that of a second comparator element ($fM2$), the total pressure of the system cancels, as does the total budget of the element in the system, to yield \citep{sossi2024moon}:

\begin{equation}
\frac{f^{vap}_{M1}}{(1-f^{vap}_{M1})} = \frac{f^{vap}_{M2}}{(1-f^{vap}_{M2})} \left(\frac{\gamma _{M1}}{\gamma _{M2}}\right) (fO_2)^\frac{\Delta{n_{M2-M1}}}{4} \exp \left(\frac{(\Delta G^o_{M2}-\Delta G^o_{M1})}{RT}\right),
\label{eq:relative_vap}
\end{equation}

where ($\Delta G^o_{M2}-\Delta G^o_{M1}$) relates to the standard state free energy of the exchange reaction between the pure components vapour and condensed phase(s), such as
  
\begin{equation}
Mn(g) + NaO_{0.5}(s,l) + \frac{1}{4}O_2(g) = Na(g) + MnO(s,l).
\label{eq:Mn-Na}
\end{equation}

Because relative entropy changes for gas-liquid or gas-solid reactions are dominated by the (ideal) gas, the quantity ($\Delta G^o_{M2}-\Delta G^o_{M1}$) is nearly independent of temperature, and exchange reactions become more discriminating (i.e., larger fractionation of $M1$ from $M2$) proportional to exp(1/$T$). While the free energy change ($\Delta G^o_{M2}-\Delta G^o_{M1}$) is readily tabulated from thermodynamic databases (e.g. JANAF), the activity coefficients for many components in silicate melts are uncertain \citep[see][]{sossi2018thermodynamics,fegley2023chemical}. However, divalent metal oxides mix near-ideally in silicate liquids \citep{oneilleggins2002,woodwade2013}, while activity coefficients ($\gamma$) for the alkali metal oxides are typically very small but relatively well characterised \citep{charles1967,mathieu2011,sossi2019evaporation}. Here, measurements of activity coefficients at a given temperature are extrapolated according to:

\begin{equation}
 \frac{\mathrm{ln} \gamma}{ \mathrm{ln} \gamma_{ref}} = \frac{T_{ref}}{T},    
 \label{eq:activity_coeff_ex}
\end{equation}

where $\gamma$ tends to 1 at infinite temperature. Values of $\gamma_{ref}$ and $T_{ref}$ are given in Table \ref{tab:evap_table}. 

\begin{table}[!ht]
    \centering
    \caption{Thermodynamic properties of evaporation reactions for selected moderately volatile elements of the form eq. \ref{eq:reaction_stoichiometry}.}
    \begin{tabular}{cccccc}
    \hline
        \textbf{} & \textbf{$\Delta$H$^o$ (kJ/mol)} & \textbf{$\Delta$S$^o$ (kJ/mol.K)} & \textbf{$n$} & \textbf{$\gamma_{ref}$$^{*}$} & Source \\ \hline
        \textbf{Li} & 419.0 & 0.141 & 1 & 0.2 & \citep{sossi2019evaporation}  \\ 
        \textbf{K} & 236.1 & 0.122 & 1 & 7.3$\times 10^{-5}$ & \citep{sossi2019evaporation,wolf2023}  \\ 
        \textbf{Na} & 267.0 & 0.121 & 1 & 5.0$\times 10^{-4}$ & \citep{sossi2019evaporation,wolf2023} \\ 
        \textbf{Mn} & 606.3 & 0.173 & 2 & 1 & \citep{kohn1994importance} \\ 
        \textbf{Mg} & 666.4 & 0.186 & 2 & 0.20 & \citep{wolf2023} \\ \hline
        \multicolumn{5}{l}{$^{*}$\footnotesize{$\gamma_{ref}$ calculated for the oxide in silicate liquid at a reference temperature of 1673 K.}}
    \end{tabular}
    \label{tab:evap_table}
\end{table}

For a canonical solar nebula gas, the $f$O$_2$ is roughly $\Delta$IW-6 during the condensation of most rock-forming elements (Fig. \ref{fig:cond_fo2}). The $f$O$_2$ during  melting and vaporisation/condensation on planets \textit{after} the dispersal of the nebular gas, reflects the composition of the evaporating/condensing material \citep[e.g.,][]{visscherfegley2013}. The FeO contents of small telluric bodies and oxybarometry of basaltic achondrites imply oxygen fugacities near $\Delta$IW-1 \citep[Fig. \ref{fig:fO2_planets}, Table \ref{tab:chemphys_prop},][]{wadhwa2008redox}. \\

\begin{figure}[!ht]
    \centering
    \includegraphics[width=1\linewidth]{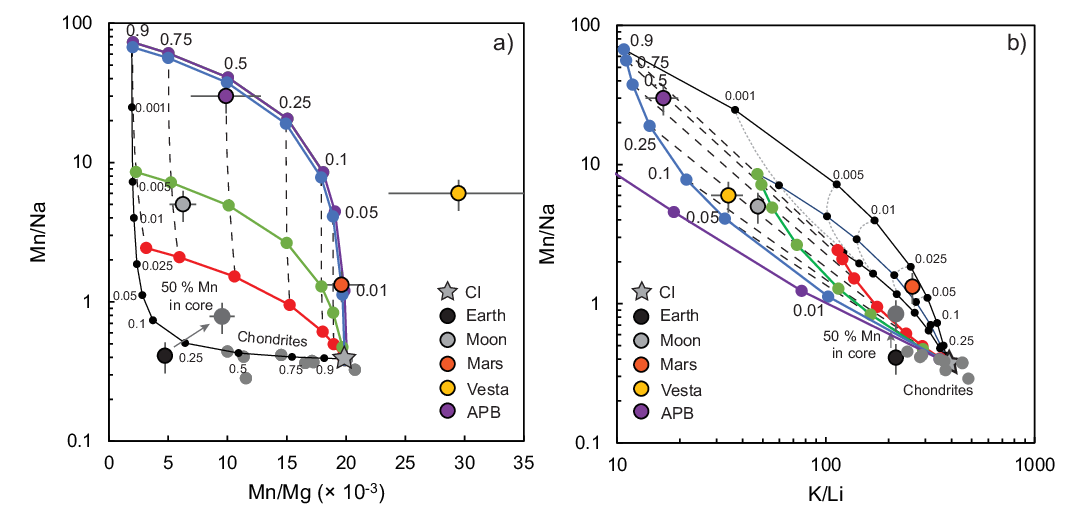}
    \caption{\textbf{a)} Mn/Na vs. Mn/Mg and \textbf{b)} Mn/Na vs. K/Li ratios among planetary bodies. The coloured curves denote the evolution of the composition of an evaporation residue calculated using eq. \ref{eq:relative_vap} at 1100 K (purple), 1500 K (blue), 2000 K (green) and 2500 K (red) at a fixed relative oxygen fugacity ($\Delta$IW-1, except for 1100 K, which was calculated at $\Delta$IW-6, see text), with numbers denoting the fraction of Mn loss undergone by the body. The black dashed lines connect equal degrees of Mn loss at different temperatures. The solid black curves trace a binary mixture of a volatile-depleted end-member with 90 \% Mn loss and a CI-like end-member. The small numbers denote the mass fraction of CI in the mixture. On a), only the volatile-depleted end-member at 1500 K is shown due to clarity, whereas in b) the end-members are taken at the three temperatures, and the dotted grey lines connect equal fractions of CI material. The grey point gives the composition of the Earth assuming 50 \% of its Mn budget resides in the core \citep[see][]{siebert2018}. The grey field denotes the compositional range of chondrites \citep{wassonkallemeyn1988,siebert2018}. }
    \label{fig:mn-mg-na-k-li}
\end{figure}

The change in the Mn/Na (Fig. \ref{fig:mn-mg-na-k-li}a) and K/Li (Fig. \ref{fig:mn-mg-na-k-li}b) ratios of evaporation residues/partial condensates are computed as a function of the Mn/Mg ratio using eq. \ref{eq:relative_vap}.
The Mn/Na ratio is not only sensitive to $f$O$_2$ \citep{oneillpalme2008}, but also to temperature at constant relative $f$O$_2$. For 25 \% Mn loss, 70 \%, 89 \%  and 98.5 \% of Na is lost at 2500 K, 2000 K and 1500 K, respectively.\add{ Therefore, at higher temperatures, the fraction of Na lost approaches that of the Mn lost. That is, evaporation exchange reactions (eq. \ref{eq:relative_vap}) become less discriminating.} Under nebular conditions (1100 K, $\Delta$IW-6), the Na loss is nearly identical to the 1500 K, $\Delta$IW-1 case (98.6 \% lost), but the resulting K/Li is 10-fold lower (compare blue and purple curves). 
The APB has lost $\sim$50 \% of its Mn, 
and has Mn/Na of $\sim$30, achievable for single-stage evaporation at $\Delta$IW-1 and 1500~K; conditions also consistent with its K/Li.  
By contrast, the Mn/Na and Mn/Mg range of chondrites is explained by mixtures of a volatile-depleted component and CI chondrites (thin black line, Fig. \ref{fig:mn-mg-na-k-li}a), equivalent to chondrules (volatile-poor) and matrix (volatile-rich) \citep[e.g.,][]{alexander2019,hellmann2020}. \\ 


The Earth has Mn/Na -- Mn/Mg ratios lower than for any chondrite, its Mn/Na being too low to have been set by evaporation alone. The degree of Na depletion ($\sim$ 80 \% lost) in the Earth can be used to infer the maximum permitted Mn loss by volatility alone. Roughly 36 \% Mn would have been lost at 2500 K, decreasing to 2\% for 1500 K or in the solar nebula. 
This implies between 1/2- to 2/3rds of the Earth's Mn budget is in the core \citep[see also][]{drake1989,ringwood1991,siebert2018}; a similar result arises when assuming Li is entirely lithophile and is as volatile as Mn. \\ 

Compositions of STBs can be reproduced by vapour loss at a single set of temperature-$f$O$_2$, favouring relatively `high' temperatures of about $1400-1800$ K and relatively high $f$O$_2$ at $\Delta$IW-1 compared to conditions in the solar nebula. Vesta is a possible exception (Fig. \ref{fig:mn-mg-na-k-li}a), perhaps estimates for its mantle Mg content are too low.  The fact that the elemental fractionations observed differ empirically from those of chondrites indicates distinct processes were responsible in setting their bulk compositions; that is, evaporation (either in a single-stage or an average of multiple events) for STBs rather than mixing among components as is relevant for chondrites \citep{alexander2019}. 
This result, coupled with the ancient ($<$ 5 Myr after CAIs) volatile depletion ages implied by the low $^{87}$Sr/$^{86}$Sr$_i$ \citep[] []{Hans2013rbsr} leads to the conclusion that there must have been localised, `oxidised' domains at the loci of planetesimal formation, even in the presence of the nebular gas. \\

On the other hand, the compositions of Mars and Earth (see also Section \ref{sec:multi-stage-loss} for Earth) cannot be reconciled with volatile loss at any given temperature.\add{ This is because the temperature inferred from Fig. \ref{fig:mn-mg-na-k-li}a differs from that deduced from \ref{fig:mn-mg-na-k-li}b. Implicit in these models is the absence of the major volatiles, such as H, C and S, which may modify the stable gas species considered in equations such as eq. \ref{eq:Mn-Na}, that assume monatomic gases. Thermodynamic models indicate the stability of hydroxide- and sulfide-bearing metallic species, in particular \citep{fegley2016solubility,ivanov2022trace}, yet a systematic investigation of how such differences influence the extent of volatile loss are yet to be undertaken.}
Nevertheless,\add{ within the framework explored here,} mixing between volatile-depleted and volatile-rich end-members is required. It is also clear from Fig. \ref{fig:mn-mg-na-k-li}, however, that these end-members cannot have been identical for Earth and Mars.  
These observations suggest that more complex, multi-component mixtures of volatile-poor and volatile-rich materials are likely needed to account for the compositions of the Earth and Mars.


\subsection{Isotopic fractionation}
\label{sec:iso_consequences}


Partitioning of an element between two phases during cosmo-/geochemical processes is often accompanied by fractionation of its isotopes. 
The estimated mantle isotopic compositions of planetary objects are determined via analyses of (ultra-)mafic rocks. On Earth, this typically involves peridotites \citep[e.g.,][]{sossi2016iron,savage2015copper}, or, for (highly) incompatible elements, basalts/komatiites \citep[e.g.,][]{dauphas2010magnesium,hibbert2012iron,jerram2020cr}. For  other planetary bodies, we are limited to basaltic samples, as there are no mantle samples. 
Consequently, particularly for achondrites, the limited number of samples \citep[cf. the angrite parent body,][]{keil2012angrites}, and their genetic relationship with their source \citep[cf. geochemical signatures in Martian shergottites,][]{borg2002mars} introduces uncertainty into the assignment of the measured isotopic signature to that of the body. 
Nonetheless, isotopic bulk compositions determined in this manner are compiled in Table \ref{tab:isos}. 


\begin{table}
\centering
    \caption{Estimated mean values of the mass-dependent, stable, isotopic compositions of planetary bodies. All values expressed in per mille deviations from the listed standard.}

\begin{tabular}{l l l l l l l l l l}   
\hline
 & APB & EPB & Moon & Mars & Earth & CC & OC & EC & References \\ \hline
\(\delta \)\textsuperscript{7/6}Li\textsubscript{L-SVEC} &  & 3.7 & 3.8 & 4.4 & 3.5 & 3.2 & 2.7 & 1.9 & 1-3 \\
95\% c.i. &  & 0.1 & 0.8 & 1.6 & 0.5 & 0.6 & 0.4 & 0.6 &  \\
\(\delta \)\textsuperscript{25/24}Mg\textsubscript{DSM-3} & -0.079 & -0.114 & -0.130 & -0.113 & -0.121 & -0.140 &  & -0.134 & 4-6 \\
95\% c.i. & 0.010 & 0.008 & 0.010 & 0.006 & 0.003 & 0.010 &  & 0.004 &  \\
\(\delta \)\textsuperscript{30/28}Si\textsubscript{NBS-28} & -0.21 & -0.42 & -0.29 & -0.48 & -0.30 & -0.47 &  & -0.69 & 7,8 \\
95\% c.i. & 0.03 & 0.03 & 0.08 & 0.03 & 0.03 & 0.03 &  & 0.05 &  \\
\(\delta \)\textsuperscript{41/39}K\textsubscript{SRM3141a} & -1.70 & 0.41 & -0.04 & -0.30 & -0.42 & -0.25 & -0.78 & -0.19 & 9-17 \\
95\% c.i. &  & 0.08 & 0.06 & 0.03 & 0.01 & 0.14 & 0.34* & 0.39* &  \\
\(\delta \)\textsuperscript{53/52}Cr\textsubscript{SRM979} &  & -0.220 & -0.216 & -0.17 & -0.124 & -0.128 & -0.102 & -0.050 & 18-22 \\
95\% c.i. &  & 0.030 & 0.020 & 0.08* & 0.030 & 0.011 & 0.006 & 0.006 &  \\
\(\delta \)\textsuperscript{57/54}Fe\textsubscript{IRMM014} & 0.190 & 0.010 & 0.08 & -0.010 & 0.050 & -0.010 &  &  & 23,24 \\
95\% c.i. & 0.020 & 0.010 & 0.03 & 0.020 & 0.010 & 0.010 &  &  &  \\
\(\delta \)\textsuperscript{65/63}Cu\textsubscript{JMC-Lyon} &  & 0.5 & 0.50 &  & 0.70 & -0.7 & -0.15 & -0.25 & 25-27 \\
95\% c.i. &  & 0.5** & 0.10 &  & 0.010 & 1.3* & 0.5* & 0.08 &  \\
\(\delta \)\textsuperscript{66/64}Zn\textsubscript{JMC-Lyon} &  & 1.1 & 1.39 & 0.237 & 0.170 & 0.36 & 0.01 & 0.23 & 28-35 \\
95\% c.i. &  & 2.2** & 0.12 & 0.030 & 0.011 & 0.08 & 0.32 & 0.07 &  \\
\(\delta \)\textsuperscript{71/69}Ga\textsubscript{IPGP} &  &  & 0.14 &  & 0.000 & -0.15 & -0.6 & -0.23 & 36-38 \\
95\% c.i. &  &  &  &  & 0.016 & 0.11 & 0.8* & 0.10 &  \\
\(\delta \)\textsuperscript{87/85}Rb\textsubscript{SRM984} & -1.2 & 1.0 & 0.03 & 0.10 & -0.13 & 0.11 & -0.12 & 0.02 & 39-42 \\
95\% c.i. &  & 1.4 & 0.03 & 0.03 & 0.06 & 0.10 & 0.20 & 0.27 &  \\
\hline
\label{tab:isos}
\end{tabular}

*Uncertainty quoted as 2s instead of 95 percent confidence interval as the data in literature do not pass Wilk's test for a normal distribution around a single population mean.

**Uncertainty quoted as 2s instead of 95 percent confidence interval because of the potential presence of a (unquantifiable) systematic uncertainty (see text).

For Mg, Si and Fe, no OC (and EC for Fe) data are shown, because they have been grouped with CC as their compositions are not statistically different.
References:
1-3: \cite{magna2006,magna2014,poggevonstrandmann2011}

4-6: \cite{hin2017magnesium,liu2023magnesium,klaver2024titanium}

7,8: \cite{armytage2011,dauphas2015}

9-17: \cite{tian2019,tian2020,kujacobson2020,wang2016,hu2022potassium,bloom2020,jiang2021,koefoed2020}

18-22: \cite{schoenberg2008cr,schoenberg2016cr,bonnand2016cr,sossi2018cr,zhu2019vesta,zhu2021chromium} 

23,24: \cite{sossi2016iron,poitrasson2019}

25-27: \cite{herzog2009,savage2015copper, dhaliwal2024}

28-35: \cite{luck2005zn,moynier2011zn,paniello2012zinc,sossi2018zinc,paquet2023origin, fang2024origin}

36-38: \cite{kato2017gallium,kato_etal2017gallium,wimpenny2022gallium}

39-42: \cite{pringlemoynier2017,nie2023rb,wang2023rubidium,wang2024rubidium}

K: chondrite data are from observed falls only to avoid potential effects of terrestrial weathering.

Zn: as for K, chondrite data are from observed falls only, with EC data deriving from EH only due to suspected metamorphism effects on the analysed EL6 falls.

\end{table}

\subsubsection{The major elements Mg, Si and Fe}
\label{sec:iso_MgSiFe}

\begin{figure}[!ht]
    \centering
    \includegraphics[width=1\linewidth]{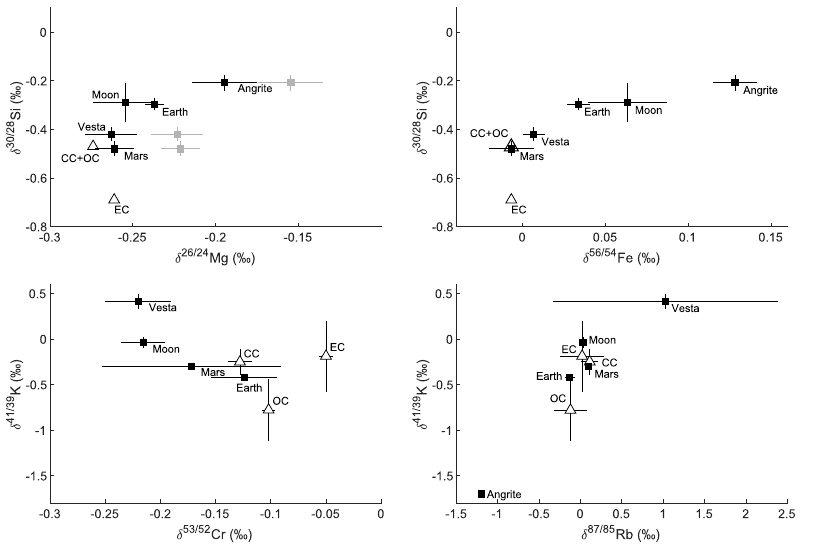}

        \caption{Mass-dependent isotopic variations among bulk differentiated (squares) and undifferentiated (triangles) planetary objects. a) $\delta ^{30/28}$Si vs. $\delta ^{26/24}$Mg. The grey points correspond to measured values pre-correction (see main text for details). b)  $\delta ^{30/28}$Si vs. $\delta ^{56/54}$Fe. c)  $\delta ^{41/39}$K vs. $\delta ^{53/52}$Cr. d)  $\delta ^{41/39}$K vs. $\delta ^{87/85}$Rb. See Table \ref{tab:isos} for data sources.
}
    \label{fig:stable_isos_binary}
\end{figure}

The estimated Mg, Si and Fe isotope ratios of bulk, differentiated planetary bodies are similar to- or higher than those of the (averages of) carbonaceous, ordinary and enstatite chondrites (Fig. \ref{fig:stable_isos_binary}a,b; Table \ref{tab:isos}).
Taking hydrothermal alteration for type 1 and 2 carbonaceous chondrites into account \citep{young1999fluid,hin2017magnesium}, estimated ratios of Mg, Si and Fe isotopes are higher in the BSE than they are in chondrites. The same applies to the Moon \citep[e.g.,][]{klaver2024titanium,armytage2012silicon,sossi2016iron}, whose isotopic compositon is indistinguishable from that of the Earth for these elements. Vesta (as estimated largely by eucrites) and Mars (via lherzolitic shergottites), on the other hand, have Fe and Si isotope ratios that overlap with carbonaceous and ordinary chondrites \citep[though heavier than ECs; Fig. \ref{fig:stable_isos_binary},
][]{dauphas2015}. Although these bodies may appear heavier in $\delta^{26/24}$Mg (light grey squares) the current estimates should be corrected downward by  $\sim$0.04‰ to account for the effect of silicate differentiation \citep{liu2022equilibrium,liu2023magnesium}, which also renders their values similar to chondrites \citep[see also][black symbols in Fig \ref{fig:stable_isos_binary}a]{young2019near}. 
\\

The Mg, Si and Fe isotope ratios in bulk planetary objects correlate (Fig. \ref{fig:stable_isos_binary}a, b), implying a control by the same cosmo-/geochemical fractionation process. 
Because Mg, Si and Fe are major elements, only reservoir separation involving significant mass fractions of these elements or with large isotope fractionation factors, could have affected their mantle isotopic ratios. This leaves core-mantle-crust separation or vapour-condensed-phase fractionation \citep{poitrasson2004iron,sossi2016iron,hin2017magnesium,young2019near}. 
Silicon and, in particular, Mg are sparingly soluble in metallic Fe during the differentiation of small telluric bodies. This factor, compounded with the fact that the Fe isotope fractionation factor between metal and silicate is small at low pressures \citep{hin2012experimental} in the absence of S \citep{kubik2022absence}, means that core formation is unlikely to have caused substantial isotopic fractionation of Fe, and, by extension, Mg and Si \citep[although debate is ongoing, see][]{bourdon2018isotope,shahar_young2020}. Despite the fact that the $\Delta$Si$_{met-sil}$ is considerably larger than that for Fe \citep{shahar2009experimentally,hin2014experimental}, the observation that the angrite and HED parent bodies, despite both being too oxidised to host Si metal in their cores (Fig. \ref{fig:fO2_planets}), have significantly heavier Si (and Fe in the case of the APB) isotopes than chondrites \citep[cf.][]{pringle2014silicon,liu2017iron}, casts further doubt over the role of core formation, making vapour-condensed phase fractionation a more likely process.


\subsubsection{Cr and the moderately volatile elements}
\label{sec:iso_MVEs}

In marked contrast to Mg, Si and Fe, the isotope ratios of Cr in bulk, differentiated planetary bodies are similar to- or lower than those in chondrites. Earth and Mars have $\delta^{53/52} $Cr within error of all types of chondrites, while the Moon and Vesta have significantly lower values (Table \ref{tab:isos}, see section \ref{sec:single-stage-loss}). The isotope ratios of MVEs (\add{the lithophiles} Zn, K and Rb, \add{and the siderophiles Cu and Ga}) \add{often} vary widely compared to those of the major elements and Cr, even within samples from a single parent body, including chondrites \citep{paniello2012zinc}. 
The isotopic ratios of K, Zn, and Rb are similar to chondritic values in the Earth and Mars, but are enriched in heavy isotopes in the Moon and Vesta (Table \ref{tab:isos}; Fig. \ref{fig:stable_isos_binary}c,d). This tendency is also seen in the scarcer data on Cu and Ga isotopes, which are chondritic in the Earth, but significantly higher in lunar and vestan rocks \citep[Table \ref{tab:isos};][]{kato2017gallium,day2019volatile,wimpenny2022gallium}. The APB, instead, appears to be enriched in light K and Rb isotopes, interpreted as reflecting partial (re)condensation of these elements \citep{hu2022potassium,wang2024rubidium}. Within current measurement precision, no significant variations occur among the Li isotope ratios of the various planetary objects \citep[cf.][]{tomascak2016advances}. \\

\begin{figure}[!ht]
    \centering
    \includegraphics[width=1\linewidth]{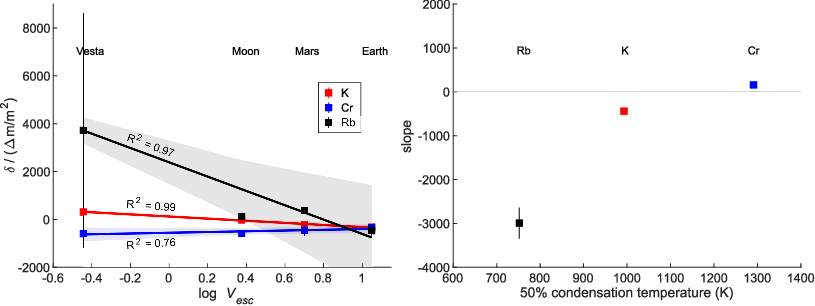}
    \caption{Mass-dependent isotopic fractionation among bulk planetary objects and its relation to the escape velocity of planetary object (a) and their volatility as parameterised by their 50\% condensation temperature (b).  Isotopic compositions have been normalised to the relative mass differences of the isotopes ($\Delta m/m^2$). }
    \label{fig:iso_vesc_slopes}
\end{figure}

None of the isotopic variations in the aforementioned MVEs correlate with those in the major elements, Mg, Si and Fe. As shown in Fig. \ref{fig:stable_isos_binary}c,d, however, they correlate with one another \citep[see also][]{wang2024rubidium}, again implying that a common process controlled their observed variation. 
As first pointed out by \cite{day_moynier2014} and extended by \cite{tian2021potassium}, higher isotopic ratios are found on bodies with lower escape velocities. We show in Fig. \ref{fig:iso_vesc_slopes}a that such correlations among the Earth, Mars, the Moon and Vesta occur for the isotopes of K, Cr and Rb \add{(of the MVE isotopic systems with data for these four planetary objects, only Zn does not yield a statistically significant correlation)}. In the case of K and Rb, data are also available for the APB \citep{hu2022potassium,wang2024rubidium}, yet do not follow the trend defined by the other planetary bodies, irrespective of the assumed mass of the APB \citep{tissot2022}. 

\add{Only three isotope systems currently display these correlations, which are based on only four data points per isotopic system, but they seem a worthy endeavour for future investigation.} Following normalisation by $\Delta m$/$m^2$, 
the variations in Fig. \ref{fig:iso_vesc_slopes}a must be caused by either differences in the fractionation factor 
between vapour and condensed phase(s) or by the extent of mass loss, which is itself proportional to the relative volatilities of the elements. In Fig. \ref{fig:iso_vesc_slopes}b, we show that the slopes of the regressions presented in Fig. \ref{fig:iso_vesc_slopes}a become steeper for increasingly volatile elements (as quantified by $T_c^{50})$. As such, differences in K, Cr and Rb isotopic ratios among the different planetary bodies are controlled by the extent of mass loss experienced by the body in volatilisation/condensation event(s), that is, a Rayleigh process with a finite element budget. Because the isotopes of Cr and the MVEs in the Earth and Mars are not fractionated from the range of chondrites, these bodies show no isotopic evidence for vapour loss of these elements of the sort preserved in the STBs, despite the fact that they also have lower MVE abundances than in most chondrites. 

\subsection{Physical causes of volatile loss}
\label{sec:physics_loss}

Planetary bodies with magma exposed at their surfaces will develop a vapour atmosphere, with the surface pressure determined by the magma temperature and its composition \citep[e.g.,][]{wolf2023}. 
If the atmosphere is no longer hydrostatic, upwards motion can drive atmospheric loss, so-called hydrodynamic escape \citep{young2019near,chao2021}. 
Whether hydrodynamic escape is efficient or not depends on the competition between thermal energy and gravity. More efficient escape occurs from small, high temperature bodies, as quantified in 
the parameter $\lambda$,

\begin{equation}
    \lambda = \frac{GM}{r} \frac{\mu}{R T_s}
    \label{eq:acc_hydro}
\end{equation}

where $G$ is the gravitational constant, $M$ and $r$ are the mass and radius of the body, $\mu$ is the atmospheric mean molar mass, $T_s$ is the surface temperature and $R$ is the gas constant. 
A large value of $\lambda$ indicates inefficient escape (gravity wins). For a diatomic gas, the critical value of $\lambda$ above which hydrodynamic escape shuts off is 2.4-3.6 \citep{Volko-etal:2011}. For an adiabatic atmosphere, the critical value can be shown to be $\gamma_{ad}/(\gamma_{ad}-1)$ \citep{ZahnlKasti:1986}, where $\gamma_{ad}$ is the ratio of the specific heat at constant pressure over constant volume. For a diatomic gas this ratio is 7/5, so the critical value of $\lambda$ in this case is 3.5. 

    \begin{figure}[!ht]
        \centering
        \includegraphics[width=0.75\linewidth]{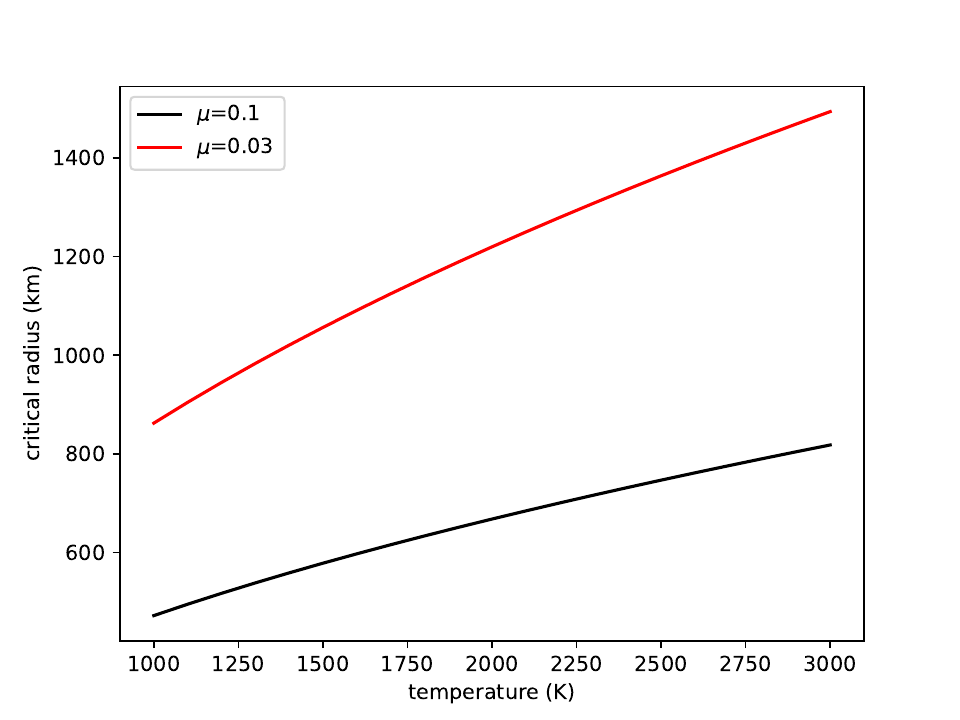}
        \caption{Radius above which hydrodynamic escape is shut off, as a function of surface temperature, for two different molar masses $\mu$ in kg/mol. Here, we assume the critical value of $\lambda$ is to be 3, and a bulk density of $\rm 4000~kg~m^{-3}$. }
        \label{fig:critical_mass}
    \end{figure}

Escape of atmospheres with a $\mu \geq 0.03$ kg/mol ($\geq$air) is only possible on $<$1000~km radii bodies (smaller than the Moon; Fig.~\ref{fig:critical_mass}). This limit increases to Mars-sized bodies for $\lambda$ = 3 for $\mu$ = 0.004 kg/mol at 2000 K. Hydrodynamic escape from the fully-formed Earth is not possible (except possibly for pure hydrogen, and even then only at very high temperatures, $\geq$3500 K). Hydrodynamic escape is impeded relative to that predicted from eq. \ref{eq:acc_hydro} if the surrounding disk has a non-zero pressure. It is also limited by the thermal energy available: heat is both radiated and advected away from the hot surface, and as the magma cools the vapour pressure drops, reducing the outwards flux of material. This limits the length of time available for escape to occur in the absence of external heat sources. There are two main sources of energy available to drive escape;
\begin{itemize}
    \item Decay of $^{26}$Al. Although this process likely caused melting and differentiation of early-formed solar system bodies, its ability to drive hydrodynamic escape is impeded by the rate at which it supplies thermal energy. A magma body radiating at 700~K \citep[similar to terrestrial lava lakes;][]{Patri-etal:2016} loses energy at a rate of 14~kW~m$^{-2}$. Conversely, $\rm ^{26}Al$ decay would produce a surface heat flux of 0.5~kW~m$^{-2}$ in a 1000~km radius body at $t_0$, and less thereafter. As a result, $\rm ^{26}$Al heating cannot produce significant mass loss in bodies of such size (we note that the contrary finding expressed in \cite{young2019near} is due to a numerical error; Young, pers. comm.).
    \item Impacts.  Sufficiently energetic impacts can cause wide-spread melting and vaporisation \citep{Nakaj-etal:2021}. Hydrodynamic loss is very sensitive to the melt temperature because it determines the surface vapour density, which in turn governs the mass loss rate through the equation:
\end{itemize}

\begin{equation}
    \frac{dN_{i}}{dt} = 4 \pi ({r_B})^2C_sn_{i}
    \label{eq:acc_hydro_massloss}
\end{equation}

where $r_B$ is the critical radius (i.e. that at which the gas outflow speed equals the sound speed), $C_s$ the sound speed at $r_B$, and $n_i$ the number density of the vapour. During conventional accretion, impact velocities are typically comparable to the escape velocity of the target body \citep{safronov1969, Agnor-etal:1999}, which depends on its mass. Higher impact velocities produce more melting and higher temperatures \citep{Nakaj-etal:2021}. But if impact velocity is controlled by target size, then the high-velocity impacts will typically only take place on large bodies, from which hydrodynamic escape is prohibited (see above). There are several potential ways in which this trade-off can be overcome. \\

The first is to appeal to some event stirring up the planetesimals so that their impact velocities are higher than would be expected based on the arguments above. For instance, \citet{hin2017magnesium} argued that high-velocity impacts driven by\add{ excited bodies on highly eccentric orbits, as induced by} gas giant migration \citep{Carte-etal:2015} yield sufficient energy to cause planetesimal melt production and vapour loss. \cite{calogero2025can} explored this hypothesis in more detail, focusing on potassium. A second mechanism states that planetary envelopes enriched in rock-forming elements  can exchange mass with the ambient nebular gas. As long as the molecular mass is dominated by that of H$_2$ (0.002 kg/mol), escape from envelopes may be efficient \citep{steinmeyer2023}.  A third is to consider impact-produced melt that is not contained within the target planet, such as an impact-generated proto-lunar disk. Here, the liquid is hot but the effective gravity of the proto-lunar disk is low, facilitating escape from the Moon, but not from the Earth-Moon system \citep{charnoz2021}.\\ 

While hydrodynamic escape is efficient in engendering mass loss, it does not itself produce large mass fractionation between isotopes in the atmosphere  \citep{hunten1987}. 
However, isotopic fractionation can occur at the melt-vapour interface, the magnitude and direction of which depends on two factors that trade off against each other;
\begin{itemize}
    \item Temperature. Higher temperatures lead to smaller isotopic fractionation factors if occurring at equilibrium (proportional to 1/$T^2$).
    \item Degree of equilibrium. If Langmuir (kinetic) fractionation prevails (i.e., $p_i~<~p_{i,eq}$), then the fractionation factor, $\alpha$ is $^{\frac{i}{j}}\alpha_{kin} = \sqrt{\frac{m_j}{m_i}}$, where $m$ is the isotopic mass. If equilibrium prevails (i.e., $p_i~=~p_{i,eq}$), then, in most instances, the gas phase is also enriched in lighter isotopes, but the fractionation factor is closer to unity \cite{young2019near}. This is because most rock-forming elements evaporate as monatomic gases \citep[e.g., K$^0$, Na$^0$, Mg$^0$, see][]{sossi2019evaporation} or as oxide species with a \textit{lower} mean oxidation state than in the condensed phase (e.g., SiO(g) vs. SiO$_2$(s,l), GeO(g) vs. GeO$_2$(s,l)).
\end{itemize}

\subsubsection{Volatile loss due to evaporation only}
\label{sec:single-stage-loss}

If evaporation mimics a Rayleigh process (see section \ref{sec:iso_MVEs}), then the degree of isotopic fractionation follows:

\begin{equation}
\frac{^{i/j}R_{res}}{^{i/j}R_0} = F^{\left(^{i/j}\alpha_{gas/cond}-1\right)},
   \label{eq:iso_rayleigh}
\end{equation}

where $R$ is the isotopic ratio of the residue (res) and the system (0), respectively and $F$ is the fraction of isotope $j$ remaining in the residue. Here, we examine the correlations between $\frac{^{i/j}R_{res}}{^{i/j}R_0}$ and $F$ fit with values of $\alpha$, and compare them to those expected from Langmuir (kinetic) and Knudsen (equilibrium) fractionation. \\

The prime case in which the expected correlation between $F$ and $\frac{^{i/j}R_{res}}{^{i/j}R_0}$ is observed is the Moon. Taking the BSE as ${^{i/j}R_0}$,  $\delta_{Moon}-\delta_{BSE}$ correlates with the degree of depletion, $F$ (Fig. \ref{fig:Moon}). The fitted values of $\alpha$ (for Rb, K, Zn and Cu) are consistent with values of $p_i/p_{i,eq}$ of 0.99, should evaporation have taken place at $\sim$2500 K \citep{niedauphas2019}, or 1 (i.e., equilibrium) for $T\sim$1300 K \citep{tartese2021,dauphas2022alkali}. Hence, the fitted value of $\alpha$ is degenerate with respect to $p_i/p_{i,eq}$ and temperature. However, Sn and Cr are anomalous, in that they preserve \textit{lighter} isotopic ratios in the Moon than in the bulk silicate Earth \citep{sossi2018cr,wang2019tin}. Because both Cr and Sn have gas species in which the metal is more strongly bound than in the condensed phase, fractionation at equilibrium can uniquely account for this observation in a vapour-condensed phase system. This interpretation, and the negative slope in Fig. \ref{fig:iso_vesc_slopes} thus support loss of Cr and Sn on STBs at relatively oxidised conditions ($\sim$IW) \citep[see section \ref{sec:chem_consequences},][]{zhu2019vesta}.

The propensity for (near-) equilibrium isotopic fractionation to have prevailed reflects; \textit{i)} the fact that vapour phase 
reactions reach equilibrium in 10\textsuperscript{-9} to 1 s \citep{fegley2020volatile}; much shorter than lifetimes of magma oceans \citep[$\sim$10$^2$--10$^7$ yr;][]{salvador2023} and \textit{ii)} $p_i/p_{i,eq} > 0.95$ when the (far-field) total pressure ($P$) exceeds $\sim$10$^{-8}$ bar \citep{young2019near,TangYoung:2020}. 
In summary, 
evidence points to near- or at equilibrium values of $p_i/p_{i,eq}$ ($>$ 0.99), at least for the Moon. \add{Such conditions are more readily achieved on planetesimals that are Moon-mass or larger \citep{young2019near}, \add{which indicates} that volatile loss on the Moon likely occurred when it was close to its present-day mass\add{, though does not require it.}} 


\subsubsection{Volatile loss due to evaporation + mixing}
\label{sec:multi-stage-loss}

\begin{figure}[!ht]
        \centering
        \includegraphics[width=0.5\linewidth]{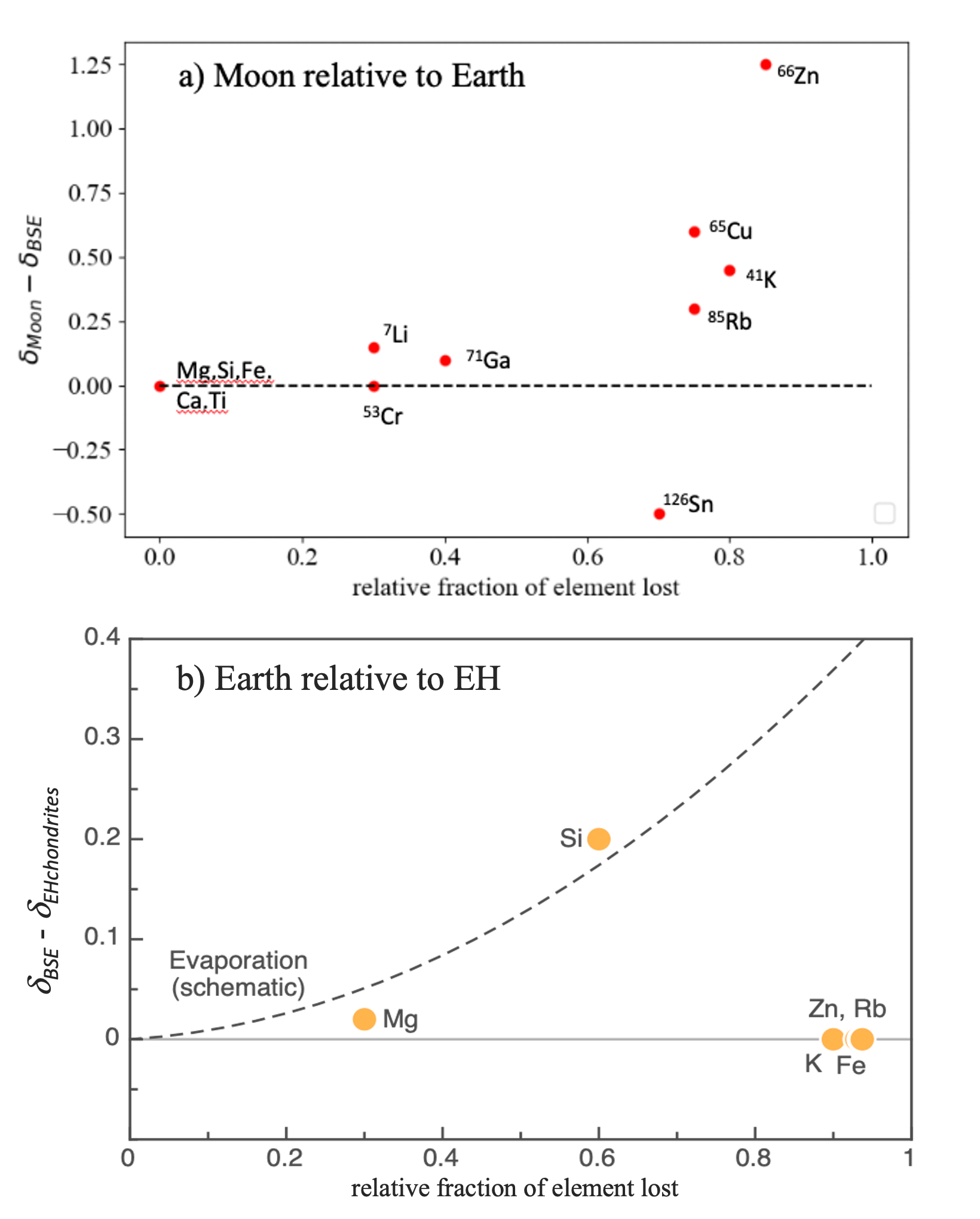}
        \caption{a) The isotopic differences between the bulk silicate Earth and Moon for the isotopes of various elements, against the depletion of those elements in the Moon relative to the bulk silicate Earth.  b) Mass-dependent isotopic fractionation in the bulk silicate Earth relative to EH chondrites for selected elements as a function of their elemental depletion (calculated as $X$/Ca$_{BSE}$/$X$/Ca$_{EH}$, where $X$ = Mg or Si).  The dashed line shows the expected increase in isotopic fractionation for larger degrees of evaporative loss.  Despite their greater degrees of depletion, the moderately volatile elements (K, Zn, Rb) do not show any isotopic deviation with respect to EH chondrites. Source data are listed Table \ref{tab:isos}. }
        \label{fig:Moon}
\end{figure}

\add{Correlations} between isotopic fractionation in a given planetary body and its escape velocity (Fig. \ref{fig:iso_vesc_slopes}a) could imply that \add{those bodies} experienced volatilisation and vapour loss at their (near-)final masses, as posited by \cite{tian2021potassium}. The calculations in section \ref{sec:physics_loss} and the observations in section~\ref{sec:single-stage-loss} support the idea that 
Vesta and the Moon \citep[see also][]{charnoz2021}, were able to lose vapour by hydrodynamic escape, even at their present day masses. Conversely, Mars, and in particular the Earth, were barely able to lose any vapour at their current masses. As opposed to the Moon and Vesta, Earth and Mars do not preserve any detectable mass-dependent isotopic fractionation relative to chondrites, yet they are clearly volatile-depleted with respect to CI chondrites (Fig. \ref{fig:Moon}b). 

Figure~\ref{fig:Moon}b shows that Mg and Si, relative to Ca and EH chondrites ($X$/Ca$_{BSE}$/$X$/Ca$_{EH}$), are slightly (32~\%) and moderately (59~\%) depleted, respectively, in the BSE and are also marginally isotopically heavy \citep{hin2017magnesium}. These figures decrease to 20 \% for Mg and 29 \% for Si when normalised to Ca and CI chondrites.
Therefore, the depletion of Si and Mg observed in the BSE relative to chondrites (both normalised to Ca) is coupled with heavy isotopic enrichment, whereas the elemental depletion in MVEs records no isotopic fractionation compared to chondrites. The simplest explanation for this duality is that the \add{Earth (and Mars) accreted bodies in which} MVEs underwent near-complete evaporative loss (or were hardly accreted at all), \delete{followed by later addition}\add{and mixed with bodies} of undepleted, potentially chondritic material \citep{hin2017magnesium,sossi2022stochastic}. On the Earth's total mass-basis, this undepleted material is insignificant 
such that the isotope ratios of the major elements (Mg, Si and Fe) remained essentially unaffected, yet the budgets of the MVEs were entirely overprinted. This implies that there must be at least two populations of smaller bodies, one isotopically heavy and volatile-poor and the other volatile-undepleted and isotopically lighter (i.e., unfractionated relative to chondrites). \\ 


To test this idea, output from an N-body model \citep{Carte-etal:2015} was analysed to examine the competition between the chemical and isotopic fractionation induced by impact-driven evaporation of K and the mixing of these bodies over time \citep{calogero2025can}. Bodies were assumed to have an initial bulk and isotopic composition identical to that of Mars. Subsequent impacts in some cases caused melting, magma ocean formation and hydrodynamic loss of K. Isotopic fractionation was assumed to occur in such cases with $\alpha$=0.999541. Other collisions did not generate melting \add{and} were \add{thus} assumed to simply mix the compositions of the two colliding objects.


\begin{figure}[!ht]
    \centering
        \includegraphics[width=1\linewidth]{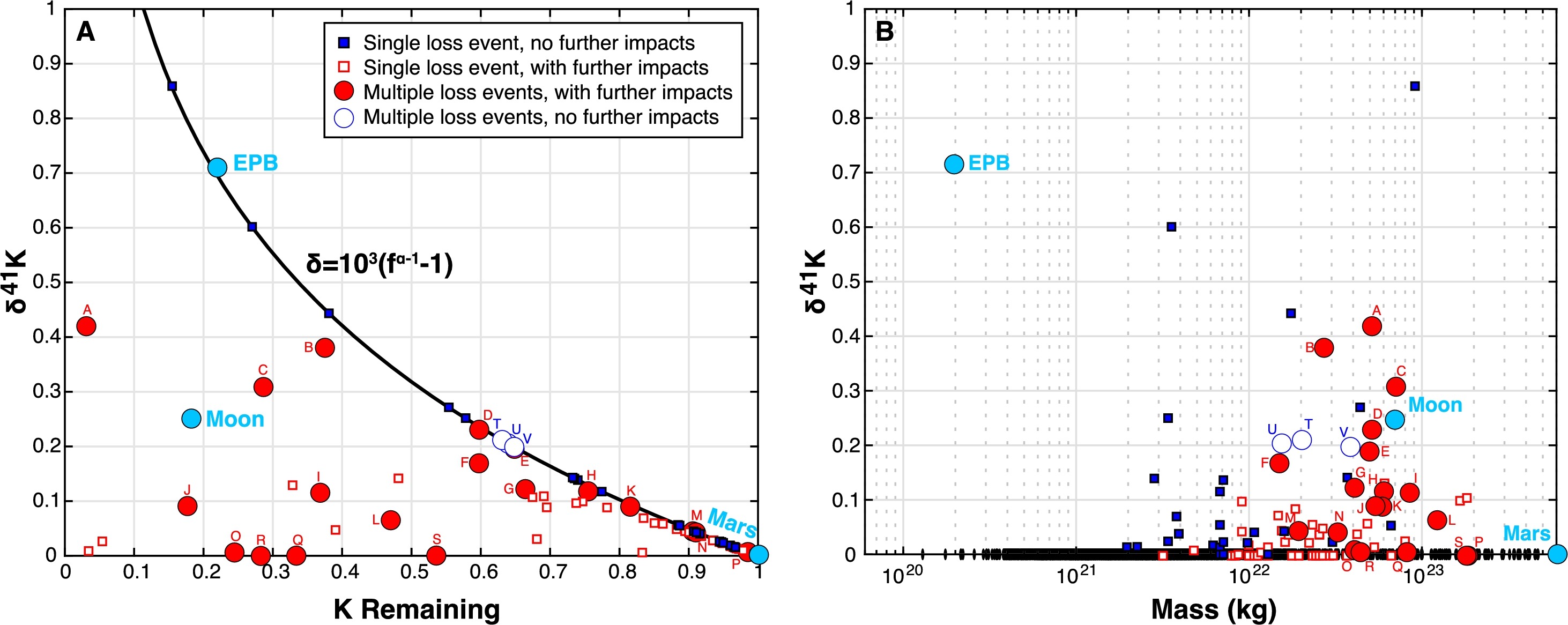}
    \caption{Calculations using output from N-body simulations \citep{Carte-etal:2015} showing the K isotope fractionation (expressed as $\delta^{41}$K) with respect to the \textbf{a)} fraction of K remaining and \textbf{b)} the mass of individual planetary bodies, assuming a Mars-like starting composition. Different symbols show whether bodies experienced multiple impact-driven evaporative events. The solid line in a) is that for evaporative loss of K governed by the Rayleigh equation with a value of $\alpha$ = 0.999541. See \cite{calogero2025can} for further details}
    \label{fig:k_iso_mixing}
\end{figure}

These simulations show that in general, as predicted by eq. \ref{eq:acc_hydro}, smaller bodies are the most isotopically fractionated (Fig. \ref{fig:k_iso_mixing}b). However, they are not necessarily the most depleted in K. Indeed, other, larger bodies, which have experienced more impacts, can be \textit{more} depleted in K, but have lower (i.e., more chondritic) $\delta^{41}$K values (e.g. bodies I and L in Fig. \ref{fig:k_iso_mixing}). This arises because the combination of early evaporative loss in small bodies followed by mixing results in larger bodies having isotopically unfractionated signatures (because more isotopically unfractionated material was added). For an Earth-sized body, this process naturally results in broadly chondritic $\delta^{41}$K and $\delta^{66}$Zn values, because the budget of these elements is entirely dominated by the addition of volatile-rich material that experienced little- to no evaporation. This is the scenario envisaged by \cite{sossi2022stochastic}, and states that, all else being equal, the compositions (both chemical and isotopic) of larger bodies will be determined to a greater extent by mixing, and those of smaller bodies by partial evaporation.

\section{Provenance of planet-building materials}
\label{sec:provenance}

The provenance of planetary building materials can be quantified using nucleosynthetic isotope anomalies. These arise through the heterogeneous distribution of presolar material in the solar protoplanetary disk and, as such, allow genetic links among and between meteorites and planets to be examined.
The isotope anomalies in a given element are quantified by their part per ten-thousand deviation ($\varepsilon$) from a terrestrial reference standard (in most cases close to but often not identical with the BSE), following internal normalisation to a chosen ratio:

\begin{equation}
    \varepsilon^iX = \left( \frac{(^iX/^jX)_{body}}{(^iX/^jX)_{std}} - 1 \right) \times 10000
    \label{eq:epsilon}
\end{equation}

where $i$ and $j$ are the numerator- and denominator isotope, respectively, and $X$ is the element of interest. Implicit in computing isotope anomalies in this way is the assumption that the chosen internal normalising ratio of all samples is terrestrial. For samples having nucleosynthetic isotope anomalies this is almost certainly not true, and so calculated $\varepsilon^iX$ values may not reflect anomalies on a particular isotope, but rather the combined effects of abundance variations of several isotopes, including those used for the internal normalisation. \\

A key observation from the nucleosynthetic isotope anomalies, as well as $\Delta^{17}$O (whose variations are not nucleosynthetic in origin), is the notion of an isotopic dichotomy between non-carbonaceous (NC) and carbonaceous chondrite (CC) type materials \citep[see][for recent reviews]{bermingham2020nc,kleine2020non,kruijer2020great}. The first indication of what is now known as the NC-CC dichotomy dates back to \citet{trinquier2007cr}, who observed that, for nucleosynthetic isotope anomalies, differentiated meteorites and carbonaceous chondrites define two distinct fields. \citet{warren2011stable} then showed that isotope anomalies in  O, Ti, Cr, and Ni, meteorites always fall into two distinct clusters, which he termed the non-carbonaceous (NC) and carbonaceous (CC) groups. Since then, the number of elements for which the NC-CC dichotomy has been identified has grown considerably, 
making it 
a fundamental characteristic of the early solar system. 
Importantly, the dichotomy was also identified for isotopes of Mo \citep{budde2016,poole2017}, in which it exists not only for chondrites but also for iron meteorites, which derive from parent bodies that accreted within the first 1 Myr of the solar system \citep{kruijer2017age}. Thus, together, these observations reveal that the dichotomy was established early, and was maintained for essentially the entire lifetime of the disk.  \\

It is now widely thought that the NC reservoir represents the inner disk, and the CC reservoir the outer disk \citep{warren2011stable,budde2016,kruijer2017age}. 
Below we 
assess whether or not CC material is the dominant source of \textit{i)} volatile elements to the Earth and \textit{ii)} oxidation relative to the solar nebula among telluric bodies. Owing to their distinct nucleosynthetic heritage, we discuss the iron-peak (section \ref{sec:iron-peak}) and heavier elements (section \ref{sec:heavy_elements}) separately.

\subsection{Iron-peak elements}
\label{sec:iron-peak}

The `iron-peak' elements have binding energies per nucleon that are higher than those of the surrounding nuclides, leading to enhanced abundances during stellar nucleosynthesis via the alpha process \citep{clayton2003isotopes}. Typically, these elements are Ti, V, Cr, Mn, Fe, Ni, Cu and Zn. Here, we also extend this classification to Si (produced by O burning) and Ca (produced by Si burning), owing to their similar masses and nucleosynthetic heritage. \\ 

In a diagram of $\varepsilon^{50}$Ti versus $\varepsilon^{54}$Cr, the composition of the BSE plots within the NC field, albeit as an endmember\add{, with excesses in neutron-rich isotopes compared to other NCs,} pointing towards the CC field \citep{trinquier2009origin}. Thus, the composition of the BSE can be produced by mixtures of the extant groups of NC and CC meteorites for these two systems. However, the choice of end-members to represent each group is non-unique.
One interpretation (hereinafter hypothesis A) first mooted by \cite{warren2011stable}, and developed by \cite{schiller2018calcium} states that the isotopic composition of the Earth evolved over time from ureilite-like (i.e., the meteorites defining the lower end of the NC trend) to its present-day composition by the addition of CI-like material. \citet{schiller2018calcium} based their arguments on anomalies in $^{48}$Ca, but the correlated nature of the isotope anomalies in Ca, Ti, and Cr \add{among the NC meteorites} means that the same argument holds for any of these elements. An alternative hypothesis (B), is built on the observation that, in many isotopic systems, including Cr and Ti, the Earth (and Moon) have compositions similar to those of enstatite chondrites (ECs) \citep[e.g.,][]{warren2011stable,dauphas2017isotopic}. In this scenario, the Earth accreted $\sim$95~\% EC-like material, which could reflect objects with a uniform, EC-like isotopic composition  \citep{dauphas2017isotopic,dauphas2024}, but also via accretion of a variety of NC materials with a range of isotopic compositions, whose average is that of the ECs \citep{burkhardt2021terrestrial}. \\

As pointed out by \citet{warren2011stable}, hypotheses A and B lead to very different CC fractions inferred for the Earth. The CI-derived fraction for each element, $X$, in the BSE can be computed by mass balance: 

\begin{equation}
    f_{X,CI} = \frac{\varepsilon^iX_{NC} - \varepsilon^iX_{BSE}}{\varepsilon^iX_{NC} - \varepsilon^iX_{CI}}
    \label{eq:nucleo_mixing}
\end{equation}

where NC denotes the isotopic composition of the NC material in Earth, which can either be ureilite- (hypothesis A) or EC-like (hypothesis B). Use of eq. \ref{eq:nucleo_mixing} implies identical element concentrations in each reservoir. \delete{If true}\add{For identical element concentrations in each reservoir}, the CI-derived fraction for an element $X$ is the same as the apparent mass fraction of CI material accreted to the Earth. This is likely to hold for RLEs (e.g., Ti) and potentially the main components (Cr, Si), but may not for volatile (e.g., Zn) and siderophile elements (e.g., Fe, Ni). This is because these elements are depleted in the BSE relative to chondrites as a result of volatile depletion \citep{steller2022nucleosynthetic,savage2022zinc} or core formation, respectively \citep{dauphas2017isotopic}. \\

\begin{figure}[!ht]
    \centering
    \includegraphics{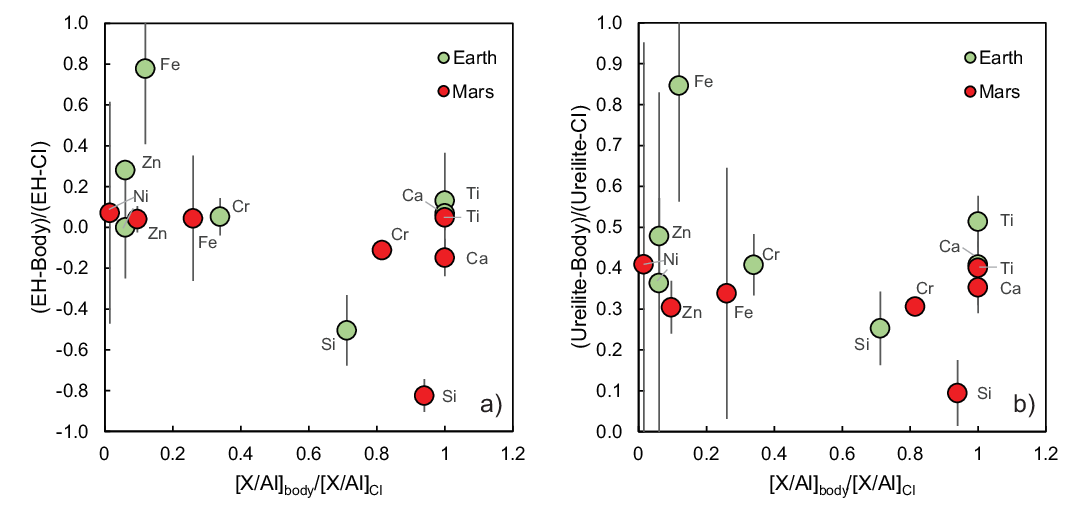}
    \caption{The fraction of each element in the bulk silicate Earth (BSE, green) and bulk silicate Mars (BSM, red) brought by the CI component in mixtures between a) EH-like and CI-like and b) ureilite-like and CI-like isotopic end-members, each with identical concentrations, as a function of their Al, CI-normalised abundances in the BSE and BSM. Note the large uncertainties for Fe and Ni, owing to the relatively small degree of isotopic variation relative to the uncertainty on the analyses. Isotope ratios used are $\mu^{30}$Si, $\varepsilon^{48}$Ca, $\varepsilon^{50}$Ti, $\varepsilon^{54}$Cr, $\varepsilon^{54}$Fe and $\varepsilon^{62}$Ni. Data from \citep{trinquier2007cr,trinquier2009origin,schiller2018calcium,schiller2020iron,burkhardt2021terrestrial,steller2022nucleosynthetic, savage2022zinc,onyett2023}.}
    \label{fig:ci_ureilite}
\end{figure}

In both hypotheses, the apparent fraction of CI-like material is near-constant for Ca, Ti, and Cr, at $\sim$5~\% for the BSE assuming an EC-like NC end-member, or $\sim$40~\% for a ureilite-like NC end-member (Fig. \ref{fig:ci_ureilite}). As noted above, this constancy of inferred CI fractions in both hypotheses reflects the correlated nature of the isotope anomalies in these elements \add{among the NC meteorites, and the position of Earth on the neutron-rich end of the NC trend}.\\

For Ni, using $\varepsilon^{62}$Ni as in Fig. \ref{fig:ci_ureilite}, the inferred CI fractions are similar to those obtained from the three aforementioned elements, but the small Ni isotope variations have larger uncertainties and would also be compatible with a wider range of CI fractions \citep{steele2012neutron,tangdauphas201460fe,Nanne2019}. For example, when using the $\varepsilon^{60}$Ni anomaly, the BSE and CI chondrites are indistinguishable within error \citep{spitzer2024}. For Fe, the $\varepsilon^{54}$Fe values of CI chondrites and the BSE are indistinguishable, leading to high apparent CI fractions in the BSE \add{(0.79$\pm$0.39 and 0.84$\pm$0.27 for an EC-CI mixture or Ureilite-CI mixture, respectively; Fig. \ref{fig:ci_ureilite}})\delete{but the uncertainty is also large, preventing an accurate estimate }\citep{schiller2020iron,hopp2022ryugu}. \add{However, CI chondrites are unlikely to have significantly contributed to the Fe budget of the BSE because, as noted by \cite{hopp2022ryugu}, the isotopic compositions of both Cr (less siderophile) and Ni (more siderophile) in CIs are distinct from the BSE. We note that the apparent contradiction can be resolved by considering that the BSE lies at the neutron-rich (i.e., $^{54}$Fe-rich) end of the NC trend, indicating that Fe is entirely consistent with an NC origin \citep[see also][]{hopp2022earth}.} \\

For Si the apparent fractions of CI-like material are systematically lower than that from other elements (and extend to negative values), which indicates that the $\varepsilon^{30}$Si$_{BSE}$ is \textit{lower} (=0) than both CI ($\varepsilon^{30}$Si = $\sim$0.3) and EC ($\varepsilon^{30}$Si = $\sim$0.1) \citep[see][]{onyett2023}. On this basis, \citet{onyett2023} argued that EC-CI mixtures are precluded as proto-Earth source material. 
However, because ECs are fractionated mass-dependently in $\delta^{30}$Si, correcting the $\varepsilon^{30}$Si for such fractionation yields \delete{no significant Si isotope anomaly in ECs}\add{a revised value of $\varepsilon^{30}$Si = 0.04$\pm$0.03}\delete{compared to the BSE} \citep{dauphas2024}\delete{again permitting EC-like material in the Earth} \add{which scarcely differs from that of the BSE and hence is broadly consistent with hypothesis B.} \add{Because such a correction is implicit on mass-dependent fractionation occurring at high temperatures, additional data to better resolve any potential differences between ECs and the BSE are essential in discerning between hypotheses A and B.} \\

\begin{figure}[!ht]
    \centering
    \includegraphics[width=0.5\linewidth]{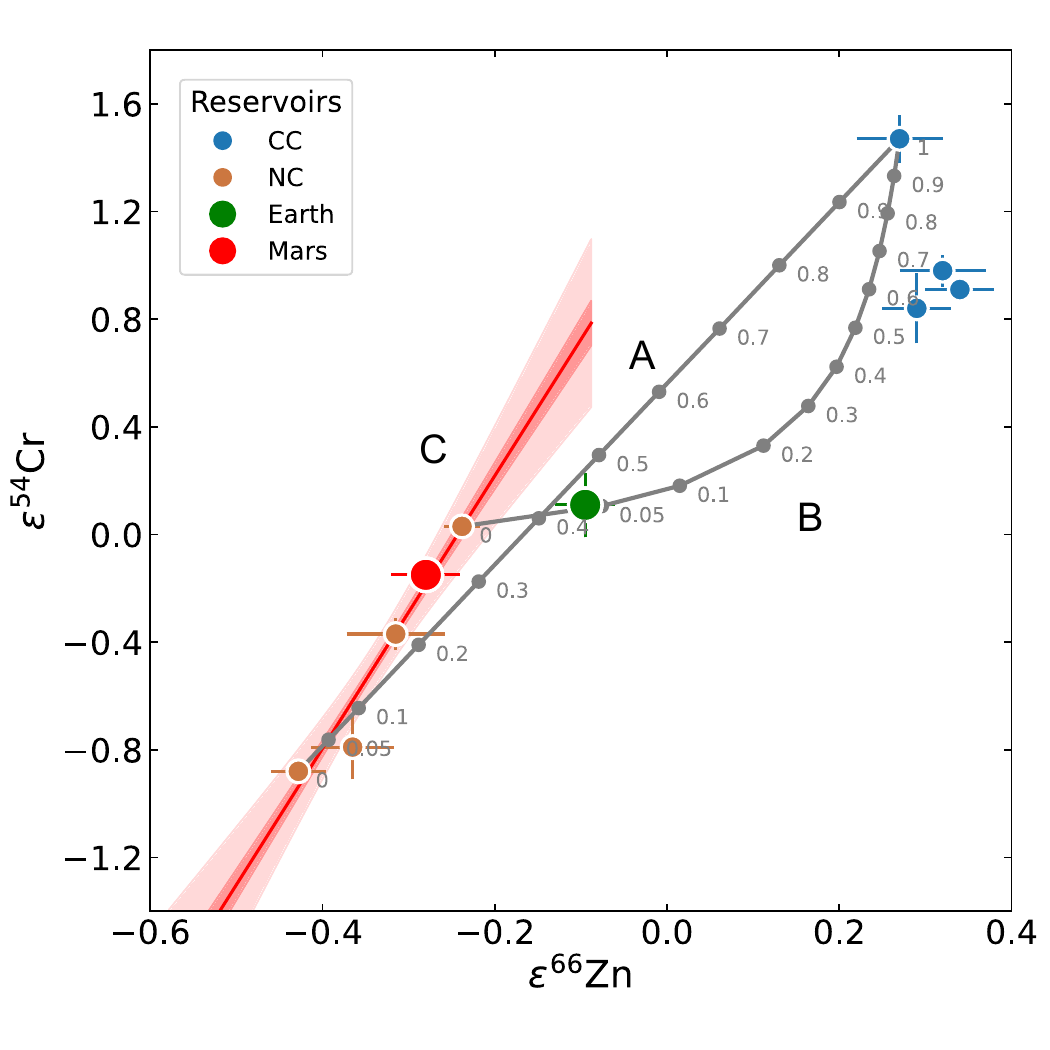}
    \caption{The $\varepsilon^{54}$Cr-$\varepsilon^{66}$Zn variation among non-carbonaceous meteorites (NC, orange), carbonaceous meteorites (CC, blue), the bulk silicate Earth (Earth, green), and bulk silicate Mars (Mars, red). The grey curves denote mixing lines between CI chondrites and, A - a ureilite-like end-member with Zn/Cr ratios identical to CI chondrites (0.12) and B - an EC-like end-member with 35 ppm Zn (Zn/Cr$_{EC}$ = 0.014). Tick marks show the mass fraction of CI-like material in the mixture.  C denotes a fit according to the York method \citep{york2004} assuming uncorrelated errors in \textit{x} and \textit{y} and its associated 1-$\sigma$ (dark red) and 2-$\sigma$ (light red) error envelope\add{, yielding $\varepsilon^{54}$Cr = (5.06$\pm$1.05)$\varepsilon^{66}$Zn + 1.23$\pm$0.33.}  All data taken from \cite{steller2022nucleosynthetic} for chondrites and the Earth, and \cite{kleine2023inner,paquet2023origin} for Mars using $^{67}$Zn/$^{64}$Zn-normalised values. }
    \label{fig:Zn-Cr}
\end{figure}

A key distinction between hypotheses A and B can \add{instead} be made using Zn isotopic composition of the BSE \citep{steller2022nucleosynthetic,savage2022zinc,martins2023nucleosynthetic}. In hypothesis A, the fraction of CI-derived Zn is 0.48$\pm$0.09 (Fig. \ref{fig:ci_ureilite}b); indistinguishable from Cr (0.41$\pm$0.07) and Ti (0.51$\pm$0.06). Hypothesis B yields a higher apparent CI-derived fraction computed from Zn (0.29$\pm$0.05) compared to Cr (0.07$\pm$0.03) and Ti (0.05$\pm$0.09; Fig. \ref{fig:ci_ureilite}a). 
In $\varepsilon^{66}$Zn versus $\varepsilon^{54}$Cr space, using the Zn concentrations of CC meteorites and the BSE, any mixing line between the CC field that passes through the BSE intersects the NC field near the \textit{isotopic} composition of the ECs with \add{an average} $\sim$35 ppm Zn \add{for the NC material} \citep[][curve `B', Fig. \ref{fig:Zn-Cr}]{kleine2023inner,kleinenimmo2024}. This implies that the BSE contains $\sim$6~\% CI-like material by mass in hypothesis B. 
Notably, this $\sim$6~\% CI chondrite-like material is identical to the CC \textit{mass} fractions inferred from the isotope anomalies in Ca, Ti, and Cr in Hypothesis B (Fig. \ref{fig:ci_ureilite}), meaning Earth would have accreted predominantly from NC material with an EC-like isotopic composition, on average, with a 6~\% contribution of CI material, which delivered $\sim30\%$ of Earth's Zn budget. Hypothesis A, in contrast to Hypothesis B, would require that the NC (ureilite) and CC (CI) end-members had the same \delete{and BSE-like} Zn/Cr ratios such that the mixing line passes through the BSE (curve `A', Fig. \ref{fig:Zn-Cr}). \add{Indeed, Zn/Cr ratios in ureilites are similar, to within a factor $\sim$2, of those in CI chondrites \citep{brugier2019zinc,zhu2022nickel}.} In this scenario, the apparent CI-like fraction in the BSE inferred from Zn matches those from Ca, Ti and Cr, but, in contrast to hypothesis B, CI chondrites must represent $\sim$40~\% of the mass of the Earth. \\

\add{Variations in oxygen isotopes, although not nucleosynthetic in origin, nevertheless provide information on a sample’s provenance. Ureilites are heterogeneous with respect to $\delta^{17}$O and $\delta^{18}$O \citep{claytonmayeda1996achondrites,kruttasch2025}, yet, unlike other NC bodies, they lie on the Carbonaceous Chondrite Anhydrous Mineral Line (CCAM), which defines a slope of unity in a three-isotope plot \citep{clayton1977}. Because CI chondrites also lie on the CCAM with $\delta^{17}$O $\sim$9 \textperthousand~and $\delta^{18}$O $\sim$16~\textperthousand~\citep{claytonmayeda1999}, any mixture of ureilite-like and CI-like material, even if a specific proportion can produce the $\Delta^{17}$O composition of the BSE, does not pass through the BSE composition \citep[$\delta^{17}$O = 2.7 \textperthousand , $\delta^{18}$O = 5.2 \textperthousand,][]{eiler2001}. As such, ureilite-CI mixtures appear not to be compatible with the O isotope composition of the BSE.}

\begin{figure}[!ht]
    \centering
    \includegraphics[width=1\linewidth]{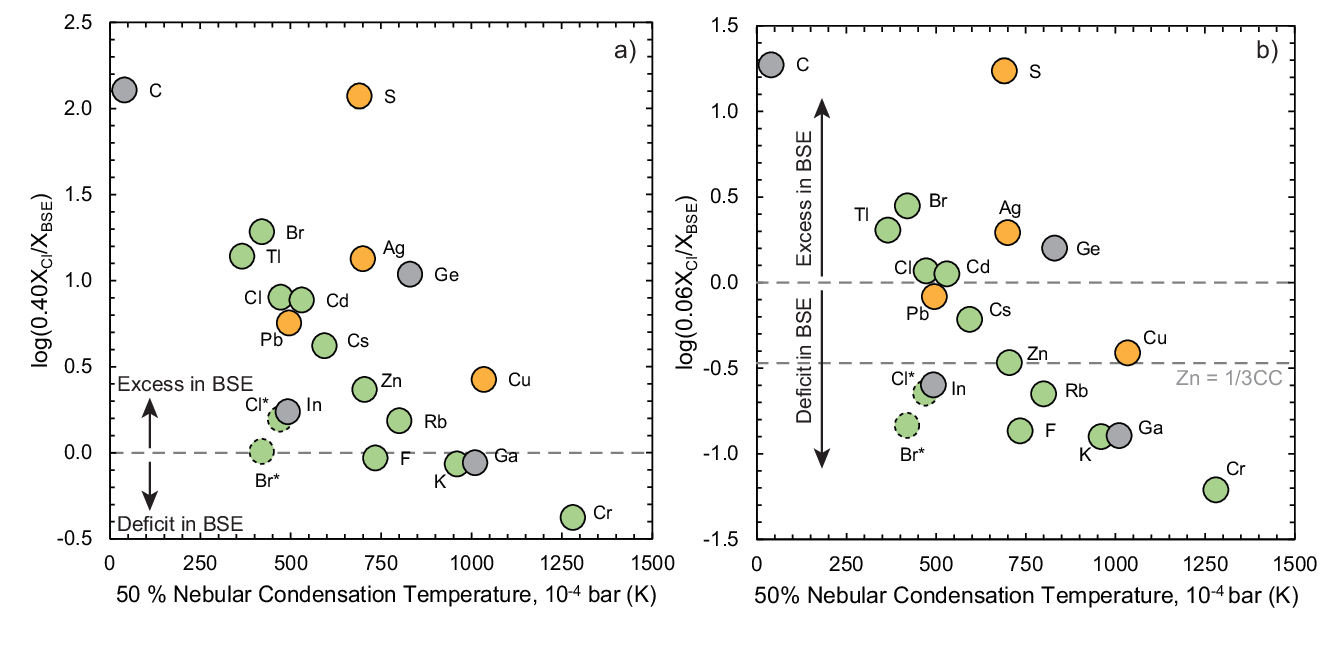}
    \caption{The log$_{10}$ of the ratio of the concentrations of a selection of moderately- and highly volatile elements for a) 40~\% and b) 6~\% (by weight) given by addition of CI chondrites to an otherwise volatile-free proto-Earth, normalised by the present-day bulk silicate Earth (BSE) abundances of these elements. Green = lithophile, grey = siderophile, orange = chalcophile. Cl* and Br* refer to the CI chondrite abundances of \cite{clay2017halogens}. 
    Values that exceed 0 indicate that these elements are more abundant in 0.40$\times$ (a) or 0.06$\times$ (b) CI chondrite than in the BSE.}
    \label{fig:ci_excess}
\end{figure}

These distinct interpretations provide a testable hypothesis, namely, which (if either) of the two scenarios is consistent with the volatile element budget of the BSE? Figure \ref{fig:ci_excess} shows the abundances of a selection of volatile elements for mixing between a volatile-free proto-Earth and a) 40~\%- and b) 6~\% by mass of CI chondrites. In the latter case, this is equivalent to $\sim$30~\% of the Zn budget of the BSE as explained above and posited by \cite{steller2022nucleosynthetic,savage2022zinc} in the framework of hypothesis B. The choice of a volatile-free proto-Earth provides upper limits for the inferred volatile element abundances. 


Delivery of 6~\% CI to an otherwise volatile-free proto-Earth (hypothesis B) would account for roughly the entire present-day budget of Pb, while the budgets of C, S, Ag, Ge, Tl and Br ($\pm$Cl, Cd) would be in excess of those observed. Because Ag, Ge, C and S are more siderophile than is Zn, their additional sequestration into the core \textit{after} the delivery of CI material could reconcile their predicted excesses \citep{mann2009,wood2014,blanchard2022}. The same argument cannot be made for Tl, Br, Cl and Cd, whose metal-silicate partition coefficients are similar to \citep[Cd;][]{wang2016} or lower than \citep[Br, Cl, Tl;][]{kuwahara2017,yang2023cl} those for Zn. 
The excess halogen problem can be resolved if the CI abundances of \citep{clay2017halogens} are adopted. Hence, hypothesis B remains plausible without significant volatile loss post-CI-addition (though additional core formation \add{after CI addition} is needed). \\

The addition of 40~\% CI chondrites by mass in hypothesis A, results in an overabundance of all volatile elements, including Zn. Therefore, the BSE must have \add{to have} undergone volatile loss \textit{after} the delivery of CI material for hypothesis A to hold, which is very difficult because thermal loss of volatiles (except perhaps hydrogen) is not energetically feasible from a body $>$10 \% of Earth's mass (see section \ref{sec:physics_loss}). Alternatively, the material was only isotopically CI-like, but had similar and BSE-like elemental ratios (such as Cr/Zn), see above. \add{A final possibility states that ureilite- and CI-like material was accreted to the Earth in constant 60:40 proportions, before mixing to form a uniform reservoir, from which a fraction of the Zn budget was lost \citep{bizzarro2025}.} To this end, sublimation of accreting pebbles in H$_2$-rich planetary envelopes was proposed in order to explain the depletion of S relative to more refractory elements \citep{steinmeyer2023}, but is yet to be tested for a larger suite of elements. Moreover, this volatile loss must not have engendered any perceptible mass-dependent isotopic fractionation \citep[section \ref{sec:iso_MVEs},][]{sossi2022stochastic}. \delete{Therefore, we consider hypothesis A as less likely.}\add{Therefore, for hypothesis A to remain valid, it requires any elemental fractionation occurring in the H$_2$-rich envelope of the growing Earth to reproduce both its smooth decline in abundances
of volatile elements with condensation temperature, and their lack of
isotopic fractionation. No such results have yet been demonstrated.} 

Finally, owing to the relatively few studies conducted on Zn isotopic variations among meteorites \citep{steller2022nucleosynthetic,savage2022zinc,martins2023nucleosynthetic}, Fig. \ref{fig:Zn-Cr} shows that the BSE \add{is displaced from} \delete{also lies within 2-$\sigma$ uncertainty of an extension of the} trend defined by NC bodies \add{and Mars}. \add{However, the 2-$\sigma$ uncertainty is sufficiently large that the BSE falls close to its extension (envelope `C'), though more data are required to provide a statistical assessment.} \add{Alternatively, the apparent deviation of the BSE from the $^{66}$Zn-$^{54}$Cr trend might reflect the late-stage addition of volatile-rich NC material having a composition beyond the NC trend (i.e., unsampled NC material). Provided this material is volatile-rich compared to the composition of the proto-Earth, it would exert a stronger control on the BSE’s isotope composition for Zn than for Cr, a potential explanation why the BSE appears to diverge from the NC trend for these two elements.} This permits the possibility that the Zn and Cr isotopic compositions of the BSE are consistent with those of an NC body that lies to higher $\varepsilon^{66}$Zn and $\varepsilon^{54}$Cr than any existing NC meteorite group.



\subsection{Heavy elements}
\label{sec:heavy_elements}

These elements, defined as those with $Z~>~26$ (Fe), were synthesised by nuclear processes distinct from those of the lighter elements. In brief, 
nuclides form by neutron capture at different rates; the $s$-process (slow neutron capture) and $r$-process (rapid neutron capture), which produce roughly equal amounts of nuclides, and the less widespread $p$-process that gives rise to proton-rich nuclei. The rates of neutron capture (slow or rapid) are determined relative to the $\beta^-$-decay rate of the target nuclide, and depend on the nuclide's neutron capture cross section, which is highly temperature-dependent \citep{schrammnorman1976}. In general, $r$-process nucleosynthesis produces more neutron-rich nuclides of a given element relative to the $s$-process, as well as higher-$Z$ elements. 
Isotopic variations among heavy elements have been reported for, for example, Sr \citep{moynier2012,Hans2013rbsr,schneider2023Sr}, Zr \citep{akram2015,render2022}, Mo \citep{dauphas2002molybdenum,burkhardt2011molybdenum}, Ru \citep{chen2010ruthenium,fischergodde2015ru}, Nd and Sm \citep{qin2011nucleo, boyet2013,burkhardt2016nd,frossard2022earth}. \\

The BSE is an end-member among \textit{all} planetary materials in the isotopic compositions observed for heavy elements, as has been shown for Mo \citep{burkhardt2011molybdenum,budde2019molybdenum}, Zr \citep{akram2015,render2022}, Nd \citep{burkhardt2016nd}, and Ru \citep{fischergoddekleine2017}. Importantly, this implies that the BSE composition is inconsistent with any mixture of NC- and CC components currently available, an interpretation that is otherwise permitted from iron-peak elements alone (see section \ref{sec:iron-peak}). 
Specifically, for isotopic variations between an iron-peak element and a heavy element (e.g., $\varepsilon^{50}$Ti vs. $\varepsilon^{96}$Zr), the NC trend is \delete{approximately perpendicular}\add{transverse} to variations among CCs \citep[see][]{burkhardt2021terrestrial,render2022,yaptissot2023,kleinenimmo2024} such that no mixture of \delete{an}\add{a known} NC and a (known) CC body passes through the terrestrial composition. \add{Consequently, the processes that lead to variability among NC bodies were distinct from those that gave rise to variation among CC bodies, and the two populations cannot be considered as sampling different portions of a continuum, as in hypothesis A.} \\ 

\begin{figure}[!ht]
    \centering
    \includegraphics[width=0.66\linewidth]{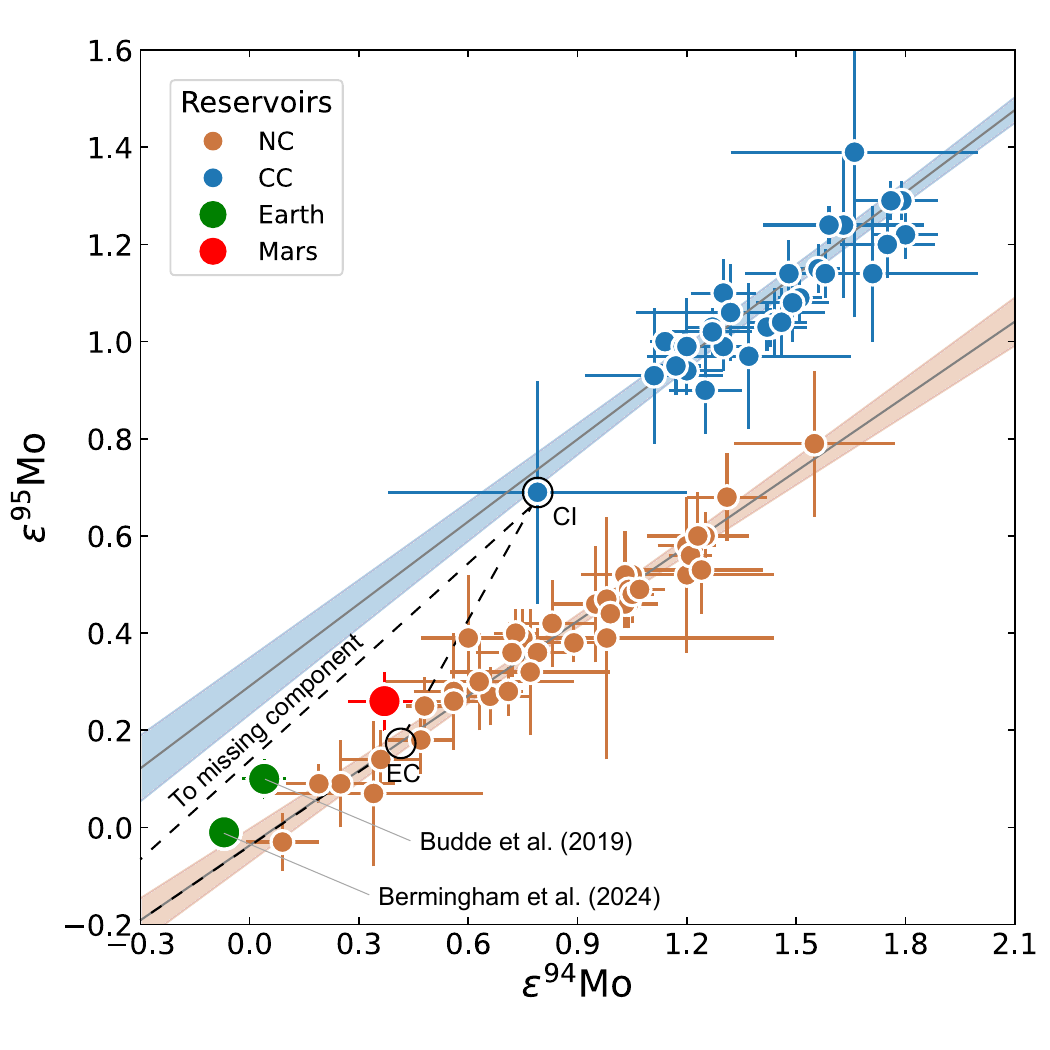}
    \caption{Compositions of CC meteorites (blue circles), NC meteorites (red circles), the Earth (green circle) and Mars (red circle) expressed in terms of their $\varepsilon^{94}$Mo and $\varepsilon^{95}$Mo isotopic compositions. Data from \cite{spitzer2025comparison}. The CC and NC trends were fit according to the York method \citep{york2004} assuming uncorrelated errors in \textit{x} and \textit{y}, and yield $\varepsilon^{95}$Mo$_{\mathrm{NC} }$ = (0.517$\pm$0.028)$\times~\varepsilon^{94}$Mo$_{\mathrm{NC}}$ - 0.045$\pm$0.026 and $\varepsilon^{95}$Mo$_{\mathrm{CC}}$ = (0.560$\pm$0.023) $\times~\varepsilon^{94}$Mo$_{\mathrm{CC}}$ + 0.298$\pm$0.037 (all uncertainties 2-$\sigma$; red and blue envelopes). Black circles indicate compositions of end-members in a three-component mixture reproducing the composition of the BSE in hypothesis B, in which the `missing component' has $\varepsilon^{94}$Mo = -0.60 and $\varepsilon^{95}$Mo = -0.30 \citep{burkhardt2021terrestrial}. In this scheme the contributions to the Mo budget of the BSE of \cite{budde2019molybdenum} can be reproduced by 50~\% missing component, 20~\% EC and 30~\% CI. The BSE value of \cite{bermingham2024} does not require a CC contribution within uncertainty\add{, but does not exclude it}.} 
    \label{fig:Mo-Mo}
\end{figure}

Isotopic variations induced by the $r$- and $s$-processes for a given isotopic ratio may often counteract one another, preventing an unequivocal assessment of their nucleosynthetic origin \citep{schneider2023Sr}. Molybdenum is an exception, because $r$- and $s$-process variations can be distinguished in $\varepsilon$-$\varepsilon$ including one of the two $p$-process Mo nuclides (i.e., $^{92}$Mo or $^{94}$Mo) \citep{burkhardt2011molybdenum}. The CC and NC meteorite groups in $\varepsilon^{94}$Mo-$\varepsilon^{95}$Mo space \citep{budde2016}, form two lines, which reflect $s$-process heterogeneities \citep{budde2016}, while the offset between the two lines reflects a near-constant $r$-process-excess in CC over NC materials (Fig. \ref{fig:Mo-Mo}).\\ 

\citet{spitzer2020}, based on Mo isotopic data representing 15 different parent bodies, including early- (i.e., iron meteorites, achondrites) and late-formed bodies (i.e., chondrites) found that the NC-line ($\varepsilon^{95}$Mo = [0.528$\pm$0.045]$\varepsilon^{94}$Mo -0.058$\pm$0.040) is slightly shallower than the CC-line ($\varepsilon^{95}$Mo = [0.596$\pm$0.006]$\varepsilon^{94}$Mo + 0.264$\pm$0.010).
More recently, \citet{bermingham2024} argued that the slope of the NC-line is shallower again ($\varepsilon^{95}$Mo = [0.517$\pm$0.042]$\varepsilon^{94}$Mo + 0.01$\pm$0.02); however, this result is not only based on a lower number of samples (six iron meteorites), but also is influenced by the Mo isotope data for two IAB irons, whose composition differs from that reported in an earlier study \citep{worsham2017}. The latest compilation of \cite{spitzer2025comparison}, yields $\varepsilon^{95}$Mo$_{\mathrm{NC} }$ = (0.517$\pm$0.028)$\times~\varepsilon^{94}$Mo$_{\mathrm{NC}}$ - 0.045$\pm$0.026 (95~\% confidence interval, Fig. \ref{fig:Mo-Mo}) based on data for 41 distinct NC bodies and including available data for IAB irons from different studies with an indistinguishable slope but a lower intercept than that of \cite{bermingham2024}.  \\ 

Based on Mo isotope data for several terrestrial igneous rocks, \citet{budde2019molybdenum} report $\varepsilon^{94}$Mo = 0.04$\pm$0.06 and $\varepsilon^{95}$Mo = 0.10$\pm$0.04 (95~\% confidence interval) for the BSE, \delete{which translates to} 
\delete{a CC source for 46$\pm$15~\% of the BSE's Mo, plotting close to, but distinct from the error envelope on the NC-line}
\add{indicating that in the $\varepsilon^{94}$Mo–$\varepsilon^{95}$Mo diagram the BSE plots between the NC- and CC-lines (Fig. \ref{fig:Mo-Mo}). From the position of the BSE in this diagram, these authors calculated that 46$\pm$15~\% of the BSE’s Mo derives from the CC reservoir.} \cite{budde2023} showed that there is some uncertainty on the BSE’s Mo isotope composition owing to non-exponential isotope fractionation, either intrinsic to the sample, during the measurement, or both, but also argued that this does not affect the fraction of CC material in the BSE, though uncertainties are larger. 
Conversely, \citet{bermingham2024}, based largely on data for molybdenites, argued that the BSE composition is different from that reported by \citet{budde2019molybdenum}, and quote $\varepsilon^{94}$Mo = -0.07$\pm$0.03 and $\varepsilon^{95}$Mo = -0.01$\pm$0.03. \add{Because this composition overlaps with that of a single IAB iron meteorite and, hence, is on the limit of resolution at 1-$\sigma$ from the NC line and within uncertainty at 2-$\sigma$ (Fig. \ref{fig:Mo-Mo}), these authors argue for a purely NC origin of the BSE’s Mo.}  

In hypothesis B, the fraction of CC-derived Mo is similar to the 30~\% CC-derived Zn in the BSE deduced by \citet{steller2022nucleosynthetic} and \citet{savage2022zinc}. \add{In this model, given that enstatite chondrites are taken as the NC end-member, the fraction of CC-derived Mo in the BSE can be calculated using $\Delta^{95}$Mo, which represents a measure of the position of a sample relative to the NC- and CC-lines. The BSE composition of \cite{bermingham2024} results in $\Delta ^{95}$Mo = 3$\pm$3 for the BSE; using $\Delta ^{95}$Mo = -7$\pm$4 for enstatite chondrites and $\Delta^{95}$Mo = 26$\pm$2 for the CC reservoir then results in 31$\pm$17\% CC-derived Mo in the BSE for hypothesis B, whereas using instead the value of \cite{budde2019molybdenum} yields 41±17\% CC-derived Mo in the BSE \citep[see][]{nimmo2024}. Therefore, hypothesis B indicates some of the BSE's Mo derived from CC material.}  
In this model, because Mo (siderophile) and Zn (volatile) are depleted in the BSE with respect to RLEs, their isotopic compositions do not reflect the bulk CC fraction in the Earth, but rather the late-stage addition of volatile-rich, CC material to the Earth \citep{kleinenimmo2024, nimmo2024}. 
In hypothesis B, the BSE's Mo isotopic composition is reconciled by a three-component mixture of CI, EC and $s$-process-enriched NC material \citep[the missing component, see Fig. \ref{fig:Mo-Mo},][]{burkhardt2021terrestrial, budde2019molybdenum}. \\

On the other hand, if the Mo BSE \delete{value}\add{interpretation} of \cite{bermingham2024} is correct, and if the isotopic composition of Zn is also taken to be consistent with that of an NC body \delete{(as permitted within 2-$\sigma$ uncertainty in Fig. \ref{fig:Zn-Cr})}, then this leads to a third\delete{, more speculative} hypothesis (C) in which the isotopic composition of the BSE is an entirely NC body whose composition is related to, but more extreme than other NC bodies. \add{It should be noted that, converting the fraction of the CC-derived Mo in the BSE from hypothesis B into an actual mass of CC bodies accreted by Earth is difficult, mainly because the Mo isotopic composition of the missing component in hypothesis B is not known precisely, and, since almost all the Mo of a body might be expected to reside in its metallic core, only a fraction of this Mo may now be recorded in the BSE. Thus, hypotheses B and C may eventually be reconciled if the mass fraction of the CC contribution to the BSE is small (i.e., within uncertainty).}
Hypothesis C naturally resolves the apparent contradiction in the CI-like Fe isotopic composition of the BSE (see section \ref{sec:iron-peak}) as well as the $s$-process excesses observed in nuclides of heavy elements such as Zr, \add{because, in both cases, the BSE is consistent with a linear extension of the NC trend \citep[see also][]{render2022}.} These ideas remain to be thoroughly tested.


\subsection{Implications for the astrophysical setting and redox state of the Earth and planets}
\label{sec:astrophys_sett}

\begin{figure}[!ht]
    \centering
    \includegraphics[width=1\linewidth]{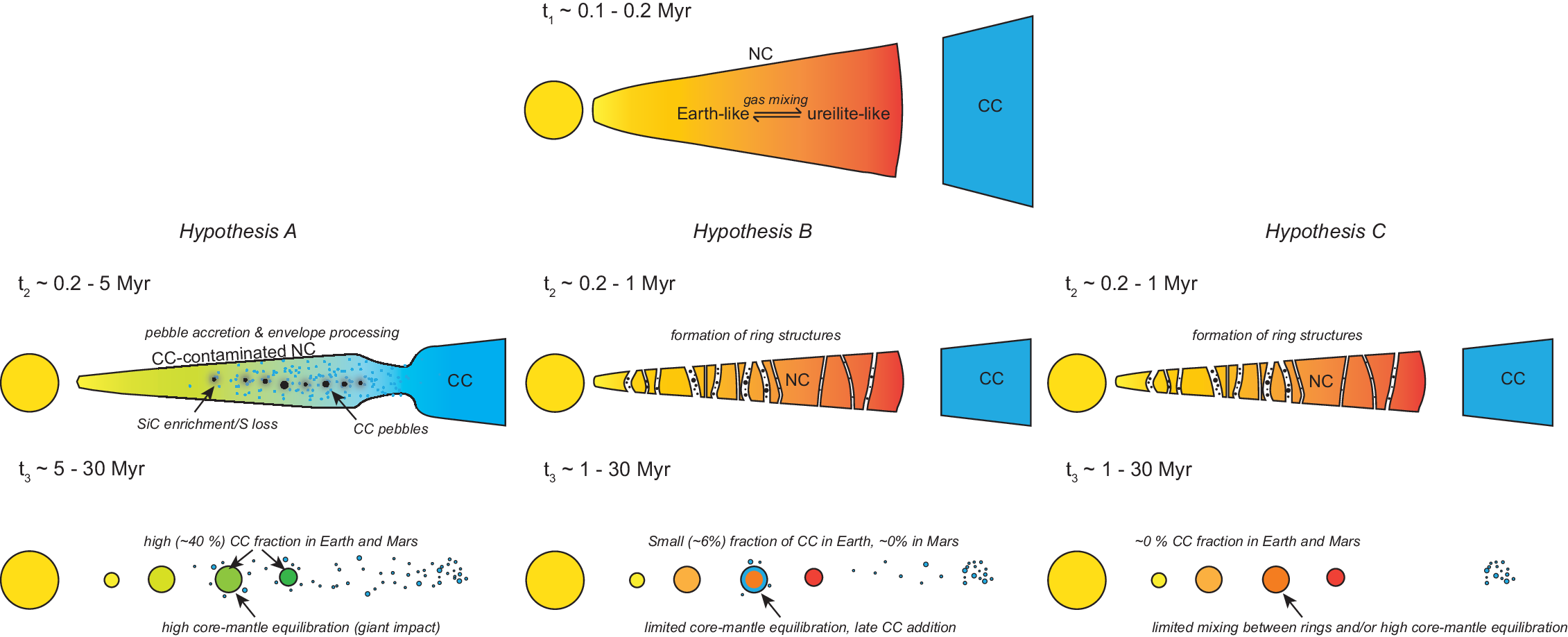}
    \caption{Schematic illustration of the astrophysical setting of planetary accretion through time.}
    \label{fig:astro_schematic}
\end{figure}

It is clear that hypothesis A requires considerable quantities of outer solar system material ($\sim$40~\%), hypothesis B requires a small fraction ($\sim$6~\%), while hypothesis C states that there is \add{essentially} none \add{within some as yet undefined uncertainty} ($\sim$0~\%). A schematic illustration as to how these scenarios could have played out spatio-temporally is shown in Fig. \ref{fig:astro_schematic}.

The initial conditions ($t_1$, Fig. \ref{fig:astro_schematic}) for all models invoke inner (NC) or outer (CC) solar system provenance, the approximate boundary between which is the present-day asteroid belt \citep{kruijer2017age}. Physical isolation of these two reservoirs is grounded in the observation that, isotopically, there is a dearth of planetary materials that show compositions intermediate to these groups. Isotopic variations in CIs, particularly in Fe \add{as well as Cr \citep{schiller2020iron,hopp2022ryugu,vankooten2024}}, have been used as a basis for the existence of a third reservoir \citep{yaptissot2023,dauphas2024}. Nevertheless, the CI reservoir is thought, initially, to have resided in the outer solar system. \\

The interpretations diverge at $t_2$ (cf. Fig. \ref{fig:astro_schematic}), where hypothesis A indicates significant inflow of CC material to the inner disk, which was initially sporadic to explain the compositions of early-formed NC planetesimals  \citep[ureilite parent body, angrite parent body,][]{schiller2018calcium} but ramped up over time, resulting in $\sim$40~\% CC in the fully-formed Earth. In hypothesis A, this reflects the inward drift of CI-like pebbles \citep[e.g.,][]{johansen2023anatomyI}. 
In hypotheses B and C, the contribution of CC material was either limited and occurred late during the accretion of the Earth ($t_3$, B) or was absent altogether ($t_3$, C). One mechanism for arresting inward drift would have been upon cooling of the disk and the formation of rings associated with pressure bumps \citep[section \ref{sec:disks},][]{whipple1972,carrera2021}\add{, compounded by the subsequent formation of giant planets \citep{lau2024sequential}} that hindered transport of condensing grains outside of heliocentrically restricted zones (Fig. \ref{fig:astro_schematic}, $t_2$). In this way, the rings could have evolved locally in terms of chemical compositions through $^{26}$Al heating and collisions within the rings, but would have undergone little mixing outside of these zones, consistent with the preservation of 
bodies in the NC trend with distinct isotopic compositions. \\

Mars does not show strong evidence for heterogeneous accretion \citep[in that its isotopic composition lies between EC and OC for most systems, see][]{dauphas2024,bizzarro2025,liebske2025mars} and also contains \textit{less} CC material than does the Earth \citep[see][at least for hypotheses A and B]{paquet2023origin,kleine2023inner}, despite its more proximal location to the presumed CC reservoir in the outer solar system. This poses a dynamical challenge, particularly because Mars formed earlier than did the Earth \citep{dauphas_pourmand2011} and would have therefore been able to accrete any CC material scattered inwards during the early growth and migration of Jupiter \citep[see][for a discussion]{kleinenimmo2024}. \add{As a result, pebble accretion is hard to reconcile with the CC-poor nature of Mars. The observations can, however, be explained if Earth but not Mars received a late CC-embryo impact, as discussed in \cite{nimmo2024}. Recent dynamical simulations \citep{branco2025dynamical} support this argument.} \\

Hypothesis C states that Mars and Earth are equally (and \add{essentially} entirely) NC, and that the Earth either i) accreted material of, on average, uniform provenance through time and/or ii) any pre-existing signature of heterogeneous accretion in the BSE was reset by perfect core-mantle equilibration, in which the isotopic composition of the mantle exchanges completely with that of the underlying core at the completion of Earth's accretion. 
Homogeneous accretion with respect to chemical composition is unlikely on the basis that increasing $f$O$_2$ during terrestrial core formation appears to be required by \cite{rubie2015accretion,rubie2025}, \citep[see discussion in section \ref{sec:internal_diff},][]{siebert2013}. \add{Although this does not \textit{per se} mandate a difference in the nucleosynthetic origin of such material, it would seem logical that chemical differences, especially in oxidation state, are linked to nucleosynthetic variations. On the other hand, $f$O$_2$ changes accompany changing $P-T$ conditions of chemically identical material (Sect.~\ref{sec:internal_diff}), such that this process cannot be used as an argument for accretion of compositionally or isotopically heterogeneous material. Therefore, while hypothesis B is consistent with many observations, hypothesis C, though promising, will require further investigation, in particular for the Zn and Mo isotope signatures of the BSE, and its ability to reproduce the BSE’s depletion in siderophile elements with homogeneous accretion.}\delete{but this does not mandate a difference in the nucleosynthetic origin of such material, even if it would seem logical that chemical differences are linked to nucleosynthetic variations. Therefore, while hypothesis B is consistent with many observations, hypothesis C cannot be ruled out and requires further investigation, in particular for Zn and Mo.}  \\

Importantly, hypothesis A appeals to CI material to oxidise the inner disk. 
Taking the bulk composition of CI chondrites, and assuming that all H associated with H$_2$O is lost while O remains behind, mixtures between an NC body (here EH) and CI chondrites are used to compute Fe/O ratios. For an Fe/O of 1.04 for EH and 0.40 for CI by mass \citep{wassonkallemeyn1988}, addition of 6~\% CI material gives a bulk Earth Fe/O ratio of 0.97, whereas 40~\% results in 0.70. The best estimate for the bulk Earth yields 1.07 (Table \ref{tab:abundances}), suggesting O must be lost, too, in hypothesis A to reproduce the composition of the Earth. This would imply mass loss of O-bearing rocky materials (e.g., the Earth's mantle), as the amount of O that binds with nominally volatile components (e.g., CO$_2$) is insignificant with respect to that stored in the mantle. Hypothesis C provides no information as to the oxidation state of the Earth, other than to imply that it must be intrinsic to the material of the inner solar system. 

\section{Summary, future directions and extension to other planets}
\label{sec:future}

Here we summarise some of the key findings from this work, and highlight possible future directions to understand terrestrial planet formation in our Solar System.

\begin{itemize}
    \item The major telluric bodies, in terms of their core-mantle ratios, are slightly- (Earth) to strongly (Moon, Mercury) non-chondritic, while small telluric bodies are more oxidised (i.e., have more Fe as FeO) than any chondrite. Consequently, the Mg and Fe concentrations of the Earth, Moon and Mercury cannot be explained by any mixture of condensed materials from a canonical solar nebula. Other processes, such as a non-solar bulk composition, physical sorting of metal/silicate grains and/or collisional erosion are required.

    \item Disk viscosities are much higher than those controlled by internal (molecular) viscosities in order to match the observed mass accretion rate of T-Tauri stars. The Sun must have lost angular momentum to the planets via early, likely magnetised outflows. The modern thinking on disks therefore states that they probably transitioned from hot and rapidly radially expanding ($>$ 1000~K inward of 1 AU) to cold and accreting very early on in their history (within a few 100~kyr). In the cold phase the snow line was inwards of 1 AU.

    \item Equilibrium and non-equilibrium condensation of the solar nebula effectively discriminates between elements according to their volatilities, owing to the limited range of temperatures over which the budget of a given element condenses (50--100 K). The compositions of small telluric bodies \textit{do} adhere to this expectation. However, the terrestrial planets do not exhibit the step-pattern expected for condensation/evaporation at a single temperature, indicating they accreted from a mixture of materials that experienced different thermal histories.
    
    \item Equilibrium condensation of the solar nebula predicts that there should be a correlation between the moderately volatile element content of the condensed grains and their oxygen content. This results from the condensation of S into FeS and additional O into FeO at similar temperatures (starting at $\sim$700 K and $\sim$550 K, respectively, independent of pressure). Because O condenses after S, this predicts that small telluric bodies and many iron meteorite parent bodies, which are almost exclusively oxidised (i.e., inferred mantle FeO contents $>$10~wt.~\%), should have their full complement of moderately volatile elements, which is not observed. Instead, this suggests either $i)$ alternative mechanisms for their oxidation and/or $ii)$ subsequent loss of highly volatile (H, C) and moderately volatile (S, Zn, etc.) elements at conditions distinct from those of the solar nebula. 
    
    \item The mantle FeO content of the Earth can be produced from initially reduced materials (all Fe as Fe$^0$), provided core-mantle equilibration occurred at mean conditions of $\sim$40~GPa and $\sim$3150~K, but requires higher Fe/O ratios ($\sim$1.06) than found in chondrites (0.4, CI to 1.04, EH) or in mixtures of equilibrium condensates ($\sim$0.81 to 0.93).

    \item Enrichment in heavy isotopes of moderately volatile elements (except for Cr, which tends to lighter values) on small telluric bodies is correlated to the extent of their elemental depletion, implying a Rayleigh process governing their loss. A correlation also exists between the isotopic fractionation and the present-day escape velocities of bodies over a range of masses (from Vesta to Earth). Evaporative loss is generally expected to result in the residues (planets) being enriched in heavier isotopes, but is not effective for bodies of Moon-mass or greater unless their atmospheres were H$_2$-dominated. 
    
    \item The inferred fractionation factors of mass-dependent stable isotope fractionation, together with the light isotope enrichment of Cr in small telluric bodies (Moon, Vesta, but not the angrite parent body) are indicative of the attainment of near-equilibrium between vapour- and condensed phase(s) during their separation. This high degree of equilibration implies that there is no means of discriminating between evaporation or condensation. 

    \item Volatile depletion from small telluric bodies (except the Moon) must have occurred within a few million years after the condensation of the first solids, and therefore likely in the presence of the nebular gas. Nevertheless, the ratios of moderately volatile elements (Mn/Na, K/Li, Mn/Mg) show evidence for evaporation under more oxidised conditions ($\Delta$IW-1) and at higher temperatures ($\sim$1400-1800~K) than in a canonical solar nebula. 

    \item The growth of planets is consistent with mixing of small bodies of variable composition through collisions to produce increasingly more massive bodies. This process reconciles the observation that small telluric bodies have step function-like abundance patterns together with isotopically heavy moderately volatile element stable isotope ratios, while more massive bodies have increasingly smooth abundance patterns, are more volatile-rich, and have chondritic stable isotope compositions. 


    \item A ureilite-like proto-Earth for Cr- and Ti isotopes implies 40~\% by mass CI chondrites to match the composition of the Earth. This delivers too many volatile elements with respect to the present-day Earth's mantle and would therefore require volatile loss thereafter. Furthermore, this loss would have had to have happened without significant fractionation of mass-dependent stable isotope compositions and the preferential loss of $s$-process depleted material to satisfy the Mo isotope composition of the BSE. Further work on volatile loss mechanisms from a protoplanet embedded in an H$_2$-rich disk is required to assess this hypothesis.

    \item An enstatite chondrite-like proto-Earth requires only 6~\% by mass CI chondrite addition, broadly in-line with the present-day abundances of the volatile elements in the BSE. So that this 6~\% of CI material delivers the $\sim$30--40~\% of the Mo and Zn budgets, the proto-Earth's mantle would have had to have had sub-CI Mo and Zn abundances (e.g. by core formation and volatile loss, respectively). This hypothesis implies the Earth accreted heterogeneously with respect to early-accreted inner (NC) material and late-accreted (CC) material. Because the BSE is an end-member with respect to the compositions of heavy elements (Zr, Mo and Ru), the NC component must contain a mixture of \add{on average} enstatite chondrite-like material and a missing component not preserved in the meteorite collection.

    \item \delete{Recent data indicate the Mo isotope composition of the BSE could be entirely NC in nature, within uncertainty. This interpretation depends on the samples used in the definition of the NC line. But if true, and \delete{when assuming the}\add{given that the }Zn\add{ and Fe} isotopic composition of the BSE is also consistent with that of an NC body, there would be no requirement for any significant CC contribution to the Earth. This observation may be satisfied by    \textit{i)} the random (stochastic) accretion with the same mean (NC) provenance, and/or \textit{ii)} differences in the provenance of material erased by perfect core-mantle equilibration thereafter.} \add{The evidence for a minor contribution of CC material to an otherwise NC Earth (hypothesis B) is based entirely on the BSE’s Zn and Mo isotope composition. For both some uncertainty exists in the estimated CC fractions, which is mostly related to uncertainties in the isotopic composition of the missing component accreted by Earth. Provided the CC fractions for these two elements are even smaller than currently estimated, as permitted by their uncertainties, the Earth may be considered an essentially pure NC body whose composition is more extreme than other NC bodies uniformly across all elements (hypothesis C).}


    \item  There must have been a population of oxidised, volatile-depleted bodies in the inner solar system, given that the Earth represents nearly 50 \% of its mass and is markedly poorer in both highly- and moderately volatile elements than chondrites. The oxidation states of these bodies must have been somewhat elevated ($\Delta$IW-1 to -2) with respect to the solar nebula ($\Delta$IW-6). The Earth and Mars obtained their moderately volatile element budgets largely ($\sim$60–100 \%, Earth) or almost entirely (Mars) from NC bodies, implying that oxidised, MVE-rich bodies were endemic to the inner disk. 
    
    \item One mechanism to prevent ingress of CC material into the inner disk is through trapping at pressure bumps\add{ and/or giant planet growth}. However, cooling of a solar composition gas to completion should still result in CI-like abundances of volatile elements, which are not observed in the terrestrial planets. As such, it \add{may} imply that relatively high temperatures were sustained in the inner disk until the dispersal of gas and/or this dispersal occurred earlier than it did in the CC-forming region.

    \item Evidence for a clear heliocentric gradient in composition preserved in the planets is incomplete, despite a decrease in core/mantle ratio with distance from the Sun. The composition of Mercury will be key to constraining the extent to which volatile elements were able to condense at small heliocentric distances. Additional constraints on its moment of inertia and core density will be instrumental to facilitate better estimates of its composition.


\end{itemize}

\textbf{Acknowledgements} \\
We thank the handling editor, Dominik Hezel, for his patience and persistence in helping us see this contribution over the finish line, and to two anonymous reviewers, who highlighted the role of dust transport, envelope processing, oxygen isotope constraints, and other factors in the cosmochemical evolution of the early solar system. This contribution was the fruit of an ISSI Meeting held in Bern in July, 2023, a follow-up held at ETH Zürich in October, 2023 and finally in IPG Paris in May, 2024. PAS is grateful to the Swiss National Science Foundation (SNSF) via an Eccellenza Professorship (203668) and
the Swiss State Secretariat for Education, Research and Innovation (SERI) under contract number MB22.00033, a SERI-funded ERC
Starting Grant ``2ATMO". FN acknowledges support from NSF-CSEDI-2054876 and NASA-EW-80NSSC18K0594. RCH acknowledges support from the European Research Council (ERC-STG-2020 grant 949417 - VapLoss) and the Italian Ministry of University and Research (NRRP M4C2 Investment 1.2 Young Researchers grant ERC-PI\_0000002 - VapourTime). \\

\textbf{Conflict of interest statement} \\
Not applicable.






\bibliographystyle{agu}
\bibliography{refs}

\end{document}